\documentclass[runningheads]{llncs}
\usepackage[T1]{fontenc}
\usepackage{url}
\usepackage{hyperref}

\usepackage{cite}
\usepackage{amsmath,amssymb,amsfonts}
\usepackage{algorithmic}
\usepackage{graphicx}
\usepackage{float}
\usepackage{longtable,array}
\usepackage{listings}
\usepackage{fontspec}
\newfontfamily\leanmono{DejaVuSansMono.ttf}[Scale=MatchLowercase]
\makeatletter
\lst@InputCatcodes
\def\lst@DefEC{%
 \lst@CCECUse \lst@ProcessLetter
  ⟨⟩·→∈∧∀βγ—∃Φ≠←∨↦α≤§₁¬…π×»«ρ∉Σ▸⇒⊢↔₂–∑λκΔ%
  ^^00}
\lst@RestoreCatcodes
\makeatother
\usepackage{textcomp}
\usepackage{xcolor}
\definecolor{leanblue}{RGB}{31,78,121}
\definecolor{leangreen}{RGB}{45,110,70}
\definecolor{leanpaper}{RGB}{247,248,250}
\lstdefinestyle{leanproofs}{
  basicstyle=\leanmono\scriptsize,
  backgroundcolor=\color{leanpaper},
  frame=single,
  rulecolor=\color{black!20},
  framesep=5pt,
  breaklines=true,
  columns=fullflexible,
  keepspaces=true,
  showstringspaces=false,
  escapeinside={(*@}{@*)},
  literate={{-}{{-}}1},
  aboveskip=7pt,
  belowskip=9pt
}
\usepackage{tikz}
\usepackage[disable]{todonotes}
\usetikzlibrary{positioning,arrows.meta}

\newcommand{\hash}{H}
\newcommand{\mr}{\mathsf{mr}}               
\newcommand{\mup}{\mathsf{up}}              
\newcommand{\oroot}{\varrho}                
\newcommand{\mmr}{\mathsf{mmr}}            
\newcommand{\mmrup}{\mathsf{mmrup}}        
\newcommand{\sk}{\mathit{sk}}
\newcommand{\pk}{\mathit{pk}}

\newcommand{\tx}{T}
\newcommand{\txid}{\mathsf{txid}}
\newcommand{\cid}{\mathsf{cid}}
\newcommand{\nul}{n}                        
\newcommand{\dg}{\mathit{dg}}               
\newcommand{\salt}{\rho}
\newcommand{\fee}{f}                       

\newcommand{\blob}{\beta}
\newcommand{\pl}{\lambda}                 
\newcommand{\cc}{\kappa}                    
\newcommand{\bid}{\mathsf{bcid}}              
\newcommand{\conf}{K}                       

\newcommand{\hdr}{\mathit{hdr}}
\newcommand{\anchor}{\hdr_{\mathrm a}}         
\newcommand{\hinc}{\hdr_{\mathrm{in}}}       
\newcommand{\blobsRoot}{\mathsf{blobsRoot}}
\newcommand{\histRoot}{\mathsf{histRoot}}

\newcommand{\InChain}{\mathsf{InChain}}
\newcommand{\Invalid}{\mathsf{InvalidBlob}}

\newcommand{\Rmem}{\Phi_{\mathrm{mempool}}}
\newcommand{\Rblob}{\Phi_{\mathrm{blob}}}
\newcommand{\Rcoin}{\Phi_{\mathrm{coin}}}
\newcommand{\pf}{\pi}

\usepackage{mdframed}
\newenvironment{relationbox}[1]{%
  \begin{mdframed}[linewidth=0.4pt,nobreak=true,
    innerleftmargin=3pt,innerrightmargin=3pt,innertopmargin=4pt,innerbottommargin=4pt,
    leftmargin=0.025\linewidth,rightmargin=0.025\linewidth,
    skipabove=8pt,skipbelow=8pt]\small\setlength{\parindent}{0pt}%
  \textbf{#1}\par\smallskip}{%
  \end{mdframed}}

\floatstyle{ruled}
\newfloat{algorithm}{tbp}{loa}
\floatname{algorithm}{Algorithm}
\newenvironment{algobox}[1]{%
  \par\vspace{6pt plus 2pt minus 2pt}\noindent\begin{minipage}{\linewidth}%
  \hrule height 0.8pt\vspace{2pt}%
  \refstepcounter{algorithm}{\sloppy\noindent\textbf{Algorithm~\thealgorithm} #1\par}\vspace{1pt}\hrule\vspace{2pt}%
  }{%
  \par\vspace{1pt}\hrule\end{minipage}\par\vspace{6pt plus 2pt minus 2pt}}

\begin{document}
\title{Compact Shielded CSV: Post-Quantum, Private, Lightweight Client-Side Validation Blockchain
}
\titlerunning{Compact Shielded CSV}


\author{Dragos Ioan Ilie\inst{1} \and
Uri Lee\inst{2} \and
Iain Stewart\inst{2} \and
Jonathan Zhu\inst{2} \and
Elliot Jones\inst{2} \and
William J. Knottenbelt\inst{2}}
\authorrunning{D.~I. Ilie et al.}
\institute{Orbs\\
\email{dragos@orbs.com} \and
Imperial College London, London, UK\\
\email{\{u.lee22,i.stewart,jonathan.zhu25,e.jones24,w.knottenbelt\}@imperial.ac.uk}}
%
%
%
\maketitle              
\vspace{.5cm}
\begin{abstract}
We propose \textsc{Compact Shielded CSV}, a private client-side validation blockchain for peer-to-peer payments designed for the post-quantum era.  
Upgrading existing blockchains to quantum-resistant cryptography substantially 
increases on-chain overhead. By keeping 
all large cryptographic artifacts off-chain, \textsc{Compact Shielded CSV}
keeps a minimal on-chain footprint independent of the size of the underlying cryptographic proofs and signatures. For a single input transaction, the on-chain footprint is just 3 hashes ($3 \times 32 \text{ bytes}$): a
nullifier, a degriefer, and a commitment to the transaction.
We introduce the \textit{degriefer}~-- a novel mechanism that enforces ownership and prevents double-spending 
using only hash commitments, eliminating the need for on-chain signatures entirely. 
These properties make \textsc{Compact Shielded CSV} a promising foundation for private, 
scalable, post-quantum digital payments.
\keywords{\sloppy post-quantum, quantum-resistant blockchain, client-side validation, zero-knowledge proofs, degriefer, PCD}
\end{abstract}

\section{Introduction}
The post-quantum era presents a concrete and structural problem for blockchain design. Quantum-resistant signature schemes and proofs are orders of magnitude larger than their classical counterparts. We introduce \textsc{Compact Shielded CSV}, designed to retain maximal privacy alike other shielded privacy blockchains, but with a very minimal on-chain footprint, enhancing scalability by minimising global validation overhead. First, we eliminate all large cryptographic objects (both proofs and signatures) from the chain~-- they are only ever needed in off-chain structures. We reduce the on-chain footprint to 3 SHA-256 hashes for a single input transaction ($2N+1$ hashes for $N$ inputs). The on-chain footprint thus remains the same independent of any size growth due to post-quantum migration of the underlying cryptographic primitives.  
Second, we introduce the \textit{degriefer}, a novel mechanism that replaces the signature, proving that the legitimate owner controls the input, and to demonstrate that a given spend attempt is invalid (to prevent \textit{griefing} by a blockchain observer).

\subsection{Motivation}

\subsubsection{The Looming Quantum Threat}
\label{sec:quantum-threat}


The arrival of a cryptographically relevant quantum computer~-- capable of breaking the public-key cryptography and ECDSA (or similar) schemes underlying most blockchain infrastructure~-- is rapidly transitioning from a theoretical \textit{if} to an engineering \textit{when} \cite{google2024qecbelowthreshold, google2023suppressing, neven2024willow, bravyi2024high}. The implications for decentralized ledgers will be catastrophic~-- such as the ability to forge digital signatures, hijack pending transactions in the mempool, and drain the users' funds.

\subsubsection{Post Quantum Cryptography}
Post-quantum cryptographic primitives will be significantly larger than their classical counterparts \cite{mallick2026quantumdisruptionsokpostquantum}. 
 Groth16, a classical ZK-proof~-- compresses proofs to a mere 128 bytes~-- and migrating to post-quantum alternatives is expected to increase roughly $125\times$ to $1,000\times$ \cite{ishai2021shorter, buterin2017starks1, buterin2018starks3}. 
 Similarly, standard ECDSA signatures are typically 64 bytes \cite{10.1007/s102070100002}. The post-quantum signature sizes are expected to increase in the region of $\sim3\times$ \cite{cryptoeprint:2020/1240} to $40\times$ \cite{raavi2021security, nist2024fips204} post-quantum, with some 
 signature schemes which are more than $100\times$ \cite{shim2023suitability, chase2017post} depending on the signature family and their parameters. Although research is on-going on post-quantum signatures and proofs, it is clear the data requirements will increase if we are to maintain existing security guarantees and properties.






\subsection{Related Work}

\paragraph{ZCash} 
ZCash \cite{hopwood2016zcash} addressed the privacy problem prevalent in transparent blockchains such as Bitcoin by using zero-knowledge proofs. A transaction, consisting of nullifiers (opaque references to unspent inputs), notes (outputs), and a ZK-proof linking the nullifiers and notes. The ZK-proof, posted on-chain, allowed all transaction details, such as the amounts, to be hidden to third-parties whilst allowing anyone to verify its validity. 





\paragraph{Client Side Validation (CSV)}

Client Side Validation, originally proposed by Todd \cite{todd2013disentangling}, shifts the responsibility of validating transactions from a global to a local responsibility. The full transaction contents and the corresponding proof of validity are kept entirely off-chain, with the proof of coin validity passed from sender to receiver. The blockchain acts primarily as a ordered sequence \cite{todd2013disentangling} of coins being spent which allows for significant reductions in computational overhead as nodes in the blockchain no longer need to validate every single transaction. In addition, this design naturally complements privacy, the full transaction data can be exchanged peer to peer rather than validated and posted globally. 

\paragraph{Shielded CSV}
Shielded CSV \cite{nick2025shielded}, a client-side validation blockchain built on top of Bitcoin, achieves remarkable reductions in on-chain data and computational overhead of the network. In Shielded CSV, coin history is conveyed peer-to-peer via \textit{recursive} zero-knowledge proofs \cite{chiesa2010proof} such that proof size remains constant regardless of transaction history length. 
On-chain, each block consists of a list of nullifiers alongside an aggregate Schnorr signature constructed using Non-Interactive Schnorr Signature Half-Aggregation with Commitments (NISSHAC), compacting multiple user signatures into a single verifiable structure and reducing on-chain data to approximately 64 bytes per transaction. 
Despite its achievements, the post-quantum version of such a system remains unclear \cite{nick2025shielded}~-- Shielded CSV's Schnorr signatures can be aggregated but are not post-quantum, and it is not clear how big a post-quantum signature with Schnorr-style aggregation capabilities will need to be.





\section{The Protocol}
\label{sec:protocol}\label{sec:objects}\label{sec:algorithms}\label{sec:security}
The principal contribution of Compact Shielded CSV is to move coin spend authorisation off chain.
In nearly every blockchain, spend authorisation is enforced by the network before transactions are accepted: Bitcoin uses signatures inside locking scripts~\cite{nakamoto2008bitcoin}, Zcash publishes a zero-knowledge proof for every shielded input~\cite{hopwood2016zcash}, and Shielded CSV includes a signature alongside nullifiers it records~\cite{nick2025shielded}.
Such authorisation data is already a substantial part of the on-chain footprint, and in a post-quantum setting it will dominate it.
To avoid this bloat, Compact Shielded CSV does not require the authorisation of a spend to be validated by the network at all, thereby removing the need for signatures or proofs to be posted on chain.
Instead, Compact Shielded CSV allows for publishing a nullifier multiple times, each paired with a \emph{degriefer}: a commitment binding the nullifier to the owner's secret key and to the spending transaction.
Off-chain validation uses the degriefer to identify the owner authorised spend among conflicting duplicates, so double spending within a chain history remains impossible, even without any on-chain verification.

The protocol therefore has two layers: a \emph{blockchain} (\S\ref{sec:consensus}) in which the network records blobs of data, and an \emph{off-chain protocol} (\S\ref{sec:csv}) that specifies how users and miners rely on off-chain structures (\S\ref{sec:structures}) to be convinced of payments.

Throughout, $\hash:\{0,1\}^*\to\{0,1\}^\lambda$ is a collision-resistant hash function, with tuples and lists serialised canonically before hashing.
We write $\mr(L)$ and $\mmr(L)$ for the Merkle~\cite{merkle1987digital} and Merkle mountain range~\cite{todd2012mmr,bunz2020flyclient} roots of a list $L$, while $\mup(\ell,k,\pf)$ and $\mmrup(\ell,k,\pf)$ compute the root from the leaf $\ell$ at index $k$ along the branch $\pf$ for each structure.

\subsection{Blockchain}
\label{sec:consensus}\label{def:chain}\label{sec:acceptance}

Since transactions are client side validated, the Compact Shielded CSV blockchain is a \emph{data-availability layer}: it maintains an ordered, append-only sequence of blocks, each recording a list of \emph{blobs}.
The ordering of blocks is left to the consensus algorithm of choice; the protocol prescribes only the structure of a blob and the manner in which the block's header commits to the list of blobs.

\paragraph{Blobs.}
A \emph{blob} is a data structure $\blob=\big(t,[(\nul_1,\dg_1),\dots,(\nul_k,\dg_k)]\big)$ with $k\ge0$ and $t,\nul_i,\dg_i\in\{0,1\}^\lambda$.
In an honest blob, $t$ is a commitment to a transaction and each pair $(\nul_i,\dg_i)$ concerns one coin that the transaction spends: the coin's \emph{nullifier}, $\nul_i$, marks the coin as spent, and the \emph{degriefer}, $\dg_i$, authorises this particular spend.
The network has no way to tell a well-formed blob from one made of random hashes, so the off-chain protocol that users and miners follow (\S\ref{sec:csv}) resolves this by verifying zero knowledge proofs.

\paragraph{Blocks.}
\label{def:hdr}
The blockchain is the sequence of blocks $B_0,B_1,\dots$, of which $B_0$ is genesis; we write $\overline{B}_h=[B_0,\dots,B_h]$ for the blockchain with tip $B_h$.
A block $B_h=(\hdr_h,\allowbreak[\blob_0,\dots,\blob_m])$ consists of a header $\hdr_h=(h,\allowbreak\ \mathsf{parent},\allowbreak\ \blobsRoot,\allowbreak\ \histRoot,\allowbreak\ \dots)$ and a list of blobs, with $\blob_0$ known as the \emph{coinbase} blob.

\paragraph{Conflicts.}
Two blobs $\blob$ and $\blob'$ \emph{conflict}, written $\mathsf{Conflict}(\blob,\blob')$, if they share a nullifier.
Since a nullifier marks one coin as spent, at most one of the blobs that conflict can be a valid spend, and an off-chain proof claiming a blob valid must show every preceding conflicting blob invalid.
The network facilitates the construction of such a proof by committing to each blob together with its list of conflicts, $\conf$, via a \emph{blob contextual identifier}:
$$\bid_{\blob,\conf}=\hash\big(t,\hash(P),\hash(\conf)\big),\qquad\text{where }\blob=(t,P).$$

\paragraph{Consensus rules.}
\label{alg:accept}\label{def:accept}\label{def:blobsroot}\label{def:conflist}\label{def:id}\label{def:record}
A node holding $\overline{B}_{h-1}$ accepts a block $B=(\hdr,[\blob_0,\dots,\blob_m])$ as $B_h$, extending its chain to $\overline{B}_h$, if and only if:
\begin{itemize}\setlength{\itemsep}{2pt}
\item[(C1)] \emph{$B$ extends the chain:} $\hdr.h=h$, $\hdr.\mathsf{parent}=\hash(\hdr_{h-1})$ and $\hdr.\histRoot=\mmr\big([\hash(\hdr_0),\dots,\hash(\hdr_{h-1})]\big)$.
\item[(C2)] \emph{No blob names a nullifier twice:} for each $\blob_i=\big(t,[(\nul_1,\dg_1),\dots,(\nul_k,\dg_k)]\big)$, $p\neq q\Rightarrow\nul_p\neq\nul_q$.
\item[(C3)] \emph{The coinbase blob carries no pairs:} $\blob_0=(t_0,[\,])$. Coinbase consumes no coins.
\item[(C4)] \emph{$\blobsRoot$ is valid:} $\hdr.\blobsRoot=\mr\big([\bid_{\blob_0,\conf^h_0},\dots,\bid_{\blob_m,\conf^h_m}]\big)$, where 
{\setlength{\abovedisplayskip}{3pt}\setlength{\belowdisplayskip}{3pt}
\[
\conf^b_j=\big[\,\bid_{\blob^{b'}_{j'},\conf^{b'}_{j'}}\ :\ B_{b'}\in\overline{B}_h,\ \blob^{b'}_{j'}\in B_{b'},\ (b',j')<(b,j),\ \mathsf{Conflict}(\blob^{b'}_{j'},\blob^b_j)\,\big]
\]}%
is the \emph{conflict list} of, $\blob^b_j$, the $j$-th blob of the $b$-th block in $\overline{B}_h$.
\end{itemize}

The conflict list $\conf$ of a blob is the list of the contextual identifiers of the earlier blobs that conflict with it, in chain order.
Those blobs precede it, so their own conflict lists, and with them their identifiers, are already fixed when the blob is included; the identifiers are thus well defined by induction over the positions in the chain, and rule (C4) gives the list exactly.

This is the entirety of the consensus rules of Compact Shielded CSV.
Notice that no proof and no signature is validated; the bulk of the computation is making sure that the conflict lists are properly recorded, which a miner supports by indexing the blobs of its chain by nullifier, so that the conflict list of a new blob can be resolved with some lookups.

\subsection{Degriefers and other structures}
\label{sec:structures}\label{def:degriefer}

The concept of \textit{degriefer} is our novel contribution to privacy chain technology. With on-chain proofs and signatures, an unspent output's nullifier cannot be replayed by a would-be \textit{griefer} who wants to block a legitimate spender's correct usage of that nullifier (the proof would be wrong). If we just drop on-chain proofs and signatures, griefing would become possible: a griefer who sees a nullifier in the mempool could place a competing usage thereof on-chain and wreck the legitimate spender's attempt to spend that output. A degriefer is an on-chain object which, like a proof or signature, achieves the avoiding of such a replay attack by a griefer; but \textit{unlike} a proof or signature, it is tiny~-- one hash!

A user holds a secret key $\sk\gets\{0,1\}^\lambda$ and the public key $\pk=\hash(\sk)$.
A degriefer binds a nullifier, under the key of its owner, to the transaction that spends it.

\begin{definition}[Degriefer]\label{def:dg}
The degriefer of a nullifier $\nul$ under a secret key $\sk$, for the transaction with identifier $t$, is $\dg_{\nul,\sk,t}=\hash(\nul,\sk,t)$.
\end{definition}

Since forming a degriefer requires $\sk$, a griefer trying to publish $\nul$ for a different transaction $t'$ cannot construct the correct degriefer and the owner can prove this off-chain when proving its own spend of $\nul$.

\paragraph{Transactions.}
A transaction $\tx=(N,O)$ is a tuple of a list $N=[\nul_1,\dots,\nul_k]$ of the nullifiers of input coins and a list $O=[o_1,\dots,o_m]$ of the output coins it creates.
Each output $o=(v,\pk,\salt)$ is a tuple of an amount value, a public key and a salt chosen randomly by the creator $\salt\gets\{0,1\}^\lambda$.
The coinbase transaction has no input coins: $N=[\,]$.
With no nullifiers to distinguish them, two coinbase transactions with the same outputs would have the same identifier in different blocks. As in Bitcoin's BIP~34~\cite{bip34}, the coinbase identifier therefore commits to the height $h$ of its block instead.
The transaction identifier is defined in the context of a list $\conf$ of blob contextual identifiers:
{\setlength{\abovedisplayskip}{4pt}\setlength{\belowdisplayskip}{4pt}%
\[
\txid_{\tx,\conf}=
\begin{cases}
\hash\big(h,\ \mr([\hash(o_1),\dots,\hash(o_m)]),\ \hash([\,])\big) & \text{if } N=[\,],\\[2pt]
\hash\big(\hash(N),\ \mr([\hash(o_1),\dots,\hash(o_m)]),\ \hash(\conf)\big) & \text{otherwise},
\end{cases}
\]}%
where $h$ is the height of the block whose coinbase $\tx$ is.\label{def:txdef}\label{def:cbtxdef}

\paragraph{Coins.}
A transaction output (also called a coin) has a unique coin identifier that commits to it without revealing the transaction or the index of the output in the transaction. The $j$-th output of the transaction with identifier $t$ has the coin identifier $\cid_{t,j}=\hash(t,\ j,\ \salt_j)$, where $\salt_j$ is the salt of the output.
The salt provides privacy to the sender by hiding the link between $\cid$ and $t$. 
Without it, a recipient observing the chain would be able to identify the transaction by hashing all transaction identifiers with feasible output indexes.
The index $j$ ensures the coin identifiers of different outputs within a transaction are distinct even if their creator reuses the salt, thereby preventing the Faerie Gold issue~\cite{hopwood2016zcash}.
\label{def:cid}

\paragraph{Nullifiers.}
The nullifier of the coin with identifier $\cid$ under the secret key $\sk$ of its owner is $\nul_{\sk,\cid}=\hash(\sk,\cid)$.\label{def:nul}
The purpose of the nullifier is to hide the recipient's spend of the coin from the coin's creator.
The creator knows $\cid$, having created the output, but without the recipient's $\sk$ it cannot compute the nullifier, and so cannot recognise on-chain when the recipient spends the coin.

\paragraph{Invalid blob witnesses.}
The structures above allow a legitimate user to invalidate any blob that conflicts with its own honest blob.
Each reason a conflicting blob may arise can be invalidated in the following way:
\begin{itemize}\setlength{\itemsep}{1pt}
\item \emph{A blob with a wrong degriefer} can be shown invalid by using the preimage of the associated nullifier ($\sk$ and $\cid$) and showing that the actual correct degriefer for the nullifier and transaction commitment (which only the owner can compute) is different than the one in the blob (a \emph{pair witness}).
\item \emph{A blob that copies a nullifier, its degriefer and the identifier $t$ from the user's blob, but not all the nullifiers $t$ commits to,} can be shown invalid by opening $t$ and showing that the hash of the nullifiers in its preimage differs from the hash of the nullifiers in the blob (a \emph{txid witness}).
\item \emph{The user's own blob, included after a conflict the user did not commit to,} can be shown invalid in the same manner: opening $t$ shows that the conflict list in its preimage differs from the conflict list the blob contextual identifier was actually computed against (again a \emph{txid witness}).
\end{itemize}
We accordingly define the predicate $\Invalid$ on a blob contextual identifier and a witness establishing one of the above.

\begin{definition}[Invalid blob]\label{def:invalid}\sloppy
A tuple $w=(\blob,\cc,u)$, with $\blob=\big(t,[(\nul_i,\dg_i)]_{i=1}^{k}\big)$, $N=[\nul_i]_{i=1}^{k}$ and $u$ a \emph{pair witness} $(p,\sk,\cid)$ or a \emph{txid witness} $(N',\oroot,\cc')$, is an invalid blob witness for a contextual identifier $\bid$ iff $\Invalid(\bid,w)$, where
{\setlength{\abovedisplayskip}{4pt}\setlength{\belowdisplayskip}{4pt}%
\begin{multline*}
\Invalid(\bid,w)\iff \bid=\hash\big(t,\ \hash([(\nul_i,\dg_i)]_{i=1}^{k}),\ \cc\big)\ \wedge{}\\
\begin{cases}
\nul_p=\nul_{\sk,\cid}\ \wedge\ \dg_p\neq\dg_{\nul_p,\sk,t} & \text{if } u=(p,\sk,\cid),\\[2pt]
t=\hash\big(\hash(N'),\oroot,\cc'\big)\ \wedge\ (N',\cc')\neq(N,\cc) & \text{if } u=(N',\oroot,\cc').
\end{cases}
\end{multline*}}%
\end{definition}

\subsection{Off-chain payment protocol}
\label{sec:csv}

Users of Compact Shielded CSV perform transactions by publishing commitments in blobs and then relying on a composable proof system to convince each other that a coin exists.
At the centre of the protocol is the \emph{coin proof}: a zero-knowledge proof that a coin of a given value exists in the chain and belongs to a public key.
It certifies that the coin was created by a valid transaction: one whose inputs were coins with valid coin proofs of their own, none of them double-spent, and whose outputs (and fee) sum to its inputs.
The proofs are recursive: a coin proof is built from the coin proofs of the transaction's inputs, which were built in turn from those of their own inputs, so that every coin proof reaches back, transfer by transfer, to the coinbase transactions that minted its value.
Verifying one proof verifies this entire history at once, at a cost independent of its length, while zero knowledge keeps all of it hidden.

We first present the zero-knowledge circuits that constitute the composable proof system, and then the protocols that users and miners follow to verify these proofs against on-chain data.

\subsubsection{Zero-knowledge circuits}
\label{sec:pcd}\label{sec:proofs}
We formalise how the proofs compose as a zero-knowledge proof-carrying data (PCD) system~\cite{chiesa2010proof,bitansky2013recursive,bunz2021proof}.
Its prover $\mathcal P$ takes private \emph{local data} $w$ and \emph{incoming messages} $z_1,z_2,\dots$, each with its proof, and outputs a \emph{message} $z$ with a succinct proof $\pf$ that $\Phi(z,w,[z_i]_i)$ holds for a fixed \emph{compliance predicate} $\Phi$ and that every step behind the incoming messages was compliant too; its verifier $\mathcal V(z,\pf)$ accepts exactly then, at a cost and proof size independent of the number of steps, and learns nothing beyond $z$.

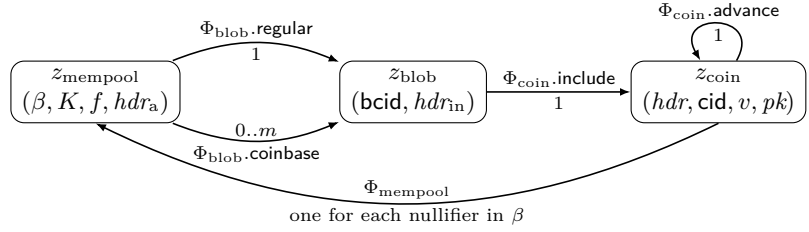
\begin{figure}[!htb]
\centering
\begin{tikzpicture}[
  msg/.style={draw, rounded corners=5pt, align=center, inner xsep=4pt, inner ysep=3pt, font=\small},
  ar/.style={-{Latex[length=1.4mm]}, semithick},
  lb/.style={font=\scriptsize, inner sep=1.5pt, align=center}
]
\node[msg] (mem) {$z_{\mathrm{mempool}}$\\$(\blob,\conf,\fee,\anchor)$};
\node[msg, right=21mm of mem] (blob) {$z_{\mathrm{blob}}$\\$(\bid,\hinc)$};
\node[msg, right=19mm of blob] (coin) {$z_{\mathrm{coin}}$\\$(\hdr,\cid,v,\pk)$};
\draw[ar] (mem) to[bend left=22] node[lb, above]{$\Rblob.\mathsf{regular}$} node[lb, below]{$1$} (blob);
\draw[ar] (mem) to[bend right=22] node[lb, below]{$\Rblob.\mathsf{coinbase}$} node[lb, above]{$0..m$} (blob);
\draw[ar] (blob) -- node[lb, above]{$\Rcoin.\mathsf{include}$} node[lb, below]{$1$} (coin);
\draw[ar] (coin.south) to[bend left=26] node[lb, above]{$\Rmem$} node[lb, below]{one for each nullifier in $\blob$} (mem.south);
\draw[ar] (coin) to[out=60, in=120, looseness=4] node[lb, above]{$\Rcoin.\mathsf{advance}$} node[lb, below]{$1$} (coin);
\end{tikzpicture}
\caption{How the circuits compose messages. Arrows carry the circuit and how many messages it consumes; $m$ is the number of blobs in a block besides the coinbase.}
\label{fig:compose}
\end{figure}

Compact Shielded CSV has three types of message, and $\Phi$ is accordingly given by three circuits, $\Rmem$, $\Rblob$ and $\Rcoin$, presented in Figure~\ref{fig:compose}. Any quantum-resistant proof system, e.g.\ a hash-based recursive one~\cite{chiesa2020fractal}, can instantiate it.

\begin{sloppypar}
The circuits ground their statements in block headers. Throughout this section we prove a block is in the history of another with: $\InChain((\hdr',\pf),\hdr)\iff\hdr'=\hdr\ \vee\ \mmrup(\hash(\hdr'),\hdr'.h,\pf)=\hdr.\histRoot$, i.e.\ $\hdr'$ is $\hdr$ itself or $\pf$ opens the history root of $\hdr$ to $\hdr'$.
\end{sloppypar}

\paragraph{Mempool message.}
\label{def:blob}\label{def:blobdef}\label{def:genuine}\label{def:forged}
The mempool message $(\blob,\conf,\fee,\anchor)$ attests that $\blob$ is a valid blob in any chain extending $\anchor$ at any position where its conflict list is $\conf$, and that the miner who includes it there collects the fee $\fee$.

\begin{relationbox}{$\Rmem$ --- message $(\blob,\conf,\fee,\anchor)$}
\textit{Incoming:} one coin message $(\hdr_i,\cid_i,v_i,\pk_i)$ per input, $1\le i\le k$.\\
\textit{Local:} the transaction $\tx=([\nul_1,\dots,\nul_k],O)$, keys $\sk_1,\dots,\sk_k$, chain proofs $\pf_1,\dots,\pf_k$, and an invalid blob witness $w_r$ per identifier $\bid_r$ in $\conf=[\bid_1,\dots,\bid_c]$.\\
\textit{Verification:}
\begin{enumerate}\setlength{\itemsep}{1pt}
\item $t=\txid_{\tx,\conf}$ --- the transaction identifier in $\blob$ is that of $\tx$ at $\conf$;
\item $\nul_i=\nul_{\sk_i,\cid_i}$ for each $i$ --- the nullifiers of $\tx$ are those of the incoming coins;
\item $\dg_i=\dg_{\nul_i,\sk_i,t}$ for each $i$ --- the degriefer is valid;
\item $\blob=\big(t,[(\nul_1,\dg_1),\dots,(\nul_k,\dg_k)]\big)$ --- the blob carries exactly $t$ and these pairs;
\item $\pk_i=\hash(\sk_i)$ for each $i$ --- the author owns the incoming coins;
\item $\InChain((\hdr_i,\pf_i),\anchor)$ for each $i$ --- the incoming coins lie in the chain of $\anchor$;
\item $\Invalid(\bid_r,w_r)$ for each $r$ --- all conflicts are invalidated;
\item $\sum_{i=1}^{k}v_i=\fee+\sum_{o_j\in O}v_j$ --- the inputs match the outputs and the fee.
\end{enumerate}
\end{relationbox}

\paragraph{Blob message.}
The blob message $(\bid,\hinc)$ attests that the blob with identifier $\bid$ is included in the block with header $\hinc$ and is valid there.
A blob message can be constructed in two ways.
For a regular blob, the user can derive it from the corresponding mempool message once a miner includes the blob in the chain.
For a coinbase blob, the miner derives it from the mempool messages of all the other blobs in the block, so its validity requires that every one of them has a valid mempool proof.
This gives miners a strong economic incentive to behave honestly, as misbehaving forfeits the entire block subsidy and all fees.

\begin{relationbox}{$\Rblob$ --- message $(\bid,\hinc)$}
\textit{Regular.} Incoming: one mempool message $(\blob,\conf,\fee,\anchor)$.\\
Local: a position $i\neq0$ and branches $\pf,\pf'$.\\
Verification:
\begin{enumerate}\setlength{\itemsep}{0pt}
\item $\bid=\bid_{\blob,\conf}$ --- the identifier is that of $\blob$ and $\conf$ from the mempool message;
\item $\mup(\bid,i,\pf)=\hinc.\blobsRoot$ --- the $\bid$ is included in the block $\hinc$;
\item $\InChain((\anchor,\pf'),\hinc)$ --- the mempool message is anchored.
\end{enumerate}
\textit{Coinbase.} Incoming: one mempool message $(\blob_i,\conf_i,\fee_i,\hdr_{{\mathrm a},i})$ per other blob of the block, $1\le i\le m$.\\
Local: the coinbase outputs $O=[o_1,\dots,o_m]$ and branches $\pf_1,\dots,\pf_m$.\\
Verification:
\begin{enumerate}\setlength{\itemsep}{0pt}
\item $t=\hash(\hinc.h,\ \mr([\hash(o_j)]_{j=1}^{m}),\ \hash([\,]))$ --- the coinbase identifier is valid;
\item $\bid=\hash(t,\hash([\,]),\hash([\,]))$ --- the coinbase blob has no pairs, no conflicts;
\item $\mr([\bid,\bid_{\blob_1,\conf_1},\dots,\bid_{\blob_m,\conf_m}])=\hinc.\blobsRoot$ --- all blobs are proved;
\item $\InChain((\hdr_{{\mathrm a},i},\pf_i),\hinc)$ for each $i$ --- mempool messages are anchored;
\item $\sum_{o_j\in O}v_j=\mathsf{subsidy}(\hinc.h)+\sum_{i=1}^{m}\fee_i$ --- the coinbase claims subsidy and fees.
\end{enumerate}
\end{relationbox}

\paragraph{Coin message.}
The coin message $(\hdr,\cid,v,\pk)$ attests that $\cid$ is a coin of value $v$ spendable by $\pk$, created by a valid transaction in the chain of $\hdr$.
The \emph{include} constructor derives a coin message from a blob message at the block that included the blob.
The \emph{advance} constructor restates a coin message at any later header of the same chain.
This lets the sender freshen a coin message to the chain tip before delivering it, so that the recipient learns no information about when the transfer happened.

\begin{relationbox}{$\Rcoin$ --- message $(\hdr,\cid,v,\pk)$}
\textit{Include.} Incoming: one blob message $(\bid,\hdr')$.\\
Local: the output's opening $(x,\cc_t,\salt,j,\pf_O)$, where $x$ and $\cc_t$ are the first and third arguments of $\hash$ in $\txid$ (\S\ref{sec:structures}), and $\pl,\cc$ the hashes beside $t$ in $\bid$.\\
Verification:
\begin{enumerate}\setlength{\itemsep}{0pt}
\item $\hdr'=\hdr$ --- blob message inclusion block matches;
\item $\bid=\hash(t,\pl,\cc)$ --- $t$ is the transaction of the blob message;
\item $t=\hash\big(x,\ \mup(\hash(v,\pk,\salt),j,\pf_O),\ \cc_t\big)$ --- $(v,\pk,\salt)$ is the $j$-th output of $t$;

\item $\cid=\hash(t,j,\salt)$ --- $\cid$ is that output's coin identifier.
\end{enumerate}
\textit{Advance.} Incoming: one coin message $(\hdr',\cid',v',\pk')$.\\
Local: a branch $\pf$.\\
Verification:
\begin{enumerate}\setlength{\itemsep}{0pt}
\item $(\cid',v',\pk')=(\cid,v,\pk)$ --- the same coin;
\item $\InChain((\hdr',\pf),\hdr)$ --- restated at a later header of the same chain.
\end{enumerate}
\end{relationbox}

\subsubsection{Payment and miner protocols}
\label{sec:payment}\label{sec:generation}\label{sec:miner}
The protocols of this section should be followed by users and miners to be able to spend their coins later.
Generally, a payment follows the following process: the sender \emph{builds} a mempool message, broadcasts it to a miner, the miner \emph{accepts} the message and includes the blob into a block, the sender \emph{delivers} coin messages to the recipients, who \emph{accept} them.
Since every proof of \S\ref{sec:pcd} is built from data only the acting party keeps, any loss of data on the wallet side is potentially damaging.

A user keeps a \emph{wallet state} $(\mathcal C,\mathcal T)$:
\begin{itemize}\setlength{\itemsep}{1pt}\sloppy
\item coins $\mathcal C\ni(\hdr_i,\cid_i,v_i,\pk_i,\pf_i,\sk_i)$, a coin message, its proof, and its key
\item pending transactions $\mathcal T\ni(\tx,\anchor)$, a transaction and the tip it was built at. This data must be kept until every nullifier of $\tx$ is spent, since $\tx$ may be needed to invalidate a conflicting blob.
\end{itemize}
A user can ask any node for the conflict list of some nullifiers. The node does not have incentive to lie about this list as an incomplete answer only makes the constructed proof not applicable to the actual conflict list, so no griefing can be performed.

\paragraph{Building a payment (sender).}
Algorithm~\ref{alg:build} builds the mempool message.

\begin{algobox}{\textsc{Build} --- sender at tip $\anchor$, paying $v'_1,\dots,v'_m$ to $\pk'_1,\dots,\pk'_m$ with fee $\fee$ from its coins $(\hdr_i,\cid_i,v_i,\pk_i,\pf_i,\sk_i)$}\label{alg:build}
\begin{algorithmic}[1]\raggedright
\STATE select coins $1,\dots,k$ with $\sum_i v_i\ge\sum_j v'_j+\fee$;\ $O\gets[(v'_j,\pk'_j,\salt_j\gets\{0,1\}^\lambda)]_{j=1}^{m}$, plus $(\sum_i v_i-\sum_j v'_j-\fee,\ \pk,\ \salt)$ as change
\STATE $N\gets[\nul_{\sk_i,\cid_i}]_{i=1}^{k}$;\ $\tx\gets(N,O)$;\ $\conf\gets$ the conflict list of $N$ after $\anchor$ (C4)
\STATE $t\gets\txid_{\tx,\conf}$;\ $\blob\gets\big(t,[(\nul_i,\dg_{\nul_i,\sk_i,t})]_{i=1}^{k}\big)$
\FOR{each $\bid_r\in\conf$, opening to a blob $\blob'_r=(t'_r,P'_r)$ at list hash $\cc_r$}
\STATE $w_r\gets$ the pair witness $(p,\sk_i,\cid_i)$ if the $p$-th pair of $P'_r$ names $\nul_i$ with a degriefer other than $\dg_{\nul_i,\sk_i,t'_r}$; else the txid witness $(N',\oroot',\hash(\conf'))$ from a pending $(\tx',\hdr'_{\mathrm a})\in\mathcal T$ with $\txid_{\tx',\conf'}=t'_r$, if it differs from $(P'_r,\cc_r)$; else \textbf{abort}: the input is spent
\ENDFOR
\STATE $\pf\gets\mathcal P\big(\Rmem$; incoming $(\hdr_i,\cid_i,v_i,\pk_i)$ with $\pf_i$; local $\tx$, $[\sk_i]_i$, branches of $\hdr_i$ into $\anchor$, $[w_r]_r\big)$;\ broadcast $(\blob,\conf,\fee,\anchor)$ with $\pf$;\ $\mathcal T\gets\mathcal T\cup\{(\tx,\anchor)\}$
\end{algorithmic}
\end{algobox}

Lines 1--3 form the transaction and its blob, with a change output to a key $\pk$ of the sender's own.
$\conf$ is non-empty exactly in the three situations of \S\ref{sec:structures}, which lines 4--6 refute from the sender's own data.
If every degriefer is correct, only the holder of the sender's keys could have formed them, so the sender itself must have broadcast a transaction $\tx'$ with identifier $t'_r$ and (assuming he followed the protocol rules) holds it in $\mathcal T$, which is why pending transactions are kept until their nullifiers are spent.
Note, after broadcasting the message, the transaction must be stored until all its nullifiers are spent as otherwise the blob might be published and impossible to both validate and invalidate.
Should the chain advance before inclusion in such a way that a new conflict arises, the sender creates a new mempool proof and broadcasts again.

\paragraph{Delivering coins (sender).}
Algorithm~\ref{alg:deliver} derives the coin messages of a pending transaction once its blob is in the chain.

\begin{algobox}{\textsc{Deliver} --- sender at tip $\hdr_{\mathsf{tip}}$, for a pending $(\tx,\anchor)\in\mathcal T$ with mempool proof $\pf$ whose blob $\blob$ is included at index $i$ of the block with $\hdr$}\label{alg:deliver}
\begin{algorithmic}[1]\raggedright
\STATE $\pf_\blob\gets\mathcal P\big(\Rblob.\mathsf{regular}$; incoming $(\blob,\conf,\fee,\anchor)$ with $\pf$; local $i$ and the branches$\big)$, a blob message $(\bid_{\blob,\conf},\hdr)$
\FOR{each output $o_j=(v_j,\pk'_j,\salt_j)$ of $\tx$}
\STATE $\pf_j\gets\mathcal P\big(\Rcoin.\mathsf{include}$; incoming $(\bid_{\blob,\conf},\hdr)$ with $\pf_\blob$; local the opening of $o_j\big)$, a coin message $(\hdr,\cid_{t,j},v_j,\pk'_j)$
\STATE $\pf_j\gets\mathcal P\big(\Rcoin.\mathsf{advance}$; incoming $(\hdr,\cid_{t,j},v_j,\pk'_j)$ with $\pf_j$; local a branch of $\hdr$ into $\hdr_{\mathsf{tip}}\big)$
\STATE send $(\hdr_{\mathsf{tip}},\cid_{t,j},v_j,\pk'_j)$ with $\pf_j$ to the holder of $\pk'_j$
\ENDFOR
\end{algorithmic}
\end{algobox}
Line 1 derives the blob message from the mempool message and its proof.
Lines 2--5 open every output into a coin message, advance it to the tip and transmit it off chain.
No secret is needed, so whoever holds the mempool proof and the output openings can \emph{deliver}.

\paragraph{Accepting a coin (recipient).}
Algorithm~\ref{alg:receive} accepts a delivered coin message into the wallet state.

\begin{algobox}{\textsc{AcceptCoin} --- recipient, on a delivery $(\hdr,\cid,v,\pk)$ with proof $\pf$}\label{alg:receive}
\begin{algorithmic}[1]\raggedright
\STATE \textbf{if} $\hdr$ is not on its chain, or $\pk\ne\hash(\sk)$ for every key $\sk$ it holds \textbf{then} reject
\STATE \textbf{if} $\mathcal V\big((\hdr,\cid,v,\pk),\pf\big)=0$ \textbf{then} reject
\STATE $\mathcal C\gets\mathcal C\cup\{(\hdr,\cid,v,\pk,\pf,\sk)\}$
\end{algorithmic}
\end{algobox}
Lines 1--2 check that the header is on the recipient's own chain, that the key is one of its own, and that the proof verifies.
This suffices: the proof certifies the creating transaction, its funding coins and, recursively, every transfer back to coinbases.
Once accepted, the coin message is stored in the user's wallet, ready to be spent as an incoming message of its own \textsc{Build}.

\paragraph{Accepting a transaction (miner).}
A miner runs the consensus protocol like every node and, on top of it, keeps a \emph{mempool} $\mathcal M$ of admitted mempool messages with their proofs.

\begin{algobox}{\textsc{AcceptTx} --- miner with mempool $\mathcal M$, on a broadcast $(\blob,\conf,\fee,\anchor)$ with proof $\pf$}\label{alg:admit}
\begin{algorithmic}[1]\raggedright
\STATE \textbf{if} $\anchor$ is not on its chain, or $\fee<0$, or $\blob$ names a nullifier twice or one already in $\mathcal M$ \textbf{then} reject
\STATE \textbf{if} $\conf\ne$ the conflict list of $\blob$ as the next blob on its chain (C4), or $\mathcal V\big((\blob,\conf,\fee,\anchor),\pf\big)=0$ \textbf{then} reject
\STATE $\mathcal M\gets\mathcal M\cup\{(\blob,\conf,\fee,\anchor,\pf)\}$
\STATE \textbf{on each new block:} remove from $\mathcal M$ every entry whose conflict list, re-derived after the new tip, differs from its $\conf$
\end{algorithmic}
\end{algobox}
Line 2 derives the conflict list the blob would have as the next blob on the miner's own chain and verifies the proof against it.
A conflict list must also account for earlier blobs in the same block, so if two admitted blobs shared a nullifier, whichever the miner placed second would get a list its sender was not aware of, therefore line 1 does not admit duplicate nullifiers in mempool.
A new block from another miner may likewise add a conflict ahead of an admitted blob, so line 4 drops every entry whose list changed: its proof no longer verifies, and its sender rebuilds at the new tip.

\paragraph{Mining (miner).}
Algorithm~\ref{alg:mine} forms a block from the mempool and, once it is in the chain, the coin message of the reward.

\begin{algobox}{\textsc{Mine} --- miner with key $\sk$, at height $h$, with mempool $\mathcal M\ni(\blob_i,\conf_i,\fee_i,\hdr_{{\mathrm a},i},\pf_i)$, $1\le i\le m$}\label{alg:mine}
\begin{algorithmic}[1]\raggedright
\STATE $O\gets[(\mathsf{subsidy}(h)+\sum_i\fee_i,\ \hash(\sk),\ \salt\gets\{0,1\}^\lambda)]$;\ $t_0\gets\hash\big(h,\ \mr([\hash(o)]_{o\in O}),\ \hash([\,])\big)$;\ $\blob_0\gets(t_0,[\,])$
\STATE $\hdr\gets$ the header at height $h$ with $\blobsRoot=\mr([\bid_{\blob_0,[\,]},\bid_{\blob_1,\conf_1},\dots,\bid_{\blob_m,\conf_m}])$ (C1)--(C4);\ publish $(\hdr,[\blob_0,\dots,\blob_m])$
\STATE once the block is in the chain: $\pf_0\gets\mathcal P\big(\Rblob.\mathsf{coinbase}$; incoming $(\blob_i,\conf_i,\fee_i,\hdr_{{\mathrm a},i})$ with $\pf_i$; local $O$ and the branches$\big)$
\STATE open the reward output into a coin message as in \textsc{Deliver}, lines 3--5, and add it to $\mathcal C$
\end{algorithmic}
\end{algobox}
Lines 1--2 form the coinbase, whose outputs sum to the subsidy and the admitted fees, and the block.
The block is published without any proof.
Once it is in the chain, lines 3--4 prove the coinbase from the mempool proofs of the blobs in the block, which the miner keeps for this purpose, and open the reward into a coin exactly as a user would.
The base case of the PCD recursion is represented by a miner constructing a block with only the coinbase, thereby its blob message consuming no incoming messages.

\subsection{Protocol Correctness Proofs}
\label{sec:correctness}\label{sec:model}\label{sec:dismissal}\label{sec:safety}\label{sec:liveness}\label{sec:lean}

We prove safety and liveness of Compact Shielded CSV in Lean~4~\cite{moura2021lean} by formalising two properties: value preservation and spendability.
\emph{Value preservation} means that the total outstanding value cannot exceed the accumulated block subsidy.
We separately prove that no double spending is possible.
\emph{Spendability} means that a valid, unspent coin cannot be trapped by a griefing blob or by an earlier stale attempt from its owner.
In this section, we highlight the main theorems proved and how the protocol was modelled, but the principal proofs are given in Appendix~\ref{app:lean}.  The correspondence table in \S\ref{app:lean:correspondence} maps the paper's protocol objects and results to their Lean names.

\paragraph{Proof abstractions.}
A blockchain is abstracted to a list of blocks containing blobs (\hyperlink{lean:blockchain}{\texttt{Blockchain}}), without any headers, Lean relying directly on the underlying data.
Instead of modelling an adversarial actor, we make our proofs under a more general framework: we state properties over all possible chains that can be constructed and are valid (\hyperlink{lean:chain-valid}{\texttt{ChainValid}}) under the blockchain consensus rules (C1)--(C4) of \S\ref{sec:acceptance}.
This setting accounts for possible griefing attacks or irrational miners choosing to include any blobs, even those without valid proofs.
Finally, we abstract the PCD circuits into validity conditions which need to be fulfilled (the privacy or succinctness aspects of the proofs are lost but are not of interest for proving correctness).  These are \hyperlink{lean:mempool-valid}{\texttt{MempoolValid}} for $\Rmem$, \hyperlink{lean:blob-valid}{\texttt{BlobMsgValid}} for $\Rblob$, and \hyperlink{lean:coin-valid}{\texttt{CoinValid}} for $\Rcoin$.

\paragraph{Value preservation (safety).}
First we show no double spending is possible. If there were two accepted $\Rblob$ proofs containing the same nullifier, by rule (C4), the later blob's conflict list contains the earlier appearance.  Check~7 of $\Rmem$ (\S\ref{def:blob}) therefore requires an invalidation witness for this earlier appearance. \hyperlink{lean:lemma1-i}{\texttt{lemma1\_i}} proves that this correctly formed earlier appearance cannot be invalidated. Therefore, both blobs could not be accepted.
This is stated by \hyperlink{lean:no-two-accepted-spends}{\texttt{no\_two\_accepted\_spends}} which shows no two accepted spends can share a nullifier. 

Secondly, we show no arbitrary inflation.  On a valid chain, the total value of coins with valid $\Rcoin$ proofs which have not been consumed is at most the accumulated block subsidy. We use an abstraction in which all accepted private proof material is visible, allowing Lean to track the inputs and outputs created and consumed (proof archive assumption \hyperlink{lean:complete-history-archive}{\texttt{CompleteHistoryArchive}}). It also uses no double spending to ensure that accepted inputs are not counted twice. This allows for any self-burning or self-griefing to happen, and the overarching no inflation states the total outstanding value is less than or equal to the total aggregate block subsidy. This is formalised by the \hyperlink{lean:no-inflation}{no-inflation theorem}.

\paragraph{Spendability (liveness).}
To show spendability, we must show a legitimate owner of a coin or coins are not prevented from spending their coins either by griefing attempts of third parties or self-griefing. We must also show that an earlier attempt by the owner cannot accidentally leave its nullifiers permanently unspendable. 

We assume that users follow \textsc{Build} (Algorithm~\ref{alg:build}). The model has no scheduler, so it does not say when a miner includes a blob. Instead, derived from chain validity, \hyperlink{lean:fair-spend-of-chain-valid}{\texttt{FairSpendOpportunity}} states that every valid chain has a valid extension including an eligible spend with the conflict list it committed to under (C4), and its $\Rmem$ proof then extends to the regular case of $\Rblob$. Eventual inclusion is the liveness of the consensus layer, outside the model. If conflicting nullifiers appeared while the transaction awaited publication, \hyperlink{lean:conflicts-mismatch}{the conflict mismatch lemma} shows the stale attempt is invalid under its new list, and \hyperlink{lean:same-txid}{the same-identifier lemma} shows that any altered copy published by a griefer is invalid too, so the nullifiers are freed for a new spend. Thus the coins are either already spent or spendable now, with no third, trapped state (\hyperlink{lean:wallet-multi-input-progress}{\texttt{wallet\_multi\_input\_progress}}), and \hyperlink{lean:payment-liveness-many}{\texttt{payment\_liveness\_many}} closes the loop of a multi-input spend, proving that valid, distinct outputs result from an intended spend. 

\section{Discussion}\label{sec:discussion}

\paragraph{Account Model.}
The protocol as presented follows the UTXO model, so a blob carries one nullifier per coin spent.
Shielded CSV~\cite{nick2025shielded} instead keeps an \emph{account} per user, whose state is a single off-chain commitment, and a transaction advances that state and publishes one nullifier per account it touches.
Our design supports the same upgrade: the nullifier would be derived from the account's previous state commitment rather than from a coin identifier, the coin proof would become a proof of the account's current state, and the degriefer and conflict machinery would carry over unchanged.
Every single-account transaction would then occupy exactly three hashes on chain regardless of how many coins it moves.

\paragraph{Proof Validation Enforcement.}

Miners are not \textit{cryptographically} required to place blobs on-chain which are valid (this increases mining efficiency). However, there is a strong \textit{cryptoeconomic} incentive, as miners forgo (at later spending time) their block reward and all fees collected if even one transaction in the block was not provably valid. This design incidentally likely makes griefing pleasantly rare in practice as miners are strongly incentivised to only place valid blobs on-chain. 

If the protocol's treatment of a miner's later spending time was relaxed to collect fees from the block's valid transactions and simply ignore any invalid ones, griefing would become cheap. The degriefer mechanism would still swing into action and protect legitimate spenders, but it would increase the computational cost of generating the zk-proofs so it is probably best on balance to make degriefing rare in practice.

\paragraph{Nullifier Index Growth.}
To derive conflict lists (C4), and to answer users' queries for the blobs and lists behind them, a node keeps an ever-growing index from each nullifier to its appearances; unlike a UTXO set, pruning is not possible, since a future blob may conflict with any past one.
Each entry is a handful of hashes, but the index grows linearly in the number of spends in the chain's history.

\paragraph{Cheaper Conflict Openings.}
The contextual identifier commits to the pairs $P$ of a blob as a plain hash of the list, so a mempool proof that refutes a conflict must open the whole pair list $P$ of the conflicting blob.
Since conflicts are rare in practice this may not be an issue, but it can be improved by splitting the identifier in two.
A separate \emph{conflict identifier}, carried in conflict lists, would commit to $P$ by a Merkle root, so that a pair witness opens a single pair with a logarithmic branch.
The contextual identifier itself never needs the pairs opened: the circuits that open it only need $t$, and a txid witness only needs the nullifiers, to compare them with those committed in $t$; it could therefore commit to $\hash(N)$ directly rather than to $P$.
This trades a Merkle root per blob on the node's side for smaller circuits on the user's side.

\paragraph{Light Clients.}
A user obtains conflict lists from a node, having to  trust it only for completeness. An omitted appearance will not be accepted by a miner but the coins are not lost.
To avoid even this inefficiency, the header could also commit to a sparse Merkle tree keyed by nullifier, each leaf listing that nullifier's appearances with their contextual identifiers.
A node could then answer a query with a branch of the tree, guaranteeing the list of conflicts is correct at that block.

\paragraph{Fast Path Transactions.}
By revealing the location of the transaction, 
senders have the option of passing unconfirmed transactions to receivers in a \textit{Fast Path} manner, such that receivers can receive the mempool proof and watch the chain for the appearance of the blob which will create the output intended for them. Similar to a transaction pre-confirmation, this method allows faster transaction experience as users do not need to wait until the blob is included and for the sender to generate another coin proof. The receiver can extend the mempool proof into a coin proof themselves. However, this requires the sender giving up the information $\txid_\tx,\ j,\ \salt_j$ which will reveal the transaction location, the number of inputs and which output the receivers output is located ($j$) as well as the $\salt$ the sender set. For most users, this level of privacy is likely the default as most of the transaction details are obscured anyway.

\paragraph{Proof of Delivery.}\label{sec:proofofdeliver}

As with all client-side validation protocols, the sender must convince the receiver that they have sent them some coins. This includes the off-chain delivery of the coin proof. As an example of the problems this may cause, a receiver could claim not to have received the proof and therefore refuse to send goods or services to the sender. This has been mentioned only in passing, or not at all, in the previous CSV works we are aware of. We regard this proof-of-delivery problem as potentially serious, and (like everyone before us) we have no good solution to it. 
Although these challenges are not strictly protocol-related, the surrounding infrastructure and how users will communicate are crucial to investigate for wide-spread adoption.


\section{Conclusion}

\paragraph{Future Works.}\label{sec:future}


In future works, we present the multi-party case such as coin-join style transactions, where multiple people collaborate to perform a single transaction. Furthermore, we explore scripting capabilities allowing multiple participants to control a single input. Together with coin-join style transactions, this enables a single transaction to open a channel between two parties, with potentially infinitely many commitments exchanged off-chain, and a transaction to settle or finalise the state of the channel on-chain. We can also introduce $m$ of $n$ spend authorisation, atomic swaps with external chains as well as a variety of other spending capabilities enabled by this structure.
We also extend the account model of \S\ref{sec:discussion} to implement smart contracts whose execution is proven entirely off-chain.

We presented Compact Shielded CSV, a private client-side validation block\-chain which thrives under post-quantum limitations. Regardless of the inevitable transition required to quantum-resistant cryptography, which will substantially increase on-chain data storage requirements, we achieve the minimal on-chain footprint of $2n+1$ hashes per transaction (where $n$ is the number of inputs being spent in the transaction). 

Our novel contribution is the \textit{degriefer}.
It removes signatures and proofs from the blockchain entirely. The large cryptographic proofs stay off chain and are exchanged peer-to-peer.
We gave the full protocol: consensus rules, zero-knowledge circuits and the user wallet and miner protocols.
We formalised it in Lean~4, proving safety, as value preservation, and liveness, as spendability.
Compact Shielded CSV is thus a promising foundation for private, scalable, and post-quantum digital payments.



\vspace{12pt}
\appendix

\section{Lean Proofs}
\label{app:lean}

The protocol proofs are written and verified in Lean~4 with Mathlib. The
Lean sources appear verbatim in \S\ref{app:lean:src}. The development formalises the protocol rules and proves
invalidation soundness, value preservation, constructive spendability, and
payment progress.

\subsection{Proof Headlines}

The Lean development proves the two properties stated in
\S\ref{sec:correctness}.  For value preservation, it proves that outstanding
value cannot exceed the accumulated block subsidy and that no coin can have two
accepted spends.  The global supply theorem uses
\texttt{CompleteHistoryArchive} to account for the private proofs and outputs
that are not visible on chain.  A block whose coinbase carries no accepted
proof creates no reward coins (\texttt{CoinbaseAccount}), so the bound is an
inequality.

For spendability, Lean constructs a valid spend from valid input coins and
produces a valid coin proof for every output.  Under degriefer unforgeability
(\texttt{DegrieferUnforgeable}) and honest \textsc{Build} behaviour
(\texttt{HonestAuthored}), a coin held by an honest wallet is either already
spent by an accepted blob of that wallet or spendable now; any altered copy of
the wallet's blob is shown invalidatable by
\texttt{mempool\_or\_invalidates\_of\_same\_txid}.
\texttt{FairSpendOpportunity} is not an assumption:
\texttt{FairSpendOpportunity.of\_chainValid} derives it from chain validity, so
every valid history has a valid extension in which the spend is accepted.  The
time until a miner includes it is left to the consensus layer and is outside
the model.  Distinct outputs are proved to have distinct coin identifiers and
nullifiers.

\subsection{Model abstractions}

The Lean model keeps the protocol rules while removing implementation details.
A blockchain is an ordered list of blocks, and each block is an ordered list of
blobs.  List positions provide heights and ordering.  Merkle trees, headers, and
branches are replaced by the data they commit to, list membership, and prefix
relations.  \texttt{BlockValid} checks one block and \texttt{ChainValid} checks
the whole chain.

The validity predicates \texttt{MempoolValid}, \texttt{BlobMsgValid} and
\texttt{CoinValid} stand for the PCD proofs.  Privacy,
succinctness, and proof-system implementation are outside the model.  Hash
collisions are treated as impossible, and each kind of protocol object is tagged
before hashing so that different identifiers cannot be confused.

\subsection{Correspondence}
\label{app:lean:correspondence}

Table~\ref{tab:correspondence} maps the paper's objects, rules and results to
their Lean names.  Each circuit is represented by an inductive predicate whose
constructor fields are the checks of the corresponding protocol rule, in
protocol order.

\begin{footnotesize}
\begin{longtable}{@{}>{\raggedright\arraybackslash}p{0.36\linewidth}>{\raggedright\arraybackslash}p{0.60\linewidth}@{}}
\caption{Correspondence between the paper and the Lean development.}\label{tab:correspondence}\\
\hline
Paper & Lean \\\hline
\endfirsthead
\hline
Paper & Lean \\\hline
\endhead
Consensus rules (C2)--(C4), \S\ref{sec:consensus} & \texttt{BlockValid}, \texttt{ChainValid}, \texttt{Conflicts} \\
Nullifier, degriefer, transaction, coin and contextual identifiers, \S\ref{sec:structures} & \texttt{nullifier}, \texttt{degriefer}, \texttt{txidRegular} / \texttt{txidCoinbase}, \texttt{cid}, \texttt{bcid} \\
Invalid blob witness, Definition~\ref{def:invalid} & \texttt{Invalidates} \\
$\Rmem$; $\Rblob$ regular / coinbase; $\Rcoin$ include / advance & \texttt{MempoolValid}; \texttt{BlobMsgValid.regular} / \texttt{.coinbase}; \texttt{CoinValid.incl} / \texttt{.advance} \\
\textsc{Build}: the blob and its mempool proof & \texttt{regularBlob}, \texttt{canonicalPairs}, \texttt{spendBlobMany} \\
\textsc{Mine} and \textsc{Deliver}: the block and its coin proofs & \texttt{spendability\_many\_from\_conflicts} \\
No double spending, \S\ref{sec:correctness} & \texttt{no\_two\_accepted\_spends} \\
Value preservation, \S\ref{sec:correctness} & \texttt{CompleteHistoryArchive.}\newline\texttt{outstanding\_value\_le\_total\_subsidy} \\
Spendability, \S\ref{sec:correctness} & \texttt{wallet\_multi\_input\_progress}, \texttt{payment\_liveness\_many} \\\hline
\end{longtable}
\end{footnotesize}

\begin{sloppypar}
\subsection{Lean proofs}
\label{app:lean:src}

The seven listings below are the Lean~4 source files, verbatim and in proof
dependency order; they can be copied and checked against Mathlib.

\end{sloppypar}

\begin{lstlisting}[style=leanproofs,title=Protocol objects]
import Mathlib.Tactic
import Mathlib.Data.Set.Basic
import Mathlib.Data.Countable.Basic

set_option linter.style.longLine false

/-! # Compact CSV — protocol objects (§2.1–§2.2 of the paper)

This module formalises the protocol objects used by the compliance predicates
and safety proofs.  Each transaction has one author, a list of input
nullifiers, a list of outputs, and a commitment to its intended conflict list.

The numbered definitions correspond to Definitions 1–6 and Algorithm 1.

## Modelling conventions

The paper's succinctness machinery is committed data rather than protocol
logic, so the model uses the following abstractions:

* `mr(L)` (Merkle root) <-> `H L`, and `up(l, k, π) = root` (branch opening)
  <-> `L[k]? = some l`;
* a block header `hdr` <-> the chain prefix whose tip is that block, so
  `InChain((hdr', π), hdr)` <-> `bc' <+: bc`;
* `histRoot` / `parent` / `h` / `len` (condition (C1) of Algorithm 1) hold by
  construction of the list representation and carry no separate hypothesis.

One deviation is deliberate.  `H` is a *single polymorphic injective* function,
so two hashes of structurally different preimages are only guaranteed distinct
when the preimages inhabit the same Lean type.  The paper's identifiers are
hashes of differently-shaped tuples, and Definition 2's two branches are not
distinguished even by arity — both are ternary — so they are collected here into
one countable preimage type `Pre`.  Its constructors are the domain separation
that a deployment would get from tagged encodings, and they make `H_inj` usable
across every identifier at once. -/

namespace CompactShieldedCSV

-- ---------------------------------------------------------------------------
-- Abstract cryptographic primitives.
-- ---------------------------------------------------------------------------

/-- Hash outputs.  Realised concretely as `Nat` (kept `irreducible` below, so the
    development still treats it abstractly), witnessing that a single
    polymorphic injective hash is consistent: hash outputs are a countable type. -/
def HashType : Type := Nat

instance HashType_countable : Countable HashType := inferInstanceAs (Countable Nat)
instance HashType_decEq : DecidableEq HashType := inferInstanceAs (DecidableEq Nat)

open Classical in
/-- `H : α → HashType`, the paper's hash function.  Injective on every countable
    domain (collision resistance); see `H_inj`. -/
noncomputable def H {α : Type} (a : α) : HashType :=
  if _h : Countable α then (Classical.choose (Countable.exists_injective_nat α) a : Nat)
  else (0 : Nat)

/-- Collision resistance: distinct preimages hash to distinct digests. -/
theorem H_inj : ∀ {α : Type} [Countable α] (x y : α), H x = H y → x = y := by
  intro α _ x y h
  have hc : Countable α := inferInstance
  simp only [H, dif_pos hc] at h
  exact (Classical.choose_spec (Countable.exists_injective_nat α)) h

/-- The block subsidy schedule, `subsidy(h)`.  Sealed, so no proof can depend on
    its value while the development stays axiom-free. -/
opaque block_subsidy : Nat → Nat := fun _ => 0

attribute [irreducible] HashType H

-- ---------------------------------------------------------------------------
-- The preimage type: domain separation for the protocol's identifiers.
-- ---------------------------------------------------------------------------

/-- Every protocol identifier is `H` of one of these.  Constructors play the role
    of the domain-separating tags a deployment would encode. -/
inductive Pre where
  /-- Definition 2, coinbase branch: `H(h, mr([H(o₁),…,H(o_m)]), H([]))`. -/
  | txidCoinbase (height : Nat) (outRoot conflictsHash : HashType)
  /-- Definition 2, regular branch: `H(H(N), mr([H(o₁),…,H(o_m)]), H(K))`. -/
  | txidRegular (nullRoot outRoot conflictsHash : HashType)
  /-- The coin identifier of §2.2: `H(txid, j, ρ)`. -/
  | cid (txid : HashType) (index : Nat) (rho : HashType)
  /-- Definition 3: `H(sk, cid)`. -/
  | nullifier (sk cid : HashType)
  /-- Definition 4: `H(n, sk, txid')`. -/
  | degriefer (n sk txid : HashType)
  /-- The contextual blob identifier: `H(txid, H(pairs), H(K))`. -/
  | bcid (txid pairsHash conflictsHash : HashType)

private def Pre.enc : Pre → Nat × Nat × List HashType
  | .txidCoinbase h r k => (0, h, [r, k])
  | .txidRegular a b k => (1, 0, [a, b, k])
  | .cid t j r => (2, j, [t, r])
  | .nullifier sk c => (3, 0, [sk, c])
  | .degriefer n sk t => (4, 0, [n, sk, t])
  | .bcid t ph kh => (5, 0, [t, ph, kh])

private lemma Pre.enc_inj : Function.Injective Pre.enc := by
  intro a b h
  cases a <;> cases b <;> simp_all [Pre.enc]

instance : Countable Pre := Pre.enc_inj.countable

-- ---------------------------------------------------------------------------
-- Outputs, transactions, coins  (§2.2).
-- ---------------------------------------------------------------------------

/-- *Output.*  A triple `o = (v, pk, ρ)`: an amount, a public key and a salt. -/
structure Output where
  amount : Nat
  pk : HashType
  rho : HashType

instance : Countable Output :=
  have h : Function.Injective (fun o : Output => (o.amount, o.pk, o.rho)) :=
    fun a b hab => by cases a; cases b; simp_all
  h.countable

instance : DecidableEq Output := fun a b => by
  rcases a with ⟨a1, a2, a3⟩; rcases b with ⟨b1, b2, b3⟩
  exact decidable_of_iff (a1 = b1 ∧ a2 = b2 ∧ a3 = b3) (by simp)

/-- *Transaction.*  `T = (N, O)`: the nullifiers of the input coins and the
    outputs created.  The coinbase transaction has `N = []`. -/
structure Tx where
  N : List HashType
  O : List Output
  /-- The sender's commitment `H(K)` to the conflict list used for this attempt. -/
  conflictsHash : HashType

/-- `mr([H(o₁),…,H(o_m)])`, the output root committed by Definition 2. -/
noncomputable def outRoot (O : List Output) : HashType := H (O.map H)

/-- Definition 2 (Transaction identifier), coinbase branch. -/
noncomputable def txidCoinbase (height : Nat) (O : List Output) : HashType :=
  H (Pre.txidCoinbase height (outRoot O) (H ([] : List HashType)))

/-- Definition 2 (Transaction identifier), regular branch.  `txid(T, K)` commits
    to the conflict list `K` at which the author intends the blob to be
    included; the chain commits to the list actually assigned, in `bcid`.  A
    blob included where the two disagree is not genuine there. -/
noncomputable def txidRegular (N : List HashType) (O : List Output)
    (conflictsHash : HashType) : HashType :=
  H (Pre.txidRegular (H N) (outRoot O) conflictsHash)

/-- **Coin identifier** (§2.2, unnumbered).  `cid_{T[j]} = H(txid_T, j, ρ_j)`.
    Note that it moves with `txid`, and so with `K`. -/
noncomputable def cid (t : HashType) (j : Nat) (rho : HashType) : HashType := H (Pre.cid t j rho)

/-- **Definition 3 (Nullifier).**  `n_{T[j]} = H(sk_j, cid_{T[j]})`. -/
noncomputable def nullifier (sk c : HashType) : HashType := H (Pre.nullifier sk c)

/-- **Definition 4 (Degriefer).**  `dg_{T[j],T'} = H(n_{T[j]}, sk_j, txid_{T'})`. -/
noncomputable def degriefer (n sk t' : HashType) : HashType := H (Pre.degriefer n sk t')

-- ---------------------------------------------------------------------------
-- Blobs  (§2.1).
-- ---------------------------------------------------------------------------

/-- **Blob** (§2.1, unnumbered).  `β_T = (txid_T, [(n₁,dg₁),…,(n_k,dg_k)])` — the
    published form of a transaction and the only object a block carries. -/
structure Blob where
  txid : HashType
  pairs : List (HashType × HashType)

instance : Countable Blob :=
  have h : Function.Injective (fun b : Blob => (b.txid, b.pairs)) :=
    fun a b hab => by cases a; cases b; simp_all
  h.countable

instance : DecidableEq Blob := fun a b => by
  rcases a with ⟨a1, a2⟩; rcases b with ⟨b1, b2⟩
  exact decidable_of_iff (a1 = b1 ∧ a2 = b2) (by simp)

/-- The nullifier list `N` a blob publishes. -/
def nullifiersOf (β : Blob) : List HashType := β.pairs.map Prod.fst

/-- **Definition 1 (Blob contextual identifier).**
    `bcid_{β_T,K} = H(txid_T, H([(n₁,dg₁),…,(n_k,dg_k)]), H(K))`. -/
noncomputable def bcid (β : Blob) (K : List HashType) : HashType :=
  H (Pre.bcid β.txid (H β.pairs) (H K))

/-- Two blobs *conflict* when they share a nullifier. -/
def sharesNullifier (γ β : Blob) : Bool :=
  (nullifiersOf γ).any (fun n => decide (n ∈ nullifiersOf β))

-- ---------------------------------------------------------------------------
-- Blockchain  (§2.1).
-- ---------------------------------------------------------------------------

/-- A block is its ordered list of blobs; blob 0 is the coinbase. -/
abbrev Block := List Blob

(*@\hypertarget{lean:blockchain}{}@*)
/-- The chain; a block's position in the list is its height. -/
abbrev Blockchain := List Block

/-- A preceding blob paired with the conflict list at its exact occurrence. -/
abbrev RecordedBlob := Blob × List HashType

/-- The conflict records already accumulated in `history` that conflict with `β`. -/
noncomputable def conflictsFrom (history : List RecordedBlob) (β : Blob) : List HashType :=
  ((history.filter (fun p => sharesNullifier p.1 β)).map
    (fun p => bcid p.1 p.2)).dedup

/-- Record a sequence in order.  Each new record commits to the conflicts that
    precede that exact occurrence, so identical blob content at two positions
    receives different records whenever its actual conflict list changes. -/
noncomputable def recordHistoryFrom : List RecordedBlob → List Blob → List RecordedBlob
  | history, [] => history
  | history, β :: rest =>
      let K := conflictsFrom history β
      recordHistoryFrom (history ++ [(β, K)]) rest

noncomputable def recordHistory (prior : List Blob) : List RecordedBlob :=
  recordHistoryFrom [] prior

/-- The occurrence-specific conflict records of preceding blobs that conflict
    with `β`, deduplicated and in chain order. -/
noncomputable def conflicts (prior : List Blob) (β : Blob) : List HashType :=
  conflictsFrom (recordHistory prior) β

lemma recordHistoryFrom_append (history : List RecordedBlob) (xs ys : List Blob) :
    recordHistoryFrom history (xs ++ ys) =
      recordHistoryFrom (recordHistoryFrom history xs) ys := by
  induction xs generalizing history with
  | nil => rfl
  | cons x xs ih =>
      simp only [List.cons_append, recordHistoryFrom]
      exact ih _

lemma recordHistoryFrom_fst (history : List RecordedBlob) (xs : List Blob) :
    (recordHistoryFrom history xs).map Prod.fst = history.map Prod.fst ++ xs := by
  induction xs generalizing history with
  | nil => simp [recordHistoryFrom]
  | cons x xs ih =>
      rw [recordHistoryFrom]
      rw [ih]
      simp

@[simp] lemma recordHistory_fst (prior : List Blob) :
    (recordHistory prior).map Prod.fst = prior := by
  simp [recordHistory, recordHistoryFrom_fst]

lemma blob_mem_of_mem_recordHistory {prior : List Blob} {β : Blob} {K : List HashType}
    (h : (β, K) ∈ recordHistory prior) : β ∈ prior := by
  have hm : β ∈ (recordHistory prior).map Prod.fst := List.mem_map.mpr ⟨(β, K), h, rfl⟩
  simpa using hm

lemma recordHistory_snoc (prior : List Blob) (β : Blob) :
    recordHistory (prior ++ [β]) =
      recordHistory prior ++ [(β, conflicts prior β)] := by
  rw [recordHistory, recordHistoryFrom_append]
  rfl

lemma mem_recordHistory_mono {xs ys : List Blob} {p : RecordedBlob}
    (h : p ∈ recordHistory xs) : p ∈ recordHistory (xs ++ ys) := by
  rw [recordHistory, recordHistoryFrom_append]
  have hkeep : ∀ (history : List RecordedBlob), p ∈ history →
      p ∈ recordHistoryFrom history ys := by
    induction ys with
    | nil => simpa [recordHistoryFrom]
    | cons y ys ih =>
        intro history hp
        rw [recordHistoryFrom]
        exact ih _ (List.mem_append_left _ hp)
  exact hkeep _ h

/-- Recording a longer list preserves the contextual identifier assigned at
    every earlier position. -/
lemma recorded_mem_of_prefix {before later : List Blob} {β : Blob}
    (h : before ++ [β] <+: later) :
    (β, conflicts before β) ∈ recordHistory later := by
  obtain ⟨tail, rfl⟩ := h
  have hm : (β, conflicts before β) ∈ recordHistory (before ++ [β]) := by
    rw [recordHistory_snoc]
    simp
  exact mem_recordHistory_mono (ys := tail) hm

/-- Every blob preceding a position: the blobs of the earlier blocks, in chain
    order, followed by the earlier blobs `L` of the block being formed. -/
def priorBlobs (prior : Blockchain) (L : List Blob) : List Blob := prior.flatten ++ L

/-- The conflict list of Definition 1, placed after the blobs `L` of the block
    being formed, on a node whose chain is `prior`.  This is what a miner derives
    in `AcceptTx` (§2.4) and every node re-derives in `AcceptBlock` (§2.1). -/
noncomputable def Conflicts (prior : Blockchain) (L : List Blob) (β : Blob) : List HashType :=
  conflicts (priorBlobs prior L) β

/-- The contextual identifier a blob is entitled to at position `i` of a block
    extending `prior` — the `bcid_{β_i,K_i}` of condition (C3). -/
noncomputable def bcidAt (prior : Blockchain) (blk : Block) (i : Nat) (β : Blob) : HashType :=
  bcid β (Conflicts prior (blk.take i) β)

/-- **Block validity** — the checks of Algorithm 1.  Conditions (C1) and (C3) hold by
    construction in this representation — heights, parents and `histRoot` are
    the list structure itself, and `blobsRoot` is derived rather than stored, so
    a block cannot disagree with its own conflict lists.  What remains is (C2):
    no blob names a nullifier twice, and the coinbase carries no pairs. -/
def BlockValid (prior : Blockchain) (blk : Block) : Prop :=
  blk ≠ [] ∧
  (∀ β ∈ blk, (nullifiersOf β).Nodup) ∧
  (∀ β, blk.head? = some β → β.pairs = [])

(*@\hypertarget{lean:chain-valid}{}@*)
/-- Every block of the chain is valid with respect to its own prefix. -/
def ChainValid (bc : Blockchain) : Prop :=
  ∀ (i : Nat) (hi : i < bc.length), BlockValid (bc.take i) bc[i]

-- ---------------------------------------------------------------------------
-- Identifier injectivity — the workhorse of every later proof.
-- ---------------------------------------------------------------------------

/-- A coin identifier determines its transaction, output index and salt. -/
lemma cid_inj {t t' : HashType} {j j' : Nat} {r r' : HashType}
    (h : cid t j r = cid t' j' r') : t = t' ∧ j = j' ∧ r = r' := by
  have := H_inj _ _ h
  cases this
  exact ⟨rfl, rfl, rfl⟩

/-- A nullifier determines the secret key and coin it was formed from. -/
lemma nullifier_inj {sk sk' c c' : HashType}
    (h : nullifier sk c = nullifier sk' c') : sk = sk' ∧ c = c' := by
  have := H_inj _ _ h
  cases this
  exact ⟨rfl, rfl⟩

/-- A contextual identifier determines the blob and the conflict list. -/
lemma bcid_inj {β γ : Blob} {K L : List HashType} (h : bcid β K = bcid γ L) :
    β = γ ∧ K = L := by
  have hp := H_inj _ _ h
  simp only [Pre.bcid.injEq] at hp
  obtain ⟨h1, h2, h3⟩ := hp
  have hpairs : β.pairs = γ.pairs := H_inj _ _ h2
  have hK : K = L := H_inj _ _ h3
  refine ⟨?_, hK⟩
  cases β; cases γ; simp_all

/-- A coinbase transaction identifier is never a regular one. -/
lemma txidCoinbase_ne_txidRegular (h : Nat) (O : List Output)
    (N : List HashType) (O' : List Output) (k : HashType) :
    txidCoinbase h O ≠ txidRegular N O' k := by
  intro hc
  have := H_inj _ _ hc
  exact Pre.noConfusion this

end CompactShieldedCSV
\end{lstlisting}
\begin{lstlisting}[style=leanproofs,title=Compliance predicates]
import Src.CompactCSV.«01_Base»

set_option linter.style.longLine false

/-! # Compact CSV — the zero-knowledge circuits (§2.2)

The three compliance predicates correspond to the three rules in Figure 1.
A transaction maps to one canonical blob, giving the recursive proof flow
`Φ_coin → Φ_mempool → Φ_blob → Φ_coin`.

As in the paper, an accepted proof is modelled by satisfaction of the
corresponding compliance predicate; privacy, succinctness and the PCD
implementation are outside the model. -/

namespace CompactShieldedCSV

-- ---------------------------------------------------------------------------
-- The canonical blob of a transaction.
-- ---------------------------------------------------------------------------

/-- One input of a transaction, as witnessed by `Φ_mempool`: the incoming coin
    message `(hdr_i, cid_i, v_i, pk_i)` together with the secret key that opens
    it.  `pk_i` is not stored: check 1 of the rule forces it to be `H(sk)`. -/
structure InputWitness where
  sk : HashType
  cid : HashType
  amount : Nat
  chain : Blockchain

/-- The `(n, dg)` pairs a transaction is entitled to publish: the *canonical*
    nullifiers and degriefers of Definitions 3 and 4.  Because `Φ_mempool`
    derives the blob from the transaction rather than from author-chosen values,
    a valid blob can carry no other pairs — this is what
    `not_invalidates_bcid_regularBlob` rests on. -/
noncomputable def canonicalPairs (T : Tx) (inputs : List InputWitness) :
    List (HashType × HashType) :=
  inputs.map (fun w =>
     (nullifier w.sk w.cid,
     degriefer (nullifier w.sk w.cid) w.sk
       (txidRegular T.N T.O T.conflictsHash)))

/-- The blob of a regular transaction, built canonically.  Genuineness
    (Definition 5) is therefore a construction here rather than a test: there is
    no separate `Genuine` predicate, because `Φ_mempool` never accepts a
    supplied blob. -/
noncomputable def regularBlob (T : Tx) (inputs : List InputWitness) : Blob :=
  { txid := txidRegular T.N T.O T.conflictsHash, pairs := canonicalPairs T inputs }

@[simp] lemma regularBlob_txid (T : Tx) (inputs : List InputWitness) :
    (regularBlob T inputs).txid = txidRegular T.N T.O T.conflictsHash := rfl

@[simp] lemma nullifiersOf_regularBlob (T : Tx) (inputs : List InputWitness) :
    nullifiersOf (regularBlob T inputs) = inputs.map (fun w => nullifier w.sk w.cid) := by
  simp [nullifiersOf, regularBlob, canonicalPairs, List.map_map, Function.comp]

-- ---------------------------------------------------------------------------
-- Invalidation witnesses (Definition 6; check 3 of Φ_mempool).
-- ---------------------------------------------------------------------------

/-- `g` *invalidates* the contextual conflict record `c` in one of two forms.

    **incorrect degriefer** — open one pair `(n, dg)` of the conflicting blob against its
    committed pair root `Δ`, where `n` is derived from the witnessed key, and
    show `dg` is not the degriefer that key would form for the record's
    transaction identifier `t`.

    **commitment mismatch** — reveal the preimage of `t` and show that either
    its nullifier-list commitment differs from the blob's list or its intended
    conflict-list commitment differs from the miner-computed list in `c`. -/
def Invalidates (c : HashType) : Prop :=
  ∃ (t pairHash conflictHash : HashType) (pairs : List (HashType × HashType)),
    c = H (Pre.bcid t pairHash conflictHash) ∧
    pairHash = H pairs ∧
    ( (∃ (sk cv dg : HashType),
         (nullifier sk cv, dg) ∈ pairs ∧
         dg ≠ degriefer (nullifier sk cv) sk t)
    ∨ (∃ (eta' rho' conflictHash' : HashType),
         t = H (Pre.txidRegular eta' rho' conflictHash') ∧
         (eta' ≠ H (pairs.map Prod.fst) ∨ conflictHash' ≠ conflictHash)) )

-- ---------------------------------------------------------------------------
-- The three compliance predicates.
-- ---------------------------------------------------------------------------

mutual

(*@\hypertarget{lean:mempool-valid}{}@*)
/-- **Φ_mempool — message `(bcid, f, hdr)`.**  Asserts that `bcid` identifies the
    blob of a transaction its author was entitled to make: it spends coins the
    author owns that exist in the chain of the anchor `hdr`, it balances against
    the fee `f`, and every conflict recorded against it is invalidated. -/
inductive MempoolValid : Blockchain → HashType → Nat → Prop
  | mk {anchor : Blockchain} {b : HashType} {f : Nat}
      (T : Tx) (inputs : List InputWitness) (K : List HashType)
      /- 2. `n_i = H(sk_i, cid_i)` — `N` names exactly the nullifiers of the
         input coins (and so has one entry per incoming coin message). -/
      (h_N : T.N = inputs.map (fun w => nullifier w.sk w.cid))
      /- incoming + 1. each input is an accepted coin message owned by the
         author: its public key is `H(sk_i)`. -/
      (h_coin : ∀ w ∈ inputs, CoinValid w.chain w.cid w.amount (H w.sk))
      /- 3. `InChain((hdr_i, π_i), hdr)` — each input coin exists in the anchor's chain. -/
      (h_inchain : ∀ w ∈ inputs, w.chain <+: anchor)
      /- 4. the transaction identifier commits to this exact conflict list. -/
      (h_K : T.conflictsHash = H K)
      /- 4. `bcid = bcid_{β_T,K}` — the message names this blob under this conflict list. -/
      (h_bcid : b = bcid (regularBlob T inputs) K)
      /- 5. every recorded conflict is invalidated. -/
      (h_invalidate : ∀ c ∈ K, Invalidates c)
      /- 6. `Σ v_i = f + Σ v_j` — the transaction balances against the fee. -/
      (h_balance : (inputs.map (·.amount)).sum = f + (T.O.map (·.amount)).sum) :
      MempoolValid anchor b f

(*@\hypertarget{lean:blob-valid}{}@*)
/-- **Φ_blob — message `(bcid, hdr)`.**  Asserts that the blob with identifier
    `bcid` is committed in the block with header `hdr` and is valid. -/
inductive BlobMsgValid : Blockchain → HashType → Prop
  /-- *Regular*: promotes a mempool message at the leaf it occupies. -/
  | regular {bc : Blockchain} {b : HashType}
      (anchor : Blockchain) (f i : Nat) (blk : Block) (beta : Blob)
      (h_i : i ≠ 0)
      (h_tip : bc.getLast? = some blk)
      (h_mem : blk[i]? = some beta)
      -- 1. `bcid' = bcid` and `up(bcid, i, π) = hdr.blobsRoot`
      (h_bcid : b = bcidAt bc.dropLast blk i beta)
      -- 2. `InChain((hdr', π'), hdr)`
      (h_anchor : anchor <+: bc)
      (h_mempool : MempoolValid anchor b f) :
      BlobMsgValid bc b
  /-- *Coinbase*: binds the whole block, collecting the fee of every other blob. -/
  | coinbase {bc : Blockchain} {b : HashType}
      (blk : Block) (cb : Blob) (O : List Output)
      (fees : List Nat) (anchors : List Blockchain)
      (h_tip : bc.getLast? = some blk)
      (h_cb : blk[0]? = some cb)
      -- 1. `t = H(hdr.h, mr([H(o₁),…]))` and `bcid = H(t, H([]), H([]))`
      (h_txid : cb.txid = txidCoinbase (bc.length - 1) O)
      (h_pairs : cb.pairs = [])
      (h_bcid : b = bcidAt bc.dropLast blk 0 cb)
      -- 2. `hdr.len = m + 1`
      (h_len : fees.length + 1 = blk.length)
      (h_lenA : anchors.length = fees.length)
      -- 3. every other blob of the block carries an admitted mempool proof,
      --    anchored in this chain
      (h_anchors : ∀ (i : Nat) (anchor : Blockchain), anchors[i]? = some anchor → anchor <+: bc)
      (h_others : ∀ (i : Nat) (beta : Blob) (fee : Nat) (anchor : Blockchain),
          blk[i + 1]? = some beta → fees[i]? = some fee → anchors[i]? = some anchor →
          MempoolValid anchor (bcidAt bc.dropLast blk (i + 1) beta) fee)
      -- 4. `Σ v_j = subsidy(hdr.h) + Σ f_i`
      (h_balance : (O.map (·.amount)).sum = block_subsidy (bc.length - 1) + fees.sum) :
      BlobMsgValid bc b

(*@\hypertarget{lean:coin-valid}{}@*)
/-- **Φ_coin — message `(hdr, cid, v, pk)`.**  Asserts that `cid` is a coin of
    value `v` owned by `pk` as of the block with header `hdr`. -/
inductive CoinValid : Blockchain → HashType → Nat → HashType → Prop
  /-- *Include*: the coin's transaction was included in the named block, opening
      the output from the blob there. -/
  | incl {bc : Blockchain} {c : HashType} {v : Nat} {pk : HashType}
      (t : HashType) (j : Nat) (rho : HashType) (O : List Output)
      (lam kap b : HashType)
      -- `t = H(x, up(H(v,pk,ρ), j, π_O))`, with `x` the height for a coinbase
      -- and `H(N)` otherwise
      (h_out : O[j]? = some ⟨v, pk, rho⟩)
      (h_t : (∃ x : Nat, t = txidCoinbase x O) ∨
        (∃ (N : List HashType) (k : HashType), t = txidRegular N O k))
      -- 2. `cid = H(t, j, ρ)`
      (h_cid : c = cid t j rho)
      -- 3. `bcid' = H(t, λ, κ)`
      (h_bcid : b = H (Pre.bcid t lam kap))
      (h_blob : BlobMsgValid bc b) :
      CoinValid bc c v pk
  /-- *Advance*: carries an existing coin message forward to a later block. -/
  | advance {bc : Blockchain} {c : HashType} {v : Nat} {pk : HashType}
      (bc' : Blockchain)
      (h_prev : CoinValid bc' c v pk)
      (h_inchain : bc' <+: bc) :
      CoinValid bc c v pk

end

-- ---------------------------------------------------------------------------
-- Elementary consequences.
-- ---------------------------------------------------------------------------

/-- A blob with no pairs conflicts with nothing, so its conflict list is empty —
    which is why the coinbase's `bcid` is `H(t, H([]), H([]))` unconditionally. -/
lemma conflicts_eq_nil_of_pairs_nil {prior : List Blob} {beta : Blob}
    (h : beta.pairs = []) : conflicts prior beta = [] := by
  simp [conflicts, conflictsFrom, sharesNullifier, nullifiersOf, h]

/-- Coin messages carry forward along the chain. -/
lemma CoinValid.mono {source target : Blockchain} (hpre : source <+: target)
    {c : HashType} {v : Nat} {pk : HashType} (h : CoinValid source c v pk) :
    CoinValid target c v pk :=
  CoinValid.advance source h hpre

end CompactShieldedCSV
\end{lstlisting}
\begin{lstlisting}[style=leanproofs,title=Invalidation guarantees]
import Src.CompactCSV.«02_Circuits»

set_option linter.style.longLine false

/-! # Protocol invalidation rules

When an invalidation witness (Definition 6) exists, and when it cannot, for a
nullifier `n = H(sk, cid)` of the holder of `sk`:

* **(i)** the blob of a transaction satisfying `Φ_mempool` is genuine at the
  list it commits to, so no witness invalidates it;
* **(ii)** under degriefer unforgeability, any on-chain blob that carries `n`
  and was not authored by the holder can be invalidated by the holder;
* **(iii)** a blob whose published nullifiers or miner-computed conflict list
  differs from the corresponding commitment in its `txid` can be invalidated.

Together these are what makes the single double-spend check of `AcceptTx`
(§2.4) work: check 3 of `Φ_mempool` has a witness for every conflicting blob
the author did not write, and for none that it did.

The three cases establish the exact invalidation guarantees used by the safety
and spendability proofs. -/

namespace CompactShieldedCSV

/-- A blob appears somewhere on the chain. -/
def OnChain (bc : Blockchain) (beta : Blob) : Prop := ∃ blk ∈ bc, beta ∈ blk

(*@\hypertarget{lean:degriefer-unforgeable}{}@*)
/-- **Degriefer unforgeability.** No party without `sk` forms a new
    `H(H(sk, cid), sk, txid_β)`. `Authored sk β` means that the holder authorised
    the transaction identifier in `β`; it also covers altered copies carrying
    that identifier and a genuine pair. This relation makes unforgeability expressible against an
    injective hash, where every digest is a total function of its preimage and
    so "the adversary cannot compute it" has no content otherwise. -/
def DegrieferUnforgeable (Authored : HashType → Blob → Prop)
    (bc : Blockchain) (sk c : HashType) : Prop :=
  ∀ (beta : Blob), OnChain bc beta →
    (nullifier sk c, degriefer (nullifier sk c) sk beta.txid) ∈ beta.pairs →
    Authored sk beta

-- ---------------------------------------------------------------------------
-- (i) a genuine blob admits no witness
-- ---------------------------------------------------------------------------

/-- The core of (i): the canonical blob of a transaction whose nullifier
    list is the one its inputs determine admits no invalidation witness.

    Both witness forms are refuted by collision resistance.  An *incorrect degriefer* witness
    must open a pair of the blob against its own pair root, but every pair the
    blob publishes is canonical, so the opened degriefer is exactly the one the
    witness claims it is not.  A *commitment mismatch* witness must exhibit a
    nullifier root or a conflict-list commitment inside the transaction
    identifier differing from the one the contextual identifier carries, but
    `txid` commits to precisely those. -/
lemma not_invalidates_bcid_regularBlob (T : Tx) (inputs : List InputWitness)
    (K : List HashType)
    (h_N : T.N = inputs.map (fun w => nullifier w.sk w.cid))
    (h_K : T.conflictsHash = H K) :
    ¬ Invalidates (bcid (regularBlob T inputs) K) := by
  rintro ⟨t, pairHash, conflictHash, pairs, hbcid, hpairs, hcase⟩
  unfold bcid at hbcid
  have hpre := H_inj _ _ hbcid
  simp only [Pre.bcid.injEq] at hpre
  obtain ⟨ht, hpairHash, hconflictHash⟩ := hpre
  -- the blob's nullifier root is the transaction's
  have hnull : nullifiersOf (regularBlob T inputs) = T.N := by
    rw [nullifiersOf_regularBlob, h_N]
  have hpairsEq : pairs = (regularBlob T inputs).pairs := by
    apply H_inj
    rw [← hpairs, ← hpairHash]
  rcases hcase with ⟨sk, cv, dg, hmem, hne⟩ |
    ⟨eta', rho', conflictHash', hteq, hne⟩
  · -- incorrect degriefer: the opened pair is canonical, contradicting `hne`
    rw [hpairsEq] at hmem
    simp only [regularBlob, canonicalPairs, List.mem_map] at hmem
    obtain ⟨w, _hw, hwe⟩ := hmem
    rw [Prod.mk.injEq] at hwe
    obtain ⟨hn, hd⟩ := hwe
    obtain ⟨hsk, hcid⟩ := nullifier_inj hn
    apply hne
    rw [← hd, hsk, hcid, ← ht]
    rfl
  · have : t = txidRegular T.N T.O T.conflictsHash := by rw [← ht]; rfl
    rw [this] at hteq
    have hp := H_inj _ _ hteq.symm
    simp only [Pre.txidRegular.injEq] at hp
    rcases hne with hneN | hneK
    · -- nullifier-list mismatch
      apply hneN
      rw [hp.1, hpairsEq, ← hnull]
      rfl
    · -- conflict-list mismatch
      apply hneK
      rw [hp.2.2, ← hconflictHash, h_K]

(*@\hypertarget{lean:lemma1-i}{}@*)
/-- **(i).**  If `β*` is the blob of a transaction satisfying `Φ_mempool` at
    the list `K` it commits to, then no witness invalidates `bcid_{β*,K}`. -/
theorem lemma1_i {anchor : Blockchain} {b : HashType} {f : Nat}
    (h : MempoolValid anchor b f) :
    ∃ (beta : Blob) (K : List HashType),
      b = bcid beta K ∧ ¬ Invalidates (bcid beta K) := by
  cases h with
  | mk T inputs K h_N _h_coin _h_inchain h_K h_bcid _h_invalidate _h_balance =>
      exact ⟨regularBlob T inputs, K, h_bcid,
        not_invalidates_bcid_regularBlob T inputs K h_N h_K⟩

-- ---------------------------------------------------------------------------
-- (ii) a blob the holder did not author can be invalidated
-- ---------------------------------------------------------------------------

(*@\hypertarget{lean:lemma1-ii}{}@*)
/-- **(ii).**  Assume degriefer unforgeability for this chain and coin.
    If an on-chain blob `β` contains `n = H(sk, cid)` and was *not* authored by
    the holder of `sk`, then the holder can invalidate `bcid_{β,K}` at any `K`.

    For the pair `(n, dg)` in `β`, either `dg ≠ H(n, sk, txid_β)` — which the
    holder demonstrates with an incorrect-degriefer witness — or `β` carries the correct
    degriefer, which unforgeability turns into authorship, contradicting the
    hypothesis. -/
theorem lemma1_ii (Authored : HashType → Blob → Prop)
    (bc : Blockchain) (sk c : HashType)
    (hunforge : DegrieferUnforgeable Authored bc sk c)
    (beta : Blob) (K : List HashType) (honchain : OnChain bc beta)
    (hmem : nullifier sk c ∈ nullifiersOf beta)
    (hnot : ¬ Authored sk beta) :
    Invalidates (bcid beta K) := by
  -- recover the pair carrying the nullifier
  obtain ⟨p, hp, hp1⟩ := List.mem_map.mp hmem
  by_cases hcanon : p.2 = degriefer (nullifier sk c) sk beta.txid
  · -- the correct degriefer is present ⇒ the holder authored it ⇒ contradiction
    exfalso
    apply hnot
    apply hunforge beta honchain
    have : p = (nullifier sk c, degriefer (nullifier sk c) sk beta.txid) := by
      rw [← hcanon, ← hp1]
    rwa [← this]
  · -- a forged or copied degriefer ⇒ incorrect-degriefer witness
    refine ⟨beta.txid, H beta.pairs, H K, beta.pairs, rfl, rfl, Or.inl ?_⟩
    refine ⟨sk, c, p.2, ?_, ?_⟩
    · have : p = (nullifier sk c, p.2) := by rw [← hp1]
      rwa [← this]
    · exact hcanon

-- ---------------------------------------------------------------------------
-- (iii) a blob whose commitments do not match its occurrence
-- ---------------------------------------------------------------------------

/-- **(iii), nullifier-list form.**  If `β` has the `txid` of a blob `β*`
    authored by the holder but a different nullifier list, then the holder can
    invalidate `bcid_{β,K}`: the identifier commits to `β*`'s list, so revealing
    its preimage exhibits the mismatch.  This is form (4) of Definition 6. -/
theorem lemma1_iii (beta : Blob) (T : Tx) (inputs : List InputWitness)
    (K : List HashType)
    (h_N : T.N = inputs.map (fun w => nullifier w.sk w.cid))
    (htxid : beta.txid = (regularBlob T inputs).txid)
    (hdiff : nullifiersOf beta ≠ T.N) :
    Invalidates (bcid beta K) := by
  refine ⟨beta.txid, H beta.pairs, H K, beta.pairs, rfl, rfl, Or.inr ?_⟩
  refine ⟨H T.N, outRoot T.O, T.conflictsHash, ?_, Or.inl ?_⟩
  · rw [htxid]; rfl
  · intro hc
    exact hdiff (H_inj _ _ hc).symm

(*@\hypertarget{lean:conflicts-mismatch}{}@*)
/-- **(iii), conflict-list form.**  A genuine authored blob is not genuine at an
    occurrence whose miner-computed conflict list differs from the one committed
    inside its transaction identifier.  The other half of form (4) of
    Definition 6.  This is why replacement transactions may use any nullifier
    set while stale occurrences remain invalidatable. -/
theorem invalidates_of_conflicts_mismatch (beta : Blob) (T : Tx)
    (inputs : List InputWitness) (K : List HashType)
    (htxid : beta.txid = (regularBlob T inputs).txid)
    (hdiff : T.conflictsHash ≠ H K) :
    Invalidates (bcid beta K) := by
  refine ⟨beta.txid, H beta.pairs, H K, beta.pairs, rfl, rfl, Or.inr ?_⟩
  refine ⟨H T.N, outRoot T.O, T.conflictsHash, ?_, Or.inr hdiff⟩
  rw [htxid]
  rfl

(*@\hypertarget{lean:same-txid}{}@*)
/-- A copy of an honestly proved transaction either retains the exact blob
    and context, or has an invalidation witness. This includes copied pairs,
    omitted inputs, altered degriefers, and stale conflict lists. -/
theorem mempool_or_invalidates_of_same_txid
    {anchor : Blockchain} {original beta : Blob}
    {intendedK K : List HashType} {fee : Nat}
    (hm : MempoolValid anchor (bcid original intendedK) fee)
    (ht : beta.txid = original.txid) :
    MempoolValid anchor (bcid beta K) fee ∨ Invalidates (bcid beta K) := by
  classical
  by_cases hinv : Invalidates (bcid beta K)
  · exact Or.inr hinv
  left
  cases hm with
  | mk T inputs K' hN hcoins hchains hK hb hchecks hbalance =>
    obtain ⟨horiginal, hlist⟩ := bcid_inj hb
    subst original
    subst intendedK
    have hcontext : T.conflictsHash = H K := by
      by_contra hne
      exact hinv (invalidates_of_conflicts_mismatch beta T inputs K ht hne)
    have hnull : nullifiersOf beta = inputs.map (fun w => nullifier w.sk w.cid) := by
      by_cases hn : nullifiersOf beta = T.N
      · exact hn.trans hN
      · exact False.elim (hinv (lemma1_iii beta T inputs K hN ht hn))
    have hpairs : beta.pairs = canonicalPairs T inputs := by
      apply List.ext_getElem
      · simpa [nullifiersOf, canonicalPairs] using congrArg List.length hnull
      · intro i hi hj
        have hw : i < inputs.length := by simpa [canonicalPairs] using hj
        have hn : beta.pairs[i].1 = nullifier inputs[i].sk inputs[i].cid := by
          have he := congrArg (fun ns => ns[i]?) hnull
          simpa [nullifiersOf, List.getElem?_eq_getElem, hi, hw] using he
        have hd : beta.pairs[i].2 =
            degriefer (nullifier inputs[i].sk inputs[i].cid) inputs[i].sk beta.txid := by
          by_contra hne
          apply hinv
          refine ⟨beta.txid, H beta.pairs, H K, beta.pairs, rfl, rfl, Or.inl ?_⟩
          refine ⟨inputs[i].sk, inputs[i].cid, beta.pairs[i].2, ?_, hne⟩
          simpa [← hn] using List.getElem_mem hi
        apply Prod.ext
        · simpa [canonicalPairs] using hn
        · simpa [canonicalPairs, ht, regularBlob] using hd
    have heq : beta = regularBlob T inputs := by
      cases beta
      exact congrArg₂ Blob.mk ht hpairs
    have hsame : K' = K := H_inj _ _ (hK.symm.trans hcontext)
    subst K'
    subst beta
    exact MempoolValid.mk T inputs K hN hcoins hchains hcontext rfl hchecks hbalance

end CompactShieldedCSV
\end{lstlisting}
\begin{lstlisting}[style=leanproofs,title=Safety and value preservation]
import Src.CompactCSV.«03_Invalidation»

set_option linter.style.longLine false

/-! # Value preservation and no double spend (Theorems 1 and 2)

`Theorem 2` is the protocol's safety core and the reason the contextual
identifier of Definition 1 is shaped as it is: two accepted spends of one coin
are impossible because the later one would have to invalidate the earlier one's
record, which `lemma1_i` forbids.

`Theorem 1` is the value-preservation side.  Outputs are off-chain, so the
value a block creates is not a function of its public data.  Two layers: the
block accounting of `BlockAccounts`, over the quantities the circuits witness;
and `outstanding_value_le_total_subsidy`, the published statement, over a
proof-only view of one chain history. -/

namespace CompactShieldedCSV

-- ---------------------------------------------------------------------------
-- Conflict-list membership.
-- ---------------------------------------------------------------------------

lemma sharesNullifier_symm {a b : Blob} (h : sharesNullifier a b = true) :
    sharesNullifier b a = true := by
  simp only [sharesNullifier, List.any_eq_true, decide_eq_true_eq] at h ⊢
  obtain ⟨n, hn, hn'⟩ := h
  exact ⟨n, hn', hn⟩

/-- A preceding blob that conflicts has its record in the conflict list. -/
lemma mem_conflicts {prior : List Blob} {beta gamma : Blob}
    {K : List HashType} (hmem : (gamma, K) ∈ recordHistory prior)
    (hshare : sharesNullifier gamma beta = true) :
    bcid gamma K ∈ conflicts prior beta := by
  unfold conflicts conflictsFrom
  rw [List.mem_dedup]
  exact List.mem_map.mpr ⟨(gamma, K), List.mem_filter.mpr ⟨hmem, hshare⟩, rfl⟩

/-- A blob with no pairs shares no nullifier with anything. -/
lemma not_sharesNullifier_of_pairs_nil {a b : Blob} (h : a.pairs = []) :
    sharesNullifier a b = false := by
  simp [sharesNullifier, nullifiersOf, h]

-- ---------------------------------------------------------------------------
-- What an accepted proof yields about invalidations.
-- ---------------------------------------------------------------------------

/-- An accepted `Φ_mempool` message invalidates every record in its conflict list,
    and — by `lemma1_i` — its own record cannot be invalidated. -/
lemma mempool_invalidations {anchor : Blockchain} {beta : Blob} {K : List HashType} {f : Nat}
    (h : MempoolValid anchor (bcid beta K) f) :
    (∀ c ∈ K, Invalidates c) ∧ ¬ Invalidates (bcid beta K) := by
  cases h with
  | mk T inputs K' h_N _h_coin _h_inchain h_K h_bcid h_invalidate _h_balance =>
      obtain ⟨hb, hK⟩ := bcid_inj h_bcid
      subst hb; subst hK
      exact ⟨h_invalidate, not_invalidates_bcid_regularBlob T inputs K h_N h_K⟩

/-- An accepted `Φ_blob` message is either a promoted mempool message for the
    same blob, or the coinbase — which carries no pairs. -/
lemma blob_mempool_or_pairless {bc : Blockchain} {beta : Blob} {K : List HashType}
    (h : BlobMsgValid bc (bcid beta K)) :
    (∃ (anchor : Blockchain) (f : Nat), MempoolValid anchor (bcid beta K) f)
      ∨ beta.pairs = [] := by
  cases h with
  | regular anchor f _i _blk _b _h_i _h_tip _h_mem _h_bcid _h_anchor h_mempool =>
      exact Or.inl ⟨anchor, f, h_mempool⟩
  | coinbase _blk cb _O _fees _anchors _h_tip _h_cb _h_txid h_pairs h_bcid
      _h_len _h_lenA _h_anchors _h_others _h_balance =>
      right
      obtain ⟨hb, _⟩ := bcid_inj h_bcid
      rw [hb]; exact h_pairs

-- ---------------------------------------------------------------------------
-- Occupying a position with an accepted proof.
-- ---------------------------------------------------------------------------

/-- `beta` occupies position `(h, i)` of `bc` and carries an accepted `Φ_blob`
    proof for the contextual identifier (C3) entitles it to there. -/
def AcceptedAt (bc : Blockchain) (h i : Nat) (beta : Blob) : Prop :=
  ∃ blk : Block, bc[h]? = some blk ∧ blk[i]? = some beta ∧
    BlobMsgValid (bc.take (h + 1)) (bcid beta (Conflicts (bc.take h) (blk.take i) beta))

/-- An accepted blob that conflicts with anything invalidates its whole conflict
    list and cannot itself be invalidated. -/
lemma acceptedAt_invalidations {bc : Blockchain} {h i : Nat} {beta gamma : Blob}
    (ha : AcceptedAt bc h i beta) (hshare : sharesNullifier beta gamma = true) :
    ∃ blk : Block, bc[h]? = some blk ∧ blk[i]? = some beta ∧
      (∀ c ∈ Conflicts (bc.take h) (blk.take i) beta, Invalidates c) ∧
      ¬ Invalidates (bcid beta (Conflicts (bc.take h) (blk.take i) beta)) := by
  obtain ⟨blk, hblk, hmem, hvalid⟩ := ha
  rcases blob_mempool_or_pairless hvalid with ⟨anchor, f, hmp⟩ | hnil
  · obtain ⟨hd, hnd⟩ := mempool_invalidations hmp
    exact ⟨blk, hblk, hmem, hd, hnd⟩
  · exact absurd hshare (by rw [not_sharesNullifier_of_pairs_nil hnil]; simp)

-- ---------------------------------------------------------------------------
-- Theorem 2 (No double spend).
-- ---------------------------------------------------------------------------

/-- A blob at an earlier block lies among the blobs preceding a later position. -/
lemma mem_priorBlobs_of_block_lt {bc : Blockchain} {h1 h2 i1 : Nat}
    {blk1 : Block} {beta1 : Blob} {L : List Blob}
    (hb : bc[h1]? = some blk1) (hm : blk1[i1]? = some beta1) (hlt : h1 < h2) :
    beta1 ∈ priorBlobs (bc.take h2) L := by
  have hblk : blk1 ∈ bc.take h2 := by
    have hb' : (bc.take h2)[h1]? = some blk1 := by
      simp [hlt, hb]
    exact List.mem_iff_getElem?.mpr ⟨h1, hb'⟩
  have hbeta : beta1 ∈ blk1 := List.mem_iff_getElem?.mpr ⟨i1, hm⟩
  exact List.mem_append_left _ (List.mem_flatten.mpr ⟨blk1, hblk, hbeta⟩)

/-- A blob earlier in the same block lies among the blobs preceding a later
    position in that block. -/
lemma mem_priorBlobs_of_blob_lt {prior : Blockchain} {i1 i2 : Nat}
    {blk : Block} {beta1 : Blob}
    (hm : blk[i1]? = some beta1) (hlt : i1 < i2) :
    beta1 ∈ priorBlobs prior (blk.take i2) := by
  have h' : (blk.take i2)[i1]? = some beta1 := by
    simp [hlt, hm]
  exact List.mem_append_right _ (List.mem_iff_getElem?.mpr ⟨i1, h'⟩)

/-- The complete prefix through a blob position is a prefix of the flattened
    chain through the end of that blob's block. -/
lemma position_prefix_endBlock {bc : Blockchain} {h i : Nat}
    {blk : Block} {beta : Blob}
    (hb : bc[h]? = some blk) (hm : blk[i]? = some beta) :
    priorBlobs (bc.take h) (blk.take i) ++ [beta] <+:
      (bc.take (h + 1)).flatten := by
  have hh : h < bc.length := (List.getElem?_eq_some_iff.mp hb).1
  have hi : i < blk.length := (List.getElem?_eq_some_iff.mp hm).1
  have hbeq : bc[h] = blk := (List.getElem?_eq_some_iff.mp hb).2
  have hieq : blk[i] = beta := (List.getElem?_eq_some_iff.mp hm).2
  have hbc : bc.take (h + 1) = bc.take h ++ [blk] := by
    rw [← List.take_append_getElem hh, hbeq]
  have hblk : blk.take i ++ [beta] <+: blk := by
    rw [← hieq, List.take_append_getElem hi]
    exact List.take_prefix _ _
  obtain ⟨tail, htail⟩ := hblk
  refine ⟨tail, ?_⟩
  unfold priorBlobs
  rw [hbc, List.flatten_append]
  simp only [List.flatten_singleton]
  simpa [List.append_assoc] using
    congrArg (fun xs => (bc.take h).flatten ++ xs) htail

/-- Every earlier position, including its blob, is a prefix of the raw blob
    sequence preceding a later position. -/
lemma position_prefix_of_before {bc : Blockchain} {h1 i1 h2 i2 : Nat}
    {blk1 blk2 : Block} {beta1 : Blob}
    (hb1 : bc[h1]? = some blk1) (hm1 : blk1[i1]? = some beta1)
    (hb2 : bc[h2]? = some blk2)
    (hbefore : h1 < h2 ∨ (h1 = h2 ∧ i1 < i2)) :
    priorBlobs (bc.take h1) (blk1.take i1) ++ [beta1] <+:
      priorBlobs (bc.take h2) (blk2.take i2) := by
  rcases hbefore with hlt | ⟨rfl, hlt⟩
  · have hpEnd := position_prefix_endBlock hb1 hm1
    have hpBlocks : bc.take (h1 + 1) <+: bc.take h2 :=
      List.take_prefix_take_left (by omega)
    obtain ⟨tailBlocks, htailBlocks⟩ := hpBlocks
    have hpFlat : (bc.take (h1 + 1)).flatten <+: (bc.take h2).flatten := by
      refine ⟨tailBlocks.flatten, ?_⟩
      rw [← htailBlocks, List.flatten_append]
    exact hpEnd.trans (hpFlat.trans (List.prefix_append _ _))
  · have hblk : blk1 = blk2 := by rw [hb1] at hb2; exact Option.some.inj hb2
    subst blk2
    have hi : i1 < blk1.length := (List.getElem?_eq_some_iff.mp hm1).1
    have hieq : blk1[i1] = beta1 := (List.getElem?_eq_some_iff.mp hm1).2
    have hpTake : blk1.take i1 ++ [beta1] <+: blk1.take i2 := by
      rw [← hieq, List.take_append_getElem hi]
      exact List.take_prefix_take_left (by omega)
    obtain ⟨tail, htail⟩ := hpTake
    refine ⟨tail, ?_⟩
    unfold priorBlobs
    rw [← htail]
    simp [List.append_assoc]

/-- **The ordered core of Theorem 2.**  A conflicting pair of accepted blobs
    cannot occupy two distinct positions: condition (C3) puts the earlier one's
    conflict record into the later one's conflict list, whose validity then
    demands an invalidation that `lemma1_i` rules out. -/
lemma no_double_spend_core {bc : Blockchain} {h1 i1 h2 i2 : Nat} {beta1 beta2 : Blob}
    (ha1 : AcceptedAt bc h1 i1 beta1) (ha2 : AcceptedAt bc h2 i2 beta2)
    (hshare : sharesNullifier beta1 beta2 = true)
    (hbefore : h1 < h2 ∨ (h1 = h2 ∧ i1 < i2)) : False := by
  obtain ⟨_blk1, hb1, hm1, _hd1, hnd1⟩ :=
    acceptedAt_invalidations ha1 hshare
  obtain ⟨blk2, hb2, _hm2, hd2, _hnd2⟩ :=
    acceptedAt_invalidations ha2 (sharesNullifier_symm hshare)
  -- the earlier blob precedes the later position
  have hprior : beta1 ∈ priorBlobs (bc.take h2) (blk2.take i2) := by
    rcases hbefore with hlt | ⟨heq, hlt⟩
    · exact mem_priorBlobs_of_block_lt hb1 hm1 hlt
    · subst heq
      have : blk2 = _blk1 := by rw [hb2] at hb1; exact Option.some.inj hb1
      subst this
      exact mem_priorBlobs_of_blob_lt hm1 hlt
  -- hence its record is in the later blob's conflict list, and must be invalidated
  have hpref : priorBlobs (bc.take h1) (_blk1.take i1) ++ [beta1] <+:
      priorBlobs (bc.take h2) (blk2.take i2) :=
    position_prefix_of_before hb1 hm1 hb2 hbefore
  have hrecord :
      (beta1, Conflicts (bc.take h1) (_blk1.take i1) beta1) ∈
        recordHistory (priorBlobs (bc.take h2) (blk2.take i2)) := by
    exact recorded_mem_of_prefix hpref
  exact hnd1 (hd2 _ (mem_conflicts hrecord hshare))

(*@\hypertarget{lean:no-double-spend}{}@*)
/-- **Theorem 2 (No double spend).**  If two blobs on a valid chain name the
    same nullifier and each has an accepted `Φ_blob` proof, then they are the
    same blob, at the same position — and therefore with the same conflict-set
    contents. -/
theorem no_double_spend {bc : Blockchain} {h1 i1 h2 i2 : Nat} {beta1 beta2 : Blob}
    (ha1 : AcceptedAt bc h1 i1 beta1) (ha2 : AcceptedAt bc h2 i2 beta2)
    (hshare : sharesNullifier beta1 beta2 = true) :
    h1 = h2 ∧ i1 = i2 ∧ beta1 = beta2 := by
  -- neither can strictly precede the other
  have hne : ¬ (h1 < h2 ∨ (h1 = h2 ∧ i1 < i2)) := fun hb =>
    no_double_spend_core ha1 ha2 hshare hb
  have hne' : ¬ (h2 < h1 ∨ (h2 = h1 ∧ i2 < i1)) := fun hb =>
    no_double_spend_core ha2 ha1 (sharesNullifier_symm hshare) hb
  have hh : h1 = h2 := by omega
  subst hh
  have hi : i1 = i2 := by omega
  subst hi
  refine ⟨rfl, rfl, ?_⟩
  obtain ⟨blk1, hb1, hm1, _⟩ := ha1
  obtain ⟨blk2, hb2, hm2, _⟩ := ha2
  have : blk1 = blk2 := by rw [hb1] at hb2; exact Option.some.inj hb2
  subst this
  rw [hm1] at hm2
  exact Option.some.inj hm2

(*@\hypertarget{lean:no-two-accepted-spends}{}@*)
/-- Two distinct accepted spends of one coin cannot exist: the same nullifier
    cannot be named by two accepted blobs at different positions.  This is the
    form the coin-proof argument uses — "coin proofs for outputs of two distinct
    spends of one coin cannot both verify". -/
theorem no_two_accepted_spends {bc : Blockchain} {h1 i1 h2 i2 : Nat}
    {beta1 beta2 : Blob} {n : HashType}
    (ha1 : AcceptedAt bc h1 i1 beta1) (ha2 : AcceptedAt bc h2 i2 beta2)
    (hn1 : n ∈ nullifiersOf beta1) (hn2 : n ∈ nullifiersOf beta2) :
    h1 = h2 ∧ i1 = i2 ∧ beta1 = beta2 := by
  refine no_double_spend ha1 ha2 ?_
  simp only [sharesNullifier, List.any_eq_true, decide_eq_true_eq]
  exact ⟨n, hn1, hn2⟩

/-- Extending a chain does not change whether a blob already accepted in a
    prefix was accepted at its original position. -/
lemma acceptedAt_of_prefix {bc bc' : Blockchain} {h i : Nat} {beta : Blob}
    (hp : bc <+: bc') (ha : AcceptedAt bc h i beta) : AcceptedAt bc' h i beta := by
  obtain ⟨tail, rfl⟩ := hp
  obtain ⟨blk, hb, hm, hv⟩ := ha
  have hh : h < bc.length := (List.getElem?_eq_some_iff.mp hb).1
  have htake : (bc ++ tail).take h = bc.take h :=
    List.take_append_of_le_length (by omega)
  have htake' : (bc ++ tail).take (h + 1) = bc.take (h + 1) :=
    List.take_append_of_le_length (by omega)
  refine ⟨blk, ?_, hm, ?_⟩
  · simpa [List.getElem?_append, hh] using hb
  · simpa [htake, htake'] using hv

/-- Validity of a chain passes to each of its prefixes. -/
lemma chainValid_of_prefix {bc bc' : Blockchain}
    (hp : bc <+: bc') (hv : ChainValid bc') : ChainValid bc := by
  obtain ⟨tail, rfl⟩ := hp
  intro i hi
  have hi' : i < (bc ++ tail).length := by simp; omega
  have htake : (bc ++ tail).take i = bc.take i :=
    List.take_append_of_le_length (by omega)
  have hget : (bc ++ tail)[i] = bc[i] := by
    simp [List.getElem_append_left hi]
  simpa [htake, hget] using hv i hi'

-- ---------------------------------------------------------------------------
-- Theorem 1 (No inflation).
-- ---------------------------------------------------------------------------

/-- Exposes the exact private data used by a particular `MempoolValid`
    constructor.  This is proof-only data: it is not available from the public
    message and is used below by the omniscient history-accounting view. -/
inductive MempoolOpens : {anchor : Blockchain} → {b : HashType} → {f : Nat} →
    MempoolValid anchor b f → Tx → List InputWitness → List HashType → Prop
  | mk {anchor b f} (T : Tx) (inputs : List InputWitness) (K : List HashType)
      (h_N : T.N = inputs.map (fun w => nullifier w.sk w.cid))
      (h_coin : ∀ w ∈ inputs, CoinValid w.chain w.cid w.amount (H w.sk))
      (h_inchain : ∀ w ∈ inputs, w.chain <+: anchor)
      (h_K : T.conflictsHash = H K)
      (h_bcid : b = bcid (regularBlob T inputs) K)
      (h_invalidate : ∀ c ∈ K, Invalidates c)
      (h_balance : (inputs.map (·.amount)).sum = f + (T.O.map (·.amount)).sum) :
      MempoolOpens
        (MempoolValid.mk T inputs K h_N h_coin h_inchain h_K h_bcid h_invalidate h_balance)
        T inputs K

/-- Opening a mempool proof recovers its own balance check, rather than taking
    a second numerical accounting premise. -/
lemma MempoolOpens.balance {anchor : Blockchain} {b : HashType} {f : Nat}
    {h : MempoolValid anchor b f} {T : Tx} {inputs : List InputWitness}
    {K : List HashType} (ho : MempoolOpens h T inputs K) :
    (inputs.map (·.amount)).sum = f + (T.O.map (·.amount)).sum := by
  cases ho with
  | mk _ _ _ _ _ _ _ _ _ h_balance => exact h_balance

lemma MempoolOpens.coins {anchor : Blockchain} {b : HashType} {f : Nat}
    {h : MempoolValid anchor b f} {T : Tx} {inputs : List InputWitness}
    {K : List HashType} (ho : MempoolOpens h T inputs K) :
    ∀ w ∈ inputs, CoinValid w.chain w.cid w.amount (H w.sk) := by
  cases ho with
  | mk _ _ _ _ h_coin _ _ _ _ _ => exact h_coin

lemma MempoolOpens.inchain {anchor : Blockchain} {b : HashType} {f : Nat}
    {h : MempoolValid anchor b f} {T : Tx} {inputs : List InputWitness}
    {K : List HashType} (ho : MempoolOpens h T inputs K) :
    ∀ w ∈ inputs, w.chain <+: anchor := by
  cases ho with
  | mk _ _ _ _ _ h_inchain _ _ _ _ => exact h_inchain

/-- Exposes the exact outputs and fee list used by a particular coinbase branch
    of `BlobMsgValid`. -/
inductive CoinbaseOpens : {bc : Blockchain} → {b : HashType} →
    BlobMsgValid bc b → List Output → List Nat → Prop
  | mk {bc b} (blk : Block) (cb : Blob) (O : List Output)
      (fees : List Nat) (anchors : List Blockchain)
      (h_tip : bc.getLast? = some blk)
      (h_cb : blk[0]? = some cb)
      (h_txid : cb.txid = txidCoinbase (bc.length - 1) O)
      (h_pairs : cb.pairs = [])
      (h_bcid : b = bcidAt bc.dropLast blk 0 cb)
      (h_len : fees.length + 1 = blk.length)
      (h_lenA : anchors.length = fees.length)
      (h_anchors : ∀ (i : Nat) (anchor : Blockchain),
        anchors[i]? = some anchor → anchor <+: bc)
      (h_others : ∀ (i : Nat) (beta : Blob) (fee : Nat) (anchor : Blockchain),
        blk[i + 1]? = some beta → fees[i]? = some fee →
        anchors[i]? = some anchor →
        MempoolValid anchor (bcidAt bc.dropLast blk (i + 1) beta) fee)
      (h_balance : (O.map (·.amount)).sum =
        block_subsidy (bc.length - 1) + fees.sum) :
      CoinbaseOpens
        (BlobMsgValid.coinbase blk cb O fees anchors h_tip h_cb h_txid h_pairs h_bcid
          h_len h_lenA h_anchors h_others h_balance)
        O fees

lemma CoinbaseOpens.balance {bc : Blockchain} {b : HashType}
    {h : BlobMsgValid bc b} {O : List Output} {fees : List Nat}
    (ho : CoinbaseOpens h O fees) :
    (O.map (·.amount)).sum = block_subsidy (bc.length - 1) + fees.sum := by
  cases ho with
  | mk _ _ _ _ _ _ _ _ _ _ _ _ _ _ h_balance => exact h_balance

/-- The value book-keeping of one block, as witnessed by its circuits: the
    coinbase output total, and for each regular blob the fee, the input total
    and the output total of the transaction `Φ_mempool` witnesses for it.
    Outputs live off-chain, so this is not a function of the block's public
    data — it is exactly the data the proofs carry. -/
structure BlockAccounts where
  height : Nat
  coinbaseOut : Nat
  fees : List Nat
  regularIn : List Nat
  regularOut : List Nat

/-- The block's circuits are satisfied: every regular transaction balances
    against its fee (check 4 of `Φ_mempool`), and the coinbase collects the
    subsidy and all fees (check 4 of the coinbase rule of `Φ_blob`). -/
def BlockAccounts.Balanced (a : BlockAccounts) : Prop :=
  a.fees.length = a.regularIn.length ∧
  a.regularIn.length = a.regularOut.length ∧
  (∀ (i : Nat) (f inV outV : Nat),
      a.fees[i]? = some f → a.regularIn[i]? = some inV → a.regularOut[i]? = some outV →
      inV = f + outV) ∧
  a.coinbaseOut = block_subsidy a.height + a.fees.sum

/-- Pointwise `in = fee + out` lifts to the sums. -/
lemma sum_eq_of_pointwise : ∀ (fs ins outs : List Nat),
    fs.length = ins.length → ins.length = outs.length →
    (∀ (i : Nat) (f inV outV : Nat),
      fs[i]? = some f → ins[i]? = some inV → outs[i]? = some outV → inV = f + outV) →
    ins.sum = fs.sum + outs.sum := by
  intro fs
  induction fs with
  | nil =>
      intro ins outs h1 h2 _
      cases ins with
      | nil =>
          cases outs with
          | nil => simp
          | cons _ _ => simp at h2
      | cons _ _ => simp at h1
  | cons f fs ih =>
      intro ins outs h1 h2 hp
      cases ins with
      | nil => simp at h1
      | cons a ins =>
          cases outs with
          | nil => simp at h2
          | cons b outs =>
              have h0 : a = f + b := hp 0 f a b (by simp) (by simp) (by simp)
              have hrest : ins.sum = fs.sum + outs.sum := by
                refine ih ins outs (by simpa using h1) (by simpa using h2) ?_
                intro i f' inV outV hf hi ho
                exact hp (i + 1) f' inV outV (by simpa using hf) (by simpa using hi)
                  (by simpa using ho)
              simp only [List.sum_cons]
              omega

/-- Under the balance conditions, the regular inputs of a block total its fees
    plus its regular outputs. -/
lemma BlockAccounts.regularIn_sum (a : BlockAccounts) (h : a.Balanced) :
    a.regularIn.sum = a.fees.sum + a.regularOut.sum := by
  obtain ⟨hlen1, hlen2, hbal, _⟩ := h
  exact sum_eq_of_pointwise a.fees a.regularIn a.regularOut hlen1 hlen2 hbal

(*@\hypertarget{lean:block-no-inflation}{}@*)
/-- **Theorem 1 (No inflation), block level.**  The total value a block creates
    equals its subsidy plus the total value it consumes: the only new value in a
    block is the subsidy. -/
theorem block_no_inflation (a : BlockAccounts) (h : a.Balanced) :
    a.coinbaseOut + a.regularOut.sum = block_subsidy a.height + a.regularIn.sum := by
  have hin : a.regularIn.sum = a.fees.sum + a.regularOut.sum := a.regularIn_sum h
  obtain ⟨_, _, _, hcb⟩ := h
  omega

(*@\hypertarget{lean:chain-no-inflation}{}@*)
/-- **Theorem 1 (No inflation), chain level.**  Summing over the blocks of a
    chain, the value created is the total subsidy plus the value consumed, so
    the coin supply is non-inflationary: the value in existence at height `N`
    exceeds what earlier spends already supplied by at most
    `∑_{i ≤ N} subsidy(i)`. -/
theorem chain_no_inflation (as : List BlockAccounts) (h : ∀ a ∈ as, a.Balanced) :
    (as.map (fun a => a.coinbaseOut + a.regularOut.sum)).sum
      = (as.map (fun a => block_subsidy a.height)).sum
        + (as.map (fun a => a.regularIn.sum)).sum := by
  induction as with
  | nil => simp
  | cons a t ih =>
      have ha : a.Balanced := h a (by simp)
      have ht : ∀ x ∈ t, x.Balanced := fun x hx => h x (by simp [hx])
      have := block_no_inflation a ha
      simp only [List.map_cons, List.sum_cons]
      rw [ih ht]
      omega

-- ---------------------------------------------------------------------------
-- Outstanding supply along one chain history.
-- ---------------------------------------------------------------------------

/-- A coin together with the value committed by its creating output. -/
structure CoinClaim where
  cid : HashType
  amount : Nat
deriving DecidableEq

private lemma outRoot_inj {O O' : List Output} (h : outRoot O = outRoot O') : O = O' := by
  unfold outRoot at h
  have hm : O.map H = O'.map H := H_inj _ _ h
  exact Function.Injective.list_map (fun x y hxy => H_inj x y hxy) hm

private lemma txid_opening_outputs_inj {t : HashType} {O O' : List Output}
    (h : (∃ x : Nat, t = txidCoinbase x O) ∨
      (∃ (N : List HashType) (k : HashType), t = txidRegular N O k))
    (h' : (∃ x : Nat, t = txidCoinbase x O') ∨
      (∃ (N : List HashType) (k : HashType), t = txidRegular N O' k)) : O = O' := by
  rcases h with ⟨x, rfl⟩ | ⟨N, k, rfl⟩
  · rcases h' with ⟨x', heq⟩ | ⟨N', k', heq⟩
    · have hp := H_inj _ _ heq
      simp only [Pre.txidCoinbase.injEq] at hp
      exact outRoot_inj hp.2.1
    · exact absurd heq (txidCoinbase_ne_txidRegular x O N' O' k')
  · rcases h' with ⟨x', heq⟩ | ⟨N', k', heq⟩
    · exact absurd heq.symm (txidCoinbase_ne_txidRegular x' O' N O k)
    · have hp := H_inj _ _ heq
      simp only [Pre.txidRegular.injEq] at hp
      exact outRoot_inj hp.2.1

private lemma CoinValid.origin : ∀ {bc : Blockchain} {c : HashType} {v : Nat}
    {pk : HashType}, CoinValid bc c v pk →
    ∃ (t : HashType) (j : Nat) (rho : HashType) (O : List Output),
      O[j]? = some ⟨v, pk, rho⟩ ∧
      ((∃ x : Nat, t = txidCoinbase x O) ∨
        (∃ (N : List HashType) (k : HashType), t = txidRegular N O k)) ∧
      c = cid t j rho
  | _, _, _, _, .incl t j rho O _lam _kap _b h_out h_t h_cid _h_bcid _h_blob =>
      ⟨t, j, rho, O, h_out, h_t, h_cid⟩
  | _, _, _, _, .advance _bc' hprev _hinchain => CoinValid.origin hprev

/-- A coin identifier fixes the value and owner carried by every accepted coin
    proof for it. -/
lemma coinValid_unique {bc bc' : Blockchain} {c : HashType} {v v' : Nat}
    {pk pk' : HashType} (h : CoinValid bc c v pk) (h' : CoinValid bc' c v' pk') :
    v = v' ∧ pk = pk' := by
  obtain ⟨t, j, rho, O, hout, ht, hcid⟩ := h.origin
  obtain ⟨t', j', rho', O', hout', ht', hcid'⟩ := h'.origin
  obtain ⟨rfl, rfl, rfl⟩ := cid_inj (hcid.symm.trans hcid')
  have hO : O = O' := txid_opening_outputs_inj ht ht'
  subst hO
  rw [hout] at hout'
  have ho := Option.some.inj hout'
  exact ⟨congrArg Output.amount ho, congrArg Output.pk ho⟩

/-- Coin claims obtained by opening every output of a transaction. -/
noncomputable def claimsOf (t : HashType) (O : List Output) : List CoinClaim :=
  O.zipIdx.map (fun p => ⟨cid t p.2 p.1.rho, p.1.amount⟩)

/-- One accepted regular transaction together with the exact private opening
    of its mempool proof. -/
structure RegularProofData where
  inclusion : Blockchain
  height : Nat
  position : Nat
  block : Block
  blob : Blob
  anchor : Blockchain
  bcid : HashType
  fee : Nat
  valid : MempoolValid anchor bcid fee
  tx : Tx
  inputs : List InputWitness
  conflicts : List HashType
  opens : MempoolOpens valid tx inputs conflicts
  accepted : BlobMsgValid inclusion bcid
  at_position : AcceptedAt inclusion height position blob
  block_at : inclusion[height]? = some block
  blob_at : block[position]? = some blob
  canonical_blob : blob = regularBlob tx inputs
  contextual_id : bcid =
    CompactShieldedCSV.bcid blob (Conflicts (inclusion.take height) (block.take position) blob)
  anchor_in_chain : anchor <+: inclusion

noncomputable def RegularProofData.created (r : RegularProofData) : List CoinClaim :=
  claimsOf (txidRegular r.tx.N r.tx.O r.tx.conflictsHash) r.tx.O

def RegularProofData.consumed (r : RegularProofData) : List CoinClaim :=
  r.inputs.map (fun w => ⟨w.cid, w.amount⟩)

private lemma inputClaims_nodup (inputs : List InputWitness)
    (hcoins : ∀ w ∈ inputs, CoinValid w.chain w.cid w.amount (H w.sk))
    (hnull : (inputs.map (fun w => nullifier w.sk w.cid)).Nodup) :
    (inputs.map (fun w => CoinClaim.mk w.cid w.amount)).Nodup := by
  induction inputs with
  | nil => simp
  | cons w ws ih =>
      simp only [List.map_cons, List.nodup_cons] at hnull ⊢
      refine ⟨?_, ih (fun x hx => hcoins x (by simp [hx])) hnull.2⟩
      intro hm
      obtain ⟨w', hw', heq⟩ := List.mem_map.mp hm
      have hw := hcoins w (by simp)
      have hw'coin := hcoins w' (by simp [hw'])
      have hcid : w.cid = w'.cid := (congrArg CoinClaim.cid heq).symm
      have hpk : H w.sk = H w'.sk := (coinValid_unique hw (hcid ▸ hw'coin)).2
      have hsk : w.sk = w'.sk := H_inj _ _ hpk
      apply hnull.1
      exact List.mem_map.mpr ⟨w', hw', by simp [hcid, hsk]⟩

lemma RegularProofData.consumed_nodup (r : RegularProofData)
    (hchain : ChainValid r.inclusion) : r.consumed.Nodup := by
  have hh : r.height < r.inclusion.length := (List.getElem?_eq_some_iff.mp r.block_at).1
  have hbvalid := hchain r.height hh
  obtain ⟨_, hall, _⟩ := hbvalid
  have hmem : r.blob ∈ r.block := List.mem_iff_getElem?.mpr ⟨r.position, r.blob_at⟩
  have hblockEq : r.inclusion[r.height] = r.block :=
    (List.getElem?_eq_some_iff.mp r.block_at).2
  have hnullBlob := hall r.blob (by simpa [hblockEq] using hmem)
  rw [r.canonical_blob, nullifiersOf_regularBlob] at hnullBlob
  exact inputClaims_nodup r.inputs r.opens.coins hnullBlob

/-- Two accepted regular entries at different chain positions consume disjoint
    coin sets.  Equal coin identifiers determine equal owners, hence equal
    nullifiers, and `no_two_accepted_spends` then forces equal positions. -/
private lemma regular_consumed_disjoint {bc : Blockchain}
    (r s : RegularProofData) (hr : r.inclusion <+: bc)
    (hs : s.inclusion <+: bc)
    (hpos : r.height ≠ s.height ∨ r.position ≠ s.position) :
    r.consumed.Disjoint s.consumed := by
  rw [List.disjoint_left]
  intro c hc hcs
  obtain ⟨w, hw, hwc⟩ := List.mem_map.mp hc
  obtain ⟨w', hw', hwc'⟩ := List.mem_map.mp hcs
  have hcid : w.cid = w'.cid := by
    have := congrArg CoinClaim.cid (hwc.trans hwc'.symm)
    simpa using this
  have hcoin := r.opens.coins w hw
  have hcoin' := s.opens.coins w' hw'
  have hpk : H w.sk = H w'.sk :=
    (coinValid_unique hcoin (hcid ▸ hcoin')).2
  have hsk : w.sk = w'.sk := H_inj _ _ hpk
  have hn : nullifier w.sk w.cid ∈ nullifiersOf r.blob := by
    rw [r.canonical_blob, nullifiersOf_regularBlob]
    exact List.mem_map.mpr ⟨w, hw, rfl⟩
  have hn' : nullifier w.sk w.cid ∈ nullifiersOf s.blob := by
    rw [s.canonical_blob, nullifiersOf_regularBlob]
    exact List.mem_map.mpr ⟨w', hw', by simp [hcid, hsk]⟩
  have hsame := no_two_accepted_spends
    (acceptedAt_of_prefix hr r.at_position)
    (acceptedAt_of_prefix hs s.at_position) hn hn'
  exact hpos.elim (fun h ↦ h hsame.1) (fun h ↦ h hsame.2.1)

lemma RegularProofData.balance (r : RegularProofData) :
    (r.consumed.map CoinClaim.amount).sum =
      r.fee + (r.created.map CoinClaim.amount).sum := by
  have h := r.opens.balance
  have hout : (r.created.map CoinClaim.amount).sum =
      (r.tx.O.map Output.amount).sum := by
    unfold RegularProofData.created claimsOf
    rw [List.map_map]
    change (r.tx.O.zipIdx.map (fun p => Output.amount p.1)).sum = _
    have hzip := congrArg (fun xs => (xs.map Output.amount).sum)
      (List.zipIdx_map_fst 0 r.tx.O)
    dsimp only at hzip
    rw [List.map_map] at hzip
    simpa [Function.comp_def] using hzip
  simpa [RegularProofData.consumed, hout] using h

/-- One accepted coinbase together with the exact private opening of its blob
    proof. -/
structure CoinbaseProofData where
  inclusion : Blockchain
  bcid : HashType
  valid : BlobMsgValid inclusion bcid
  outputs : List Output
  fees : List Nat
  opens : CoinbaseOpens valid outputs fees

def CoinbaseProofData.height (c : CoinbaseProofData) : Nat := c.inclusion.length - 1

noncomputable def CoinbaseProofData.created (c : CoinbaseProofData) : List CoinClaim :=
  claimsOf (txidCoinbase c.height c.outputs) c.outputs

lemma CoinbaseProofData.balance (c : CoinbaseProofData) :
    (c.created.map CoinClaim.amount).sum = block_subsidy c.height + c.fees.sum := by
  have h := c.opens.balance
  have hout : (c.created.map CoinClaim.amount).sum =
      (c.outputs.map Output.amount).sum := by
    unfold CoinbaseProofData.created claimsOf
    rw [List.map_map]
    change (c.outputs.zipIdx.map (fun p => Output.amount p.1)).sum = _
    have hzip := congrArg (fun xs => (xs.map Output.amount).sum)
      (List.zipIdx_map_fst 0 c.outputs)
    dsimp only at hzip
    rw [List.map_map] at hzip
    simpa [Function.comp_def] using hzip
  simpa [CoinbaseProofData.height, hout] using h

(*@\hypertarget{lean:coinbase-account}{}@*)
/-- A block's reward may have no accepted proof. Its regular payments and fees
    are still accounted for; an invalid reward creates no accepted coins. -/
structure CoinbaseAccount where
  inclusion : Blockchain
  fees : List Nat
  proof : Option CoinbaseProofData
  agrees : ∀ c ∈ proof, c.inclusion = inclusion ∧ c.fees = fees

def CoinbaseAccount.height (c : CoinbaseAccount) : Nat := c.inclusion.length - 1

noncomputable def CoinbaseAccount.created (c : CoinbaseAccount) : List CoinClaim :=
  match c.proof with | none => [] | some p => p.created

lemma CoinbaseAccount.balance (c : CoinbaseAccount) :
    (c.created.map CoinClaim.amount).sum ≤ block_subsidy c.height + c.fees.sum := by
  cases hp : c.proof with
  | none => simp [CoinbaseAccount.created, hp]
  | some p =>
    obtain ⟨hi, hf⟩ := c.agrees p (by simp [hp])
    have hb := p.balance
    rw [CoinbaseProofData.height, hi, hf] at hb
    simpa [CoinbaseAccount.created, CoinbaseAccount.height, hp] using Nat.le_of_eq hb

/-- The accepted proof material for one block. `hfees` states only the
    structural correspondence between the coinbase fee slots and the regular
    proofs; all numerical balance equations come from the proof openings. -/
structure ProofBlock where
  coinbase : CoinbaseAccount
  regulars : List RegularProofData
  hfees : coinbase.fees = regulars.map RegularProofData.fee
  hinclusion : ∀ r ∈ regulars, r.inclusion = coinbase.inclusion
  hheight : ∀ r ∈ regulars, r.height = coinbase.height
  positions_nodup : (regulars.map RegularProofData.position).Nodup

noncomputable def ProofBlock.created (b : ProofBlock) : List CoinClaim :=
  b.coinbase.created ++ b.regulars.flatMap RegularProofData.created

def ProofBlock.consumed (b : ProofBlock) : List CoinClaim :=
  b.regulars.flatMap RegularProofData.consumed

/-- C2 removes repeated inputs within one transaction; exact accepted
    positions and no double spend remove repetitions across transactions in
    the same accounted block. -/
private lemma ProofBlock.consumed_nodup {bc : Blockchain} (b : ProofBlock)
    (hb : b.coinbase.inclusion <+: bc) (hchain : ChainValid bc) :
    b.consumed.Nodup := by
  rw [ProofBlock.consumed, List.nodup_flatMap]
  constructor
  · intro r hr
    have hrincl : r.inclusion <+: bc := by
      rw [b.hinclusion r hr]
      exact hb
    exact r.consumed_nodup (chainValid_of_prefix hrincl hchain)
  · have hp : List.Pairwise
        (fun r s : RegularProofData ↦ r.position ≠ s.position) b.regulars := by
      exact (List.pairwise_map).mp b.positions_nodup
    exact hp.imp_of_mem (by
      intro r s hrmem hsmem hne
      apply regular_consumed_disjoint r s
      · rw [b.hinclusion r hrmem]
      · rw [b.hinclusion s hsmem]
      · exact Or.inr hne)

/-- Accounted blocks at different heights consume disjoint coin sets. -/
private lemma proofBlock_consumed_disjoint {bc : Blockchain}
    (b d : ProofBlock) (hb : b.coinbase.inclusion <+: bc)
    (hd : d.coinbase.inclusion <+: bc)
    (hheight : b.coinbase.height ≠ d.coinbase.height) :
    b.consumed.Disjoint d.consumed := by
  rw [List.disjoint_left]
  intro c hc hcd
  obtain ⟨r, hr, hrc⟩ := List.mem_flatMap.mp hc
  obtain ⟨s, hs, hsc⟩ := List.mem_flatMap.mp hcd
  have hdis : r.consumed.Disjoint s.consumed := by
    apply regular_consumed_disjoint (bc := bc) r s
    · rw [b.hinclusion r hr]
      exact hb
    · rw [d.hinclusion s hs]
      exact hd
    · left
      rw [b.hheight r hr, d.hheight s hs]
      exact hheight
  exact (List.disjoint_left.mp hdis hrc) hsc

private lemma regulars_balance : ∀ rs : List RegularProofData,
    (rs.flatMap RegularProofData.consumed |>.map CoinClaim.amount).sum =
      (rs.map RegularProofData.fee).sum +
        (rs.flatMap RegularProofData.created |>.map CoinClaim.amount).sum := by
  intro rs
  induction rs with
  | nil => simp
  | cons r rs ih =>
      simp only [List.flatMap_cons, List.map_append, List.sum_append,
        List.map_cons, List.sum_cons]
      have hr := r.balance
      omega

lemma ProofBlock.balance (b : ProofBlock) :
    (b.created.map CoinClaim.amount).sum ≤
      block_subsidy b.coinbase.height + (b.consumed.map CoinClaim.amount).sum := by
  have hcb := b.coinbase.balance
  have hrs := regulars_balance b.regulars
  rw [b.hfees] at hcb
  simp only [ProofBlock.created, ProofBlock.consumed, List.map_append,
    List.sum_append]
  omega

/-- The omniscient proof-only view of one chain history.  Real users need not
    possess it: it collects the accepted off-chain proof material solely to
    state the global supply invariant. -/
structure HistoryAccounts where
  chain : Blockchain
  blocks : List ProofBlock
  /-- Canonical height-indexed enumeration: accounted blocks occur once, in
      chain order, and only at actual heights of this history. -/
  heights : List.Sublist (blocks.map (fun b => b.coinbase.height))
    (List.range chain.length)
  /-- Every proof block's inclusion chain is the corresponding prefix of the
      represented history. -/
  blocks_on_chain : ∀ b ∈ blocks, b.coinbase.inclusion <+: chain
  /-- Omniscience is definitional: the collected outputs are exactly the coin
      claims for which an accepted coin proof exists at this history's tip. -/
  created_complete : ∀ c : CoinClaim,
    (∃ pk, CoinValid chain c.cid c.amount pk) ↔
      c ∈ (blocks.flatMap ProofBlock.created).dedup

noncomputable def HistoryAccounts.created (a : HistoryAccounts) : List CoinClaim :=
  a.blocks.flatMap ProofBlock.created

/-- Distinct created coins. Repeated occurrences of the same identifier collapse
    to one claim; by `coinValid_unique` equal identifiers carry equal values, so
    nothing is merged away. -/
noncomputable def HistoryAccounts.createdDistinct (a : HistoryAccounts) : List CoinClaim :=
  a.created.dedup

def HistoryAccounts.consumed (a : HistoryAccounts) : List CoinClaim :=
  a.blocks.flatMap ProofBlock.consumed

/-- The global consumed list is duplicate-free by C2 within each transaction,
    no double spend between accepted positions, unique positions within each
    block, and unique heights across the accounted history. -/
private lemma HistoryAccounts.consumed_nodup (a : HistoryAccounts)
    (hchain : ChainValid a.chain) :
    a.consumed.Nodup := by
  rw [HistoryAccounts.consumed, List.nodup_flatMap]
  constructor
  · intro b hb
    exact b.consumed_nodup (a.blocks_on_chain b hb) hchain
  · have hp : List.Pairwise
        (fun b d : ProofBlock ↦ b.coinbase.height ≠ d.coinbase.height) a.blocks := by
      exact (List.pairwise_map).mp (a.heights.nodup List.nodup_range)
    exact hp.imp_of_mem (by
      intro b d hb hd hheight
      exact proofBlock_consumed_disjoint b d
        (a.blocks_on_chain b hb) (a.blocks_on_chain d hd) hheight)

/-- The subsidy accumulated by a chain history. -/
def totalSubsidy (bc : Blockchain) : Nat :=
  (List.range bc.length |>.map block_subsidy).sum

/-- Coins created on this history that have not been consumed by an accepted
    spend on this history.  Thus a frozen or self-burned coin remains
    outstanding even when it is no longer spendable. -/
noncomputable def HistoryAccounts.outstanding (a : HistoryAccounts) : List CoinClaim :=
  a.createdDistinct.filter (fun c => c ∉ a.consumed)

/-- A claim is created on this history when an accepted coin proof for it is
    available at the tip of the history.  The owner is existential because it
    is irrelevant to value accounting. -/
def CreatedOn (bc : Blockchain) (c : CoinClaim) : Prop :=
  ∃ pk, CoinValid bc c.cid c.amount pk

/-- One entry of the omniscient off-chain proof archive.  It contains the
    actual accepted coin proof, not a trusted numerical balance. -/
structure CoinProofRecord (bc : Blockchain) where
  claim : CoinClaim
  pk : HashType
  valid : CoinValid bc claim.cid claim.amount pk

(*@\hypertarget{lean:complete-history-archive}{}@*)
/-- All accepted off-chain material for one finite history.  `coverage` says
    that every coin proof at the tip occurs in the archive; `claims_eq` ties
    those proof objects to the outputs opened by the accepted block proofs. -/
structure CompleteHistoryArchive where
  chain : Blockchain
  blocks : List ProofBlock
  heights : List.Sublist (blocks.map (fun b ↦ b.coinbase.height))
    (List.range chain.length)
  blocks_on_chain : ∀ b ∈ blocks, b.coinbase.inclusion <+: chain
  coinProofs : List (CoinProofRecord chain)
  coverage : ∀ (c : CoinClaim) (pk : HashType),
    CoinValid chain c.cid c.amount pk →
      ∃ r ∈ coinProofs, r.claim = c
  claims_eq : coinProofs.map CoinProofRecord.claim =
    (blocks.flatMap ProofBlock.created).dedup

/-- A complete archive canonically constructs the accounting view used by the
    no-inflation theorem. -/
noncomputable def CompleteHistoryArchive.accounts
    (m : CompleteHistoryArchive) : HistoryAccounts where
  chain := m.chain
  blocks := m.blocks
  heights := m.heights
  blocks_on_chain := m.blocks_on_chain
  created_complete := by
    intro c
    constructor
    · rintro ⟨pk, hv⟩
      obtain ⟨r, hr, rfl⟩ := m.coverage c pk hv
      rw [← m.claims_eq]
      exact List.mem_map.mpr ⟨r, hr, rfl⟩
    · intro hc
      rw [← m.claims_eq] at hc
      obtain ⟨r, hr, hclaim⟩ := List.mem_map.mp hc
      subst hclaim
      exact ⟨r.pk, r.valid⟩

/-- Every accounted input carries an actual coin proof that can be advanced
    along the recorded prefixes to the omniscient history tip. -/
private lemma HistoryAccounts.consumed_createdOn (a : HistoryAccounts)
    : ∀ c ∈ a.consumed, CreatedOn a.chain c := by
  intro c hc
  obtain ⟨b, hb, hbc⟩ := List.mem_flatMap.mp hc
  obtain ⟨r, hr, hrc⟩ := List.mem_flatMap.mp hbc
  obtain ⟨w, hw, hwc⟩ := List.mem_map.mp hrc
  refine ⟨H w.sk, ?_⟩
  have hp : w.chain <+: a.chain :=
    (r.opens.inchain w hw).trans
      (r.anchor_in_chain.trans ((b.hinclusion r hr) ▸ a.blocks_on_chain b hb))
  have hcoin := r.opens.coins w hw
  have hadv : CoinValid a.chain w.cid w.amount (H w.sk) :=
    CoinValid.advance w.chain hcoin hp
  cases hwc
  exact hadv

/-- A complete accounting trace for one chain history.  The block accounts
    carry the actual accepted regular and coinbase proof openings.  Omniscient
    completeness says exactly that every coin proof at the history tip occurs
    in the distinct created-coin view.  Input coverage and non-duplication are
    derived below from those proofs, C2, and no double spend. -/
def HistoryAccounts.Sound (a : HistoryAccounts) : Prop :=
  ChainValid a.chain

/-- The canonical accounts produced from a complete archive are sound exactly
    when their represented chain satisfies the consensus rules. -/
lemma CompleteHistoryArchive.accounts_sound (m : CompleteHistoryArchive)
    (hchain : ChainValid m.chain) : m.accounts.Sound := hchain

private lemma sum_filter_add_sum_filter_not (xs : List CoinClaim)
    (p : CoinClaim → Bool) :
    (xs.filter p |>.map CoinClaim.amount).sum +
      (xs.filter (fun x => !p x) |>.map CoinClaim.amount).sum =
        (xs.map CoinClaim.amount).sum := by
  induction xs with
  | nil => simp
  | cons x xs ih =>
      simp only [List.filter_cons, List.map_cons, List.sum_cons]
      split <;> simp_all <;> omega

private lemma sum_filter_mem_eq_of_nodup_subset {xs ys : List CoinClaim}
    (hxs : xs.Nodup) (hys : ys.Nodup) (hsub : ∀ y ∈ ys, y ∈ xs) :
    (xs.filter (fun x => x ∈ ys) |>.map CoinClaim.amount).sum =
      (ys.map CoinClaim.amount).sum := by
  have hp : List.Perm (xs.filter (fun x => x ∈ ys)) ys := by
    apply (List.perm_ext_iff_of_nodup (hxs.filter _) hys).2
    intro x
    simp only [List.mem_filter, decide_eq_true_eq]
    exact ⟨fun h => h.2, fun hy => ⟨hsub x hy, hy⟩⟩
  exact (hp.map CoinClaim.amount).sum_eq

private lemma sum_dedup_le : ∀ xs : List CoinClaim,
    (xs.dedup.map CoinClaim.amount).sum ≤ (xs.map CoinClaim.amount).sum := by
  intro xs
  induction xs with
  | nil => simp
  | cons x xs ih =>
      by_cases hx : x ∈ xs
      · simp [hx]
        exact ih.trans (Nat.le_add_left _ _)
      · simp [hx]
        omega

private lemma sum_le_of_sublist {xs ys : List Nat} (h : List.Sublist xs ys) : xs.sum ≤ ys.sum := by
  induction h with
  | slnil => simp
  | cons a h ih =>
      simp
      exact ih.trans (Nat.le_add_left _ _)
  | cons₂ a h ih =>
      simp
      exact ih

private lemma HistoryAccounts.subsidies_le (a : HistoryAccounts) :
    (a.blocks.map (fun b => block_subsidy b.coinbase.height)).sum ≤
      totalSubsidy a.chain := by
  let hs := a.blocks.map (fun b => b.coinbase.height)
  have hcalc : (hs.map block_subsidy).sum ≤
      (List.range a.chain.length |>.map block_subsidy).sum := by
    exact sum_le_of_sublist (a.heights.map block_subsidy)
  simpa [hs, totalSubsidy, List.map_map, Function.comp_def] using hcalc

private lemma proofBlocks_balance : ∀ bs : List ProofBlock,
    (bs.flatMap ProofBlock.created |>.map CoinClaim.amount).sum ≤
      (bs.map (fun b => block_subsidy b.coinbase.height)).sum +
        (bs.flatMap ProofBlock.consumed |>.map CoinClaim.amount).sum := by
  intro bs
  induction bs with
  | nil => simp
  | cons b bs ih =>
      simp only [List.flatMap_cons, List.map_append, List.sum_append,
        List.map_cons, List.sum_cons]
      have hb := b.balance
      omega

/-- **No inflation, outstanding-supply form.**  On each sound chain history,
    the total value of distinct coins created by accepted transactions and not
    consumed by an accepted transaction on that history cannot exceed the
    subsidy accumulated on that history.

    Competing forks instantiate `a` separately: a coin may be consumed on one
    history and outstanding on another, but values from the histories are never
    added together. -/
theorem outstanding_value_le_total_subsidy (a : HistoryAccounts) (h : a.Sound) :
    (a.outstanding.map CoinClaim.amount).sum ≤ totalSubsidy a.chain := by
  have hchainValid : ChainValid a.chain := h
  have hconsumedOn := a.consumed_createdOn
  have hconsumed := a.consumed_nodup hchainValid
  have hsubset : ∀ c ∈ a.consumed, c ∈ a.createdDistinct := by
    intro c hc
    exact (a.created_complete c).mp (hconsumedOn c hc)
  have hsubsidies := a.subsidies_le
  have haccounts := proofBlocks_balance a.blocks
  have hcreated : a.createdDistinct.Nodup := List.nodup_dedup _
  have hcreatedLe :
      (a.createdDistinct.map CoinClaim.amount).sum ≤
        (a.created.map CoinClaim.amount).sum := sum_dedup_le a.created
  change (a.createdDistinct.map CoinClaim.amount).sum ≤
    ((a.blocks.flatMap ProofBlock.created).map CoinClaim.amount).sum at hcreatedLe
  change (a.blocks.flatMap ProofBlock.consumed).Nodup at hconsumed
  change (a.blocks.flatMap ProofBlock.created |>.map CoinClaim.amount).sum ≤
      (a.blocks.map (fun b => block_subsidy b.coinbase.height)).sum +
        (a.blocks.flatMap ProofBlock.consumed |>.map CoinClaim.amount).sum at haccounts
  have hsupplySplit := sum_filter_add_sum_filter_not
    a.createdDistinct
    (fun c => c ∉ a.blocks.flatMap ProofBlock.consumed)
  have hspent :
      (a.createdDistinct.filter
        (fun c => !decide (c ∉ a.blocks.flatMap ProofBlock.consumed)) |>.map
          CoinClaim.amount).sum =
        ((a.blocks.flatMap ProofBlock.consumed).map CoinClaim.amount).sum := by
    have hfilter := sum_filter_mem_eq_of_nodup_subset hcreated hconsumed hsubset
    have heq :
        a.createdDistinct.filter
            (fun c => !decide (c ∉ a.blocks.flatMap ProofBlock.consumed)) =
          a.createdDistinct.filter
            (fun c => c ∈ a.blocks.flatMap ProofBlock.consumed) := by
      apply List.filter_congr
      intro c _hc
      by_cases hm : c ∈ a.blocks.flatMap ProofBlock.consumed <;> simp [hm]
    rw [heq]
    exact hfilter
  unfold HistoryAccounts.outstanding HistoryAccounts.consumed
  rw [hspent] at hsupplySplit
  omega

(*@\hypertarget{lean:no-inflation}{}@*)
/-- End-to-end archive form: a valid chain together with its complete accepted
    off-chain proof archive has outstanding value bounded by its subsidy. -/
theorem CompleteHistoryArchive.outstanding_value_le_total_subsidy
    (m : CompleteHistoryArchive) (hchain : ChainValid m.chain) :
    (m.accounts.outstanding.map CoinClaim.amount).sum ≤ totalSubsidy m.chain :=
  _root_.CompactShieldedCSV.outstanding_value_le_total_subsidy
    m.accounts (m.accounts_sound hchain)

end CompactShieldedCSV
\end{lstlisting}
\begin{lstlisting}[style=leanproofs,title=Constructive spendability]
import Src.CompactCSV.«04_Safety»

set_option linter.style.longLine false

/-! # Theorem 3 — Spendability

The paper's assumptions, in order:

* *Valid input* — the holder has an accepted `Φ_coin` proof and balances the
  outputs and fee;
* *No prior on-chain authorisation* — `NeverAuthorisedSpend`;
* *Degriefer unforgeability* — `DegrieferUnforgeable`.

Part (i) says an accepted blob naming the holder's nullifier was authored by
the holder.  Part (ii) is constructive: it builds the block `Mine` forms from
the holder's `Build` and shows it is whole-block valid with a coin proof for
every output.

Part (ii) follows because `Φ_mempool` derives every nullifier--degriefer pair
from the transaction's witnessed inputs.  Consequently a valid blob cannot mix
canonical pairs with author-chosen ones. -/

namespace CompactShieldedCSV

/-- **No prior on-chain authorisation.**  The holder has not authorised an
    on-chain spend of this coin.  A transaction it constructed or broadcast but
    that was never included does not count — it is not on the chain. -/
def NeverAuthorisedSpend (Authored : HashType → Blob → Prop)
    (bc : Blockchain) (sk c : HashType) : Prop :=
  ∀ beta, OnChain bc beta → nullifier sk c ∈ nullifiersOf beta → ¬ Authored sk beta

/-- Every blob of a block of `bc` is on `bc`. -/
lemma onChain_of_mem_flatten {bc : Blockchain} {gamma : Blob}
    (h : gamma ∈ bc.flatten) : OnChain bc gamma := by
  obtain ⟨blk, hblk, hg⟩ := List.mem_flatten.mp h
  exact ⟨blk, hblk, hg⟩

/-- Conflict-list membership, reflected. -/
lemma mem_conflicts_iff {prior : List Blob} {beta : Blob} {cr : HashType} :
    cr ∈ conflicts prior beta ↔ ∃ γ K, (γ, K) ∈ recordHistory prior ∧
      sharesNullifier γ beta = true ∧ cr = bcid γ K := by
  unfold conflicts conflictsFrom
  rw [List.mem_dedup, List.mem_map]
  constructor
  · rintro ⟨⟨γ, K⟩, hγ, rfl⟩
    obtain ⟨h1, h2⟩ := List.mem_filter.mp hγ
    exact ⟨γ, K, h1, h2, rfl⟩
  · rintro ⟨γ, K, h1, h2, rfl⟩
    exact ⟨(γ, K), List.mem_filter.mpr ⟨h1, h2⟩, rfl⟩

-- ---------------------------------------------------------------------------
-- Theorem 3 (i).
-- ---------------------------------------------------------------------------

/-- **Theorem 3 (i).**  Every blob with an accepted `Φ_blob` proof naming
    `H(sk, cid)` was authored by the holder of `sk`.

    The extracted `Φ_mempool` vertex carries the *canonical* degriefers of
    Definition 4 — the rule derives them from the transaction rather than
    accepting author-chosen values — so unforgeability identifies the author. -/
theorem spendability_i (Authored : HashType → Blob → Prop)
    (bc : Blockchain) (sk c : HashType) (beta : Blob) (K : List HashType)
    (hunforge : DegrieferUnforgeable Authored bc sk c)
    (honchain : OnChain bc beta)
    (hvalid : BlobMsgValid bc (bcid beta K))
    (hmem : nullifier sk c ∈ nullifiersOf beta) :
    Authored sk beta := by
  rcases blob_mempool_or_pairless hvalid with ⟨anchor, f, hmp⟩ | hnil
  · cases hmp with
    | mk T inputs K' h_N _h_coin _h_inchain _h_K h_bcid _h_invalidate _h_balance =>
        obtain ⟨hb, _⟩ := bcid_inj h_bcid
        subst hb
        -- the nullifier comes from one of the inputs, whose pair is canonical
        rw [nullifiersOf_regularBlob, List.mem_map] at hmem
        obtain ⟨w, hw, hwe⟩ := hmem
        obtain ⟨hsk, hcid⟩ := nullifier_inj hwe
        apply hunforge _ honchain
        have hpair : (nullifier w.sk w.cid,
            degriefer (nullifier w.sk w.cid) w.sk
              (txidRegular T.N T.O T.conflictsHash))
            ∈ (regularBlob T inputs).pairs := by
          simp only [regularBlob, canonicalPairs, List.mem_map]
          exact ⟨w, hw, rfl⟩
        rwa [hsk, hcid] at hpair
  · exfalso
    rw [nullifiersOf, hnil] at hmem
    simp at hmem

-- ---------------------------------------------------------------------------
-- The holder's Build.
-- ---------------------------------------------------------------------------

/-- The transaction the holder builds: one input coin, the chosen outputs. -/
noncomputable def spendTx (sk c : HashType) (O : List Output) (K : List HashType) : Tx :=
  { N := [nullifier sk c], O := O, conflictsHash := H K }

/-- Its single input witness, anchored at the tip. -/
noncomputable def spendInputs (sk c : HashType) (v : Nat) (bc : Blockchain) :
    List InputWitness :=
  [{ sk := sk, cid := c, amount := v, chain := bc }]

/-- The blob `Build` forms, with the canonical pairs of Definitions 3 and 4. -/
noncomputable def spendBlob (sk c : HashType) (v : Nat) (bc : Blockchain)
    (O : List Output) (K : List HashType) : Blob :=
  regularBlob (spendTx sk c O K) (spendInputs sk c v bc)

/-- Multi-input BUILD: every input witness belongs to the same transaction
    author, while individual coins may use different secret keys. -/
noncomputable def spendTxMany (inputs : List InputWitness)
    (O : List Output) (K : List HashType) : Tx :=
  { N := inputs.map (fun w => nullifier w.sk w.cid),
    O := O, conflictsHash := H K }

noncomputable def spendBlobMany (inputs : List InputWitness)
    (O : List Output) (K : List HashType) : Blob :=
  regularBlob (spendTxMany inputs O K) inputs

/-- Conflict lists depend only on the published nullifiers, not on the txid or
    degriefers of the candidate blob. -/
lemma conflicts_congr_nullifiers {prior : List Blob} {a b : Blob}
    (h : nullifiersOf a = nullifiersOf b) :
    conflicts prior a = conflicts prior b := by
  unfold conflicts conflictsFrom
  congr 2
  apply List.filter_congr
  intro p _hp
  simp only [sharesNullifier]
  rw [h]

/-- A pairless coinbase adds no conflicts before a regular payment. -/
lemma Conflicts_pairless_prefix (bc : Blockchain) (cb beta : Blob)
    (hcb : cb.pairs = []) : Conflicts bc [cb] beta = Conflicts bc [] beta := by
  simp only [Conflicts, priorBlobs, List.append_nil, conflicts]
  rw [recordHistory_snoc]
  simp [conflictsFrom, List.filter_append, not_sharesNullifier_of_pairs_nil hcb]

/-- The actual conflict list to which the one-input `Build` commits.  The probe
    uses an arbitrary provisional list; only its nullifiers affect the result. -/
noncomputable def spendConflictList (bc : Blockchain) (cb : Blob)
    (sk c : HashType) (v : Nat) (O : List Output) : List HashType :=
  Conflicts bc [cb] (spendBlob sk c v bc O [])

noncomputable def spendConflictListMany (bc : Blockchain) (cb : Blob)
    (inputs : List InputWitness) (O : List Output) : List HashType :=
  Conflicts bc [cb] (spendBlobMany inputs O [])

@[simp] lemma nullifiersOf_spendBlobMany (inputs : List InputWitness)
    (O : List Output) (K : List HashType) :
    nullifiersOf (spendBlobMany inputs O K) =
      inputs.map (fun w => nullifier w.sk w.cid) := by
  simp [spendBlobMany]

@[simp] lemma spendBlobMany_txid (inputs : List InputWitness)
    (O : List Output) (K : List HashType) :
    (spendBlobMany inputs O K).txid =
      txidRegular (inputs.map (fun w => nullifier w.sk w.cid)) O (H K) := rfl

@[simp] lemma nullifiersOf_spendBlob (sk c : HashType) (v : Nat) (bc : Blockchain)
    (O : List Output) (K : List HashType) :
    nullifiersOf (spendBlob sk c v bc O K) = [nullifier sk c] := by
  simp [spendBlob, spendInputs, spendTx]

@[simp] lemma spendBlob_txid (sk c : HashType) (v : Nat) (bc : Blockchain)
    (O : List Output) (K : List HashType) :
    (spendBlob sk c v bc O K).txid = txidRegular [nullifier sk c] O (H K) := rfl

/-- The coinbase of the block, collecting the subsidy and the single fee. -/
noncomputable def coinbaseOuts (height fee : Nat) (pkM rhoM : HashType) : List Output :=
  [{ amount := block_subsidy height + fee, pk := pkM, rho := rhoM }]

noncomputable def coinbaseBlob (height fee : Nat) (pkM rhoM : HashType) : Blob :=
  { txid := txidCoinbase height (coinbaseOuts height fee pkM rhoM), pairs := [] }

@[simp] lemma coinbaseBlob_pairs (height fee : Nat) (pkM rhoM : HashType) :
    (coinbaseBlob height fee pkM rhoM).pairs = [] := rfl

/-- A conflicting blob of a single-nullifier spend carries that nullifier. -/
lemma mem_nullifiersOf_of_shares {beta gamma : Blob} {n : HashType}
    (hnull : nullifiersOf beta = [n])
    (h : sharesNullifier gamma beta = true) : n ∈ nullifiersOf gamma := by
  simp only [sharesNullifier, List.any_eq_true, decide_eq_true_eq, hnull,
    List.mem_singleton] at h
  obtain ⟨m, hm, rfl⟩ := h
  exact hm

/-- A conflict with a multi-input BUILD shares the canonical nullifier of at
    least one of its witnessed inputs. -/
lemma input_of_shares_spendBlobMany {inputs : List InputWitness}
    {O : List Output} {K : List HashType} {gamma : Blob}
    (h : sharesNullifier gamma (spendBlobMany inputs O K) = true) :
    ∃ w ∈ inputs, nullifier w.sk w.cid ∈ nullifiersOf gamma := by
  simp only [sharesNullifier, List.any_eq_true, decide_eq_true_eq,
    nullifiersOf_spendBlobMany, List.mem_map] at h
  obtain ⟨n, hn, w, hw, rfl⟩ := h
  exact ⟨w, hw, hn⟩

/-- **Check 3 of `Φ_mempool` is satisfiable.**  Every conflict recorded against
    the holder's spend can be invalidated: a conflicting blob carries the coin's
    nullifier, the holder never authorised such a blob, so `lemma1_ii` applies. -/
lemma all_conflicts_invalidatable (Authored : HashType → Blob → Prop)
    (bc : Blockchain) (sk c : HashType) (beta : Blob) (prior : List Blob)
    (hnull : nullifiersOf beta = [nullifier sk c])
    (hunforge : DegrieferUnforgeable Authored bc sk c)
    (hnever : NeverAuthorisedSpend Authored bc sk c)
    (hsub : ∀ γ ∈ prior, sharesNullifier γ beta = true → OnChain bc γ) :
    ∀ cr ∈ conflicts prior beta, Invalidates cr := by
  intro cr hcr
  obtain ⟨γ, K, hγ, hshare, rfl⟩ := mem_conflicts_iff.mp hcr
  have hn : nullifier sk c ∈ nullifiersOf γ := mem_nullifiersOf_of_shares hnull hshare
  have hγprior : γ ∈ prior := blob_mem_of_mem_recordHistory hγ
  have honchain : OnChain bc γ := hsub γ hγprior hshare
  exact lemma1_ii Authored bc sk c hunforge γ K honchain hn (hnever γ honchain hn)

/-- Check 5 for a multi-input BUILD.  Each conflicting blob shares some input
    nullifier, whose owner can provide the same invalidation witness used in
    the single-input proof. -/
lemma all_conflicts_invalidatable_many (Authored : HashType → Blob → Prop)
    (bc : Blockchain) (inputs : List InputWitness) (O : List Output)
    (beta : Blob) (prior : List Blob)
    (hnull : nullifiersOf beta = inputs.map (fun w ↦ nullifier w.sk w.cid))
    (hunforge : ∀ w ∈ inputs, DegrieferUnforgeable Authored bc w.sk w.cid)
    (hnever : ∀ w ∈ inputs, NeverAuthorisedSpend Authored bc w.sk w.cid)
    (hsub : ∀ γ ∈ prior, sharesNullifier γ beta = true → OnChain bc γ) :
    ∀ cr ∈ conflicts prior beta, Invalidates cr := by
  intro cr hcr
  obtain ⟨γ, K, hγ, hshare, rfl⟩ := mem_conflicts_iff.mp hcr
  have hshare' : sharesNullifier γ (spendBlobMany inputs O []) = true := by
    have heq : sharesNullifier γ beta =
        sharesNullifier γ (spendBlobMany inputs O []) := by
      simp only [sharesNullifier]
      rw [hnull, nullifiersOf_spendBlobMany]
    rw [← heq]
    exact hshare
  obtain ⟨w, hw, hn⟩ := input_of_shares_spendBlobMany hshare'
  have hγprior : γ ∈ prior := blob_mem_of_mem_recordHistory hγ
  have honchain : OnChain bc γ := hsub γ hγprior hshare
  exact lemma1_ii Authored bc w.sk w.cid (hunforge w hw) γ K honchain hn
    (hnever w hw γ honchain hn)

-- ---------------------------------------------------------------------------
-- Whole-block validity.
-- ---------------------------------------------------------------------------

/-- A block satisfying the checks of Algorithm 1, all of whose blobs carry
    accepted `Φ_blob` proofs — what `Mine`/`AcceptBlock` produce. -/
def WholeBlockValid (bc' : Blockchain) (blk : Block) : Prop :=
  BlockValid bc'.dropLast blk ∧
  ∀ (i : Nat) (beta : Blob), blk[i]? = some beta →
    BlobMsgValid bc' (bcidAt bc'.dropLast blk i beta)

-- ---------------------------------------------------------------------------
-- Theorem 3 (ii).
-- ---------------------------------------------------------------------------

/-- Multi-input form of Theorem 3(ii).  All input proofs are checked by the
    same mempool proof, the transaction commits to their complete nullifier
    list, and C2 is supplied by `hnullNodup`. -/
theorem spendability_many_from_conflicts
    (bc : Blockchain) (inputs : List InputWitness) (fee : Nat)
    (O : List Output) (cb beta : Blob) (pkM rhoM : HashType)
    (hcbpairs : cb.pairs = [])
    (hcbtxid : cb.txid = txidCoinbase bc.length
      (coinbaseOuts bc.length fee pkM rhoM))
    (hbeta : beta = spendBlobMany inputs O
      (spendConflictListMany bc cb inputs O))
    (hcoins : ∀ w ∈ inputs, CoinValid w.chain w.cid w.amount (H w.sk))
    (hinchain : ∀ w ∈ inputs, w.chain <+: bc)
    (hnullNodup : (inputs.map (fun w ↦ nullifier w.sk w.cid)).Nodup)
    (hinvalidate : ∀ cr ∈ Conflicts bc [cb] beta, Invalidates cr)
    (hbalance : (inputs.map InputWitness.amount).sum =
      fee + (O.map Output.amount).sum) :
    WholeBlockValid (bc ++ [[cb, beta]]) [cb, beta] ∧
    (∀ (j : Nat) (o : Output), O[j]? = some o →
      CoinValid (bc ++ [[cb, beta]]) (cid beta.txid j o.rho) o.amount o.pk) := by
  have hdrop : (bc ++ [[cb, beta]]).dropLast = bc := by simp
  have hlast : (bc ++ [[cb, beta]]).getLast? = some [cb, beta] := by simp
  have hlenc : (bc ++ [[cb, beta]]).length - 1 = bc.length := by simp
  have hnull : nullifiersOf beta =
      inputs.map (fun w ↦ nullifier w.sk w.cid) := by rw [hbeta]; simp
  have hK : spendConflictListMany bc cb inputs O = Conflicts bc [cb] beta := by
    unfold spendConflictListMany Conflicts
    apply conflicts_congr_nullifiers
    rw [hbeta]
    simp
  have hmempool : MempoolValid bc
      (bcid beta (Conflicts bc [cb] beta)) fee := by
    refine MempoolValid.mk
      (spendTxMany inputs O (spendConflictListMany bc cb inputs O))
      inputs (Conflicts bc [cb] beta) rfl hcoins hinchain ?_ ?_ hinvalidate ?_
    · simp [spendTxMany, hK]
    · rw [hbeta]
      rfl
    · simpa [spendTxMany] using hbalance
  have hbcid1 : bcidAt (bc ++ [[cb, beta]]).dropLast [cb, beta] 1 beta =
      bcid beta (Conflicts bc [cb] beta) := by rw [hdrop]; rfl
  have hblobSpend : BlobMsgValid (bc ++ [[cb, beta]])
      (bcidAt (bc ++ [[cb, beta]]).dropLast [cb, beta] 1 beta) := by
    rw [hbcid1]
    exact BlobMsgValid.regular bc fee 1 [cb, beta] beta one_ne_zero hlast (by simp)
      hbcid1.symm (List.prefix_append _ _) hmempool
  have hblobCb : BlobMsgValid (bc ++ [[cb, beta]])
      (bcidAt (bc ++ [[cb, beta]]).dropLast [cb, beta] 0 cb) := by
    refine BlobMsgValid.coinbase [cb, beta] cb (coinbaseOuts bc.length fee pkM rhoM)
      [fee] [bc] hlast (by simp) ?_ hcbpairs rfl (by simp) rfl ?_ ?_ ?_
    · rw [hlenc]; exact hcbtxid
    · intro i anchor ha
      match i with
      | 0 =>
          simp only [List.getElem?_cons_zero, Option.some.injEq] at ha
          subst ha
          exact List.prefix_append _ _
      | (n + 1) => simp at ha
    · intro i b' f' anchor hb hf hanc
      match i with
      | 0 =>
          simp only [List.getElem?_cons_succ, List.getElem?_cons_zero,
            Option.some.injEq] at hb hf hanc
          subst hb; subst hf; subst hanc
          rw [hbcid1]
          exact hmempool
      | (n + 1) => simp at hb
    · rw [hlenc]
      simp [coinbaseOuts]
  refine ⟨⟨?_, ?_⟩, ?_⟩
  · rw [hdrop]
    refine ⟨by simp, ?_, ?_⟩
    · intro b hb
      simp only [List.mem_cons, List.not_mem_nil, or_false] at hb
      rcases hb with rfl | rfl
      · rw [nullifiersOf, hcbpairs]
        simp
      · rw [hnull]
        exact hnullNodup
    · intro b hb
      simp only [List.head?_cons, Option.some.injEq] at hb
      subst hb
      exact hcbpairs
  · intro i b' hb
    match i with
    | 0 =>
        simp only [List.getElem?_cons_zero, Option.some.injEq] at hb
        subst hb
        exact hblobCb
    | 1 =>
        simp only [List.getElem?_cons_succ, List.getElem?_cons_zero,
          Option.some.injEq] at hb
        subst hb
        exact hblobSpend
    | (n + 2) => simp at hb
  · intro j o hj
    refine CoinValid.incl beta.txid j o.rho O (H beta.pairs)
      (H (Conflicts bc [cb] beta)) (bcid beta (Conflicts bc [cb] beta))
      ?_ ?_ rfl rfl ?_
    · rcases o with ⟨amt, pk, rho⟩
      exact hj
    · exact Or.inr ⟨inputs.map (fun w ↦ nullifier w.sk w.cid),
        H (spendConflictListMany bc cb inputs O), by rw [hbeta]; rfl⟩
    · rw [← hbcid1]
      exact hblobSpend

/-- The single-input construction is the singleton instance of the general
    multi-input construction. -/
theorem spendability_ii_from_conflicts (_Authored : HashType → Blob → Prop)
    (bc : Blockchain) (sk c : HashType) (v fee : Nat) (O : List Output)
    (cb beta : Blob) (pkM rhoM : HashType)
    (hcbpairs : cb.pairs = [])
    (hcbtxid : cb.txid = txidCoinbase bc.length (coinbaseOuts bc.length fee pkM rhoM))
    (hbeta : beta = spendBlob sk c v bc O (spendConflictList bc cb sk c v O))
    (hcoin : CoinValid bc c v (H sk))
    (hinvalidate : ∀ cr ∈ Conflicts bc [cb] beta, Invalidates cr)
    (hbalance : v = fee + (O.map (·.amount)).sum) :
    WholeBlockValid (bc ++ [[cb, beta]]) [cb, beta] ∧
    (∀ (j : Nat) (o : Output), O[j]? = some o →
        CoinValid (bc ++ [[cb, beta]]) (cid beta.txid j o.rho) o.amount o.pk) := by
  apply spendability_many_from_conflicts bc (spendInputs sk c v bc) fee O cb beta
    pkM rhoM hcbpairs hcbtxid
  · simpa [spendBlob, spendBlobMany, spendTx, spendTxMany,
      spendConflictList, spendConflictListMany] using hbeta
  · intro w hw
    simp only [spendInputs, List.mem_singleton] at hw
    subst w
    exact hcoin
  · intro w hw
    simp only [spendInputs, List.mem_singleton] at hw
    subst w
    exact List.prefix_refl _
  · simp [spendInputs]
  · exact hinvalidate
  · simpa [spendInputs] using hbalance

/-- Theorem 3(ii) in its original no-prior-authorisation form. -/
theorem spendability_ii (Authored : HashType → Blob → Prop)
    (bc : Blockchain) (sk c : HashType) (v fee : Nat) (O : List Output)
    (cb beta : Blob) (pkM rhoM : HashType)
    (hcbpairs : cb.pairs = [])
    (hcbtxid : cb.txid = txidCoinbase bc.length (coinbaseOuts bc.length fee pkM rhoM))
    (hbeta : beta = spendBlob sk c v bc O (spendConflictList bc cb sk c v O))
    (hcoin : CoinValid bc c v (H sk))
    (hunforge : DegrieferUnforgeable Authored bc sk c)
    (hnever : NeverAuthorisedSpend Authored bc sk c)
    (hbalance : v = fee + (O.map (·.amount)).sum) :
    WholeBlockValid (bc ++ [[cb, beta]]) [cb, beta] ∧
    (∀ (j : Nat) (o : Output), O[j]? = some o →
        CoinValid (bc ++ [[cb, beta]]) (cid beta.txid j o.rho) o.amount o.pk) := by
  have hnull : nullifiersOf beta = [nullifier sk c] := by rw [hbeta]; simp
  have hsub : ∀ γ ∈ priorBlobs bc [cb], sharesNullifier γ beta = true → OnChain bc γ := by
    intro γ hγ hshare
    rcases List.mem_append.mp hγ with hg | hg
    · exact onChain_of_mem_flatten hg
    · have hgc : γ = cb := by simpa using hg
      subst hgc
      rw [not_sharesNullifier_of_pairs_nil hcbpairs] at hshare
      exact absurd hshare (by simp)
  have hinvalidate : ∀ cr ∈ Conflicts bc [cb] beta, Invalidates cr :=
    all_conflicts_invalidatable Authored bc sk c beta _ hnull hunforge hnever hsub
  exact spendability_ii_from_conflicts Authored bc sk c v fee O cb beta pkM rhoM
    hcbpairs hcbtxid hbeta hcoin hinvalidate hbalance

/-- **Multi-input spendability.** A single author may spend any finite list of
    owned coins, including coins controlled by different keys.  The only C2
    side condition is that the resulting canonical nullifiers are distinct. -/
theorem spendability_many (Authored : HashType → Blob → Prop)
    (bc : Blockchain) (inputs : List InputWitness) (fee : Nat)
    (O : List Output) (pkM rhoM : HashType)
    (hcoins : ∀ w ∈ inputs, CoinValid w.chain w.cid w.amount (H w.sk))
    (hinchain : ∀ w ∈ inputs, w.chain <+: bc)
    (hnullNodup : (inputs.map (fun w ↦ nullifier w.sk w.cid)).Nodup)
    (hunforge : ∀ w ∈ inputs,
      DegrieferUnforgeable Authored bc w.sk w.cid)
    (hnever : ∀ w ∈ inputs,
      NeverAuthorisedSpend Authored bc w.sk w.cid)
    (hbalance : (inputs.map InputWitness.amount).sum =
      fee + (O.map Output.amount).sum) :
    ∃ (cb beta : Blob),
      beta = spendBlobMany inputs O (spendConflictListMany bc cb inputs O) ∧
      WholeBlockValid (bc ++ [[cb, beta]]) [cb, beta] ∧
      (∀ (j : Nat) (o : Output), O[j]? = some o →
        CoinValid (bc ++ [[cb, beta]]) (cid beta.txid j o.rho) o.amount o.pk) := by
  let cb := coinbaseBlob bc.length fee pkM rhoM
  let beta := spendBlobMany inputs O (spendConflictListMany bc cb inputs O)
  have hcbpairs : cb.pairs = [] := by simp [cb]
  have hnull : nullifiersOf beta =
      inputs.map (fun w ↦ nullifier w.sk w.cid) := by simp [beta]
  have hsub : ∀ γ ∈ priorBlobs bc [cb], sharesNullifier γ beta = true →
      OnChain bc γ := by
    intro γ hγ hshare
    rcases List.mem_append.mp hγ with hg | hg
    · exact onChain_of_mem_flatten hg
    · have hgc : γ = cb := by simpa using hg
      subst hgc
      rw [not_sharesNullifier_of_pairs_nil hcbpairs] at hshare
      simp at hshare
  have hinvalidate : ∀ cr ∈ Conflicts bc [cb] beta, Invalidates cr :=
    all_conflicts_invalidatable_many Authored bc inputs O beta _ hnull
      hunforge hnever hsub
  exact ⟨cb, beta, rfl, spendability_many_from_conflicts bc inputs fee O cb beta pkM rhoM
    hcbpairs (by rfl) (by rfl) hcoins hinchain hnullNodup hinvalidate hbalance⟩

/-- Distinct output positions give distinct coin identifiers, so the outputs of
    one spend cannot collapse onto a single coin — no faerie-gold collisions. -/
theorem output_cids_injective (t : HashType) (j j' : Nat) (r r' : HashType)
    (h : j ≠ j') : cid t j r ≠ cid t j' r' := fun hc => h (cid_inj hc).2.1

end CompactShieldedCSV
\end{lstlisting}
\begin{lstlisting}[style=leanproofs,title=Payment progress]
import Src.CompactCSV.«05_Spendability»

set_option linter.style.longLine false

/-! # Theorem 4 — Payment liveness (output spendability)

Theorem 3 builds the intended blob; fairness eventually puts that raw blob on
the chain.  Inclusion and acceptance are deliberately separate: miners may
include stale, conflicting, or otherwise invalid blobs.

The second half of Theorem 4 is the no-limbo dichotomy: a coin the wallet holds
a proof for is either already spent by that wallet or spendable right now. -/

namespace CompactShieldedCSV

(*@\hypertarget{lean:fair-spend-opportunity}{}@*)
/-- **Fair spend opportunity.** On a valid current history, a proof anchored
    in that history and matching its current conflict list has an accepted
    extension opportunity. This finite-chain model states existence of an
    extension; it does not model a scheduler or a bound on inclusion time. -/
def FairSpendOpportunity (bc : Blockchain) : Prop :=
  ChainValid bc ∧
  ∀ (beta : Blob) (intendedK : List HashType) (anchor : Blockchain) (fee : Nat),
    anchor <+: bc → intendedK = Conflicts bc [] beta →
    MempoolValid anchor (bcid beta intendedK) fee →
    (nullifiersOf beta).Nodup →
    ∃ (future : Blockchain) (h i : Nat) (blk : Block),
      ChainValid future ∧ bc <+: future ∧
      future[h]? = some blk ∧ blk[i]? = some beta ∧ i ≠ 0 ∧
      anchor <+: future.take (h + 1) ∧
      Conflicts (future.take h) (blk.take i) beta = intendedK

(*@\hypertarget{lean:fair-spend-accepted}{}@*)
/-- A fair spend opportunity promotes the eligible mempool proof to an
    accepted occurrence on the future chain. -/
theorem FairSpendOpportunity.accepted {bc : Blockchain}
    (hfair : FairSpendOpportunity bc)
    (beta : Blob) (intendedK : List HashType) (anchor : Blockchain) (fee : Nat)
    (hcompatible : anchor <+: bc) (hcurrent : intendedK = Conflicts bc [] beta)
    (hmempool : MempoolValid anchor (bcid beta intendedK) fee)
    (hwellformed : (nullifiersOf beta).Nodup) :
    ∃ (future : Blockchain) (h i : Nat),
      ChainValid future ∧ bc <+: future ∧ AcceptedAt future h i beta := by
  obtain ⟨future, h, i, blk, hchain, hpre, hblk, hbeta, hi,
      hanchor, hconflicts⟩ :=
    hfair.2 beta intendedK anchor fee hcompatible hcurrent hmempool hwellformed
  have hh : h < future.length := (List.getElem?_eq_some_iff.mp hblk).1
  have htip : (future.take (h + 1)).getLast? = some blk := by
    rw [List.getLast?_eq_getElem?]
    have hlen : (future.take (h + 1)).length = h + 1 := by
      rw [List.length_take]
      omega
    rw [hlen]
    simp only [Nat.add_sub_cancel]
    rw [List.getElem?_take]
    simp only [Nat.lt_succ_self, if_pos]
    exact hblk
  have hdrop : (future.take (h + 1)).dropLast = future.take h := by
    rw [List.dropLast_eq_take]
    have hlen : (future.take (h + 1)).length = h + 1 := by
      rw [List.length_take]
      omega
    rw [hlen]
    simp only [Nat.add_sub_cancel, List.take_take]
    congr 1
    omega
  refine ⟨future, h, i, hchain, hpre, blk, hblk, hbeta, ?_⟩
  have hmempoolActual : MempoolValid anchor
      (bcid beta (Conflicts (future.take h) (blk.take i) beta)) fee := by
    rw [hconflicts]
    exact hmempool
  have hbcid : bcid beta (Conflicts (future.take h) (blk.take i) beta) =
      bcidAt (future.take (h + 1)).dropLast blk i beta := by
    rw [hdrop]
    rfl
  exact BlobMsgValid.regular anchor fee i blk beta hi htip hbeta hbcid
    hanchor hmempoolActual

(*@\hypertarget{lean:fair-spend-of-chain-valid}{}@*)
/-- The extension condition is satisfiable on every valid history: inclusion
    immediately after a pairless coinbase witnesses it. -/
theorem FairSpendOpportunity.of_chainValid {bc : Blockchain} (hc : ChainValid bc) :
    FairSpendOpportunity bc := by
  refine ⟨hc, ?_⟩
  intro beta K anchor fee hanchor hcurrent _hm hnodup
  let cb : Blob := ⟨H ([] : List HashType), []⟩
  refine ⟨bc ++ [[cb, beta]], bc.length, 1, [cb, beta], ?_,
    List.prefix_append _ _, by simp, by simp, one_ne_zero, ?_, ?_⟩
  · intro j hj
    by_cases hold : j < bc.length
    · simpa [BlockValid, List.getElem_append_left hold] using hc j hold
    · have heq : j = bc.length := by simp only [List.length_append, List.length_singleton] at hj; omega
      subst j
      simp only [List.getElem_append_right (by omega : bc.length ≤ bc.length),
        Nat.sub_self, List.getElem_cons_zero]
      refine ⟨by simp, ?_, ?_⟩
      · intro b hb
        simp only [List.mem_cons, List.not_mem_nil, or_false] at hb
        rcases hb with rfl | rfl
        · simp [cb, nullifiersOf]
        · exact hnodup
      · intro b hb
        simpa [cb] using congrArg (fun x => x.map Blob.pairs) hb
  · have hlen : bc.length + 1 = (bc ++ [[cb, beta]]).length := by simp
    rw [hlen, List.take_length]
    exact hanchor.trans (List.prefix_append bc [[cb, beta]])
  · simpa using (Conflicts_pairless_prefix bc cb beta rfl).trans hcurrent.symm

/-- The wallet has an on-chain blob of its own spending this coin. -/
def AuthoredSpendOnChain (Authored : HashType → Blob → Prop)
    (bc : Blockchain) (sk c : HashType) : Prop :=
  ∃ beta, OnChain bc beta ∧ nullifier sk c ∈ nullifiersOf beta ∧ Authored sk beta

-- ---------------------------------------------------------------------------
-- Theorem 4, first half.
-- ---------------------------------------------------------------------------

(*@\hypertarget{lean:payment-liveness-many}{}@*)
/-- Multi-input accepted payment liveness.  The nonempty premise excludes the
    degenerate zero-input regular blob, whose empty pair list is intentionally
    indistinguishable from the pairless branch at the generic extraction
    boundary. -/
theorem payment_liveness_many (Authored : HashType → Blob → Prop)
    (bc : Blockchain) (inputs : List InputWitness) (fee : Nat)
    (O : List Output) (pkM rhoM : HashType)
    (hinputs : inputs ≠ [])
    (hcoins : ∀ w ∈ inputs, CoinValid w.chain w.cid w.amount (H w.sk))
    (hinchain : ∀ w ∈ inputs, w.chain <+: bc)
    (hnullNodup : (inputs.map (fun w ↦ nullifier w.sk w.cid)).Nodup)
    (hunforge : ∀ w ∈ inputs,
      DegrieferUnforgeable Authored bc w.sk w.cid)
    (hnever : ∀ w ∈ inputs,
      NeverAuthorisedSpend Authored bc w.sk w.cid)
    (hbalance : (inputs.map InputWitness.amount).sum =
      fee + (O.map Output.amount).sum)
    (hfair : FairSpendOpportunity bc) :
    ∃ (cb beta : Blob) (future : Blockchain) (h i : Nat),
      WholeBlockValid (bc ++ [[cb, beta]]) [cb, beta] ∧
      ChainValid future ∧ bc <+: future ∧ AcceptedAt future h i beta ∧
      (∀ (j : Nat) (o : Output), O[j]? = some o →
        CoinValid future (cid beta.txid j o.rho) o.amount o.pk) ∧
      (∀ (j j' : Nat) (r r' : HashType), j ≠ j' →
        cid beta.txid j r ≠ cid beta.txid j' r') := by
  obtain ⟨cb, beta, hbeta, hwhole, houtputs⟩ := spendability_many Authored bc inputs fee O
    pkM rhoM hcoins hinchain hnullNodup hunforge hnever hbalance
  have hblob : BlobMsgValid (bc ++ [[cb, beta]])
      (bcid beta (Conflicts bc [cb] beta)) := by
    simpa [bcidAt] using hwhole.2 1 beta (by simp)
  have hpairs : beta.pairs ≠ [] := by
    intro hempty
    apply hinputs
    rw [hbeta] at hempty
    simpa [spendBlobMany, regularBlob, canonicalPairs] using hempty
  obtain ⟨anchor, includedFee, hmempool⟩ :=
    (blob_mempool_or_pairless hblob).resolve_right hpairs
  have hK : spendConflictListMany bc cb inputs O = Conflicts bc [cb] beta := by
    unfold spendConflictListMany Conflicts
    apply conflicts_congr_nullifiers
    rw [hbeta]
    simp
  have hmempoolCurrent : MempoolValid bc (bcid beta (Conflicts bc [cb] beta)) fee := by
    refine MempoolValid.mk
      (spendTxMany inputs O (spendConflictListMany bc cb inputs O))
      inputs (Conflicts bc [cb] beta) rfl hcoins hinchain ?_ ?_
      (mempool_invalidations hmempool).1 ?_
    · simp [spendTxMany, hK]
    · rw [hbeta]; rfl
    · simpa [spendTxMany] using hbalance
  have hcbpairs : cb.pairs = [] := hwhole.1.2.2 cb rfl
  obtain ⟨future, h, i, hchain, hpre, haccept⟩ :=
    hfair.accepted beta (Conflicts bc [cb] beta) bc fee (List.prefix_refl _)
      (Conflicts_pairless_prefix bc cb beta hcbpairs) hmempoolCurrent
      (hwhole.1.2.1 beta (by simp))
  have hfutureOutputs : ∀ (j : Nat) (o : Output), O[j]? = some o →
      CoinValid future (cid beta.txid j o.rho) o.amount o.pk := by
    intro j o hj
    obtain ⟨blk, _hblk, _hbeta, hvalid⟩ := haccept
    apply CoinValid.mono (List.take_prefix _ _)
    refine CoinValid.incl beta.txid j o.rho O (H beta.pairs)
      (H (Conflicts (future.take h) (blk.take i) beta))
      (bcid beta (Conflicts (future.take h) (blk.take i) beta)) ?_ ?_ rfl rfl hvalid
    · rcases o with ⟨amount, pk, rho⟩
      exact hj
    · exact Or.inr ⟨inputs.map (fun w ↦ nullifier w.sk w.cid),
        H (spendConflictListMany bc cb inputs O), by
          rw [hbeta]
          rfl⟩
  exact ⟨cb, beta, future, h, i, hwhole, hchain, hpre, haccept, hfutureOutputs,
    fun j j' r r' hne ↦ output_cids_injective _ j j' r r' hne⟩

(*@\hypertarget{lean:payment-liveness}{}@*)
/-- **Theorem 4 (payment liveness).** The one-input paper statement is
    the singleton instance of accepted multi-input payment liveness. -/
theorem payment_liveness (Authored : HashType → Blob → Prop)
    (bc : Blockchain) (sk c : HashType) (v fee : Nat) (O : List Output)
    (pkM rhoM : HashType)
    (hcoin : CoinValid bc c v (H sk))
    (hunforge : DegrieferUnforgeable Authored bc sk c)
    (hnever : NeverAuthorisedSpend Authored bc sk c)
    (hbalance : v = fee + (O.map (·.amount)).sum)
    (hfair : FairSpendOpportunity bc) :
    ∃ (cb beta : Blob) (future : Blockchain) (h i : Nat),
      WholeBlockValid (bc ++ [[cb, beta]]) [cb, beta] ∧
      ChainValid future ∧ bc <+: future ∧ AcceptedAt future h i beta ∧
      (∀ (j : Nat) (o : Output), O[j]? = some o →
         CoinValid future (cid beta.txid j o.rho) o.amount o.pk) ∧
      (∀ (j j' : Nat) (r r' : HashType), j ≠ j' →
         cid beta.txid j r ≠ cid beta.txid j' r') := by
  apply payment_liveness_many Authored bc (spendInputs sk c v bc) fee O pkM rhoM
  · simp [spendInputs]
  · intro w hw
    simp only [spendInputs, List.mem_singleton] at hw
    subst w
    exact hcoin
  · intro w hw
    simp only [spendInputs, List.mem_singleton] at hw
    subst w
    exact List.prefix_refl _
  · simp [spendInputs]
  · intro w hw
    simp only [spendInputs, List.mem_singleton] at hw
    subst w
    exact hunforge
  · intro w hw
    simp only [spendInputs, List.mem_singleton] at hw
    subst w
    exact hnever
  · simpa [spendInputs] using hbalance
  · exact hfair

-- ---------------------------------------------------------------------------
-- Theorem 4, second half — the no-limbo dichotomy.
-- ---------------------------------------------------------------------------

/-- **Theorem 4 (output spendability).**  Every coin for which the wallet holds
    a proof is either already spent on-chain by that wallet, or spendable at the
    present tip — there is no third state in which the coin is stuck.

    The second disjunct is Theorem 3(ii) applied to the case where no
    wallet-authored spend is on the chain: whatever invalid blobs an adversary
    publishes naming this nullifier, `lemma1_ii` invalidates each of them, so the
    cost a griefer imposes on the holder is a proof, not a coin. -/
theorem coin_spent_or_spendable (Authored : HashType → Blob → Prop)
    (bc : Blockchain) (sk c : HashType) (v fee : Nat) (O : List Output)
    (pkM rhoM : HashType)
    (hcoin : CoinValid bc c v (H sk))
    (hunforge : DegrieferUnforgeable Authored bc sk c)
    (hbalance : v = fee + (O.map (·.amount)).sum) :
    AuthoredSpendOnChain Authored bc sk c
    ∨ ∃ (cb beta : Blob),
        WholeBlockValid (bc ++ [[cb, beta]]) [cb, beta] ∧
        (∀ (j : Nat) (o : Output), O[j]? = some o →
           CoinValid (bc ++ [[cb, beta]]) (cid beta.txid j o.rho) o.amount o.pk) := by
  by_cases hsp : AuthoredSpendOnChain Authored bc sk c
  · exact Or.inl hsp
  · right
    have hnever : NeverAuthorisedSpend Authored bc sk c := by
      intro beta honchain hmem hauth
      exact hsp ⟨beta, honchain, hmem, hauth⟩
    exact ⟨coinbaseBlob bc.length fee pkM rhoM,
      spendBlob sk c v bc O
        (spendConflictList bc (coinbaseBlob bc.length fee pkM rhoM) sk c v O),
      spendability_ii Authored bc sk c v fee O
        (coinbaseBlob bc.length fee pkM rhoM)
        (spendBlob sk c v bc O
          (spendConflictList bc (coinbaseBlob bc.length fee pkM rhoM) sk c v O)) pkM rhoM
        rfl rfl rfl hcoin hunforge hnever hbalance⟩

/-- Combining with fair spend opportunity: absent a wallet-authored spend
    already on the chain, the freshly constructed payment is accepted on a
    valid future extension. -/
theorem coin_spent_or_paid (Authored : HashType → Blob → Prop)
    (bc : Blockchain) (sk c : HashType) (v fee : Nat) (O : List Output)
    (pkM rhoM : HashType)
    (hcoin : CoinValid bc c v (H sk))
    (hunforge : DegrieferUnforgeable Authored bc sk c)
    (hbalance : v = fee + (O.map (·.amount)).sum)
    (hfair : FairSpendOpportunity bc) :
    AuthoredSpendOnChain Authored bc sk c
    ∨ ∃ (beta : Blob) (future : Blockchain) (h i : Nat),
        ChainValid future ∧ bc <+: future ∧ AcceptedAt future h i beta ∧
        (∀ (j : Nat) (o : Output), O[j]? = some o →
           CoinValid future
             (cid beta.txid j o.rho) o.amount o.pk) := by
  by_cases hsp : AuthoredSpendOnChain Authored bc sk c
  · exact Or.inl hsp
  · right
    have hnever : NeverAuthorisedSpend Authored bc sk c := by
      intro beta honchain hmem hauth
      exact hsp ⟨beta, honchain, hmem, hauth⟩
    obtain ⟨cb, beta, future, h, i, _hwhole, hchain, hpre, haccept, hcoins, _⟩ :=
      payment_liveness Authored bc sk c v fee O pkM rhoM hcoin hunforge hnever hbalance hfair
    exact ⟨beta, future, h, i, hchain, hpre, haccept, hcoins⟩

/-- For a recipient holding the corresponding secret key, distinct output
    positions also give distinct *nullifiers* — the faerie-gold exclusion of the
    closing remark of Theorem 4. -/
theorem output_nullifiers_injective (t sk : HashType) (j j' : Nat) (r r' : HashType)
    (h : j ≠ j') : nullifier sk (cid t j r) ≠ nullifier sk (cid t j' r') := by
  intro hc
  exact output_cids_injective t j j' r r' h (nullifier_inj hc).2

end CompactShieldedCSV
\end{lstlisting}
\begin{lstlisting}[style=leanproofs,title=Honest-wallet progress]
import Src.CompactCSV.«06_Liveness»

set_option linter.style.longLine false

/-! # Honest-wallet spendability

The result states that a wallet-held coin was validly spent or can be spent
now.

The transaction identifier commits to the conflict list expected by the
sender. Consequently an honestly authored occurrence has only two outcomes:
it is accepted at the list it committed to, or its contextual identifier has
an invalidation witness because the chain assigned it a different list. This
therefore permits replacement transactions with any nullifier set. -/

namespace CompactShieldedCSV

/-- A successful protocol spend is an accepted blob proof on a chain whose
    public block rules hold. -/
def ValidlyAcceptedAt (bc : Blockchain) (h i : Nat) (beta : Blob) : Prop :=
  ChainValid bc ∧ AcceptedAt bc h i beta

/-- `Wallet sk c` means that this wallet owns coin `c` under key `sk`. -/
abbrev WalletCoins := HashType → HashType → Prop

/-- Degriefer unforgeability for every coin held by the wallet. -/
def WalletUnforgeable (Authored : HashType → Blob → Prop)
    (Wallet : WalletCoins) (bc : Blockchain) : Prop :=
  ∀ sk c, Wallet sk c → DegrieferUnforgeable Authored bc sk c

lemma take_succ_getLast {α : Type} {l : List α} {b : α} {i : Nat}
    (h : l[i]? = some b) : (l.take (i + 1)).getLast? = some b := by
  have hi : i < l.length := (List.getElem?_eq_some_iff.mp h).1
  rw [List.getLast?_eq_getElem?]
  have hlen : (l.take (i + 1)).length = i + 1 := by rw [List.length_take]; omega
  rw [hlen]
  simp only [Nat.add_sub_cancel]
  rw [List.getElem?_take]
  simp only [Nat.lt_succ_self, if_pos]
  exact h

lemma dropLast_take_succ {α : Type} {l : List α} {n : Nat} (h : n < l.length) :
    (l.take (n + 1)).dropLast = l.take n := by
  have hlen : (l.take (n + 1)).length = n + 1 := by rw [List.length_take]; omega
  rw [List.dropLast_eq_take, hlen]
  simp only [Nat.add_sub_cancel, List.take_take]
  congr 1
  omega

/-- Honest `Build` behaviour. An authorised transaction identifier comes with
    the wallet's original blob and mempool proof. The observed blob may be an
    altered copy; neither its validity nor its invalidation is assumed. -/
def HonestAuthored (Authored : HashType → Blob → Prop)
    (Wallet : WalletCoins) (bc : Blockchain) : Prop :=
  ∀ (sk c : HashType) (beta : Blob) (K : List HashType),
    Wallet sk c →
    nullifier sk c ∈ nullifiersOf beta →
    (beta, K) ∈ recordHistory bc.flatten →
    Authored sk beta →
    ∃ (h i : Nat) (blk : Block) (intendedK : List HashType)
      (anchor : Blockchain) (fee : Nat) (original : Blob),
      bc[h]? = some blk ∧ blk[i]? = some beta ∧
      K = Conflicts (bc.take h) (blk.take i) beta ∧
      i ≠ 0 ∧ anchor <+: bc.take (h + 1) ∧
      beta.txid = original.txid ∧
      MempoolValid anchor (bcid original intendedK) fee

(*@\hypertarget{lean:honest-accepted-or-invalidates}{}@*)
lemma HonestAuthored.accepted_or_invalidates
    {Authored : HashType → Blob → Prop} {Wallet : WalletCoins} {bc : Blockchain}
    (hhonest : HonestAuthored Authored Wallet bc)
    (sk c : HashType) (beta : Blob) (K : List HashType)
    (hwallet : Wallet sk c) (hn : nullifier sk c ∈ nullifiersOf beta)
    (hrecord : (beta, K) ∈ recordHistory bc.flatten) (hauth : Authored sk beta) :
    (∃ h i, AcceptedAt bc h i beta) ∨ Invalidates (bcid beta K) := by
  obtain ⟨h, i, blk, intendedK, anchor, fee, original, hblk, hbeta,
    hactual, hi, hanchor, htxid, hm⟩ := hhonest sk c beta K hwallet hn hrecord hauth
  rcases mempool_or_invalidates_of_same_txid (K := K) hm htxid with hm | hinv
  · left
    have hh : h < bc.length := (List.getElem?_eq_some_iff.mp hblk).1
    have hb : bcid beta K = bcidAt (bc.take (h + 1)).dropLast blk i beta := by
      rw [dropLast_take_succ hh, hactual]
      rfl
    have hvalid := BlobMsgValid.regular anchor fee i blk beta hi
      (take_succ_getLast hblk) hbeta hb hanchor hm
    exact ⟨h, i, blk, hblk, hbeta, by simpa only [hactual] using hvalid⟩
  · exact Or.inr hinv

(*@\hypertarget{lean:spendability-honest}{}@*)
/-- **Spendability (honest-wallet form).** A valid coin held by an honest
    wallet is either already validly spent by that wallet or can form a valid
    spend now. No restriction on overlap between successive nullifier sets is
    required. -/
theorem spendability_honest (Authored : HashType → Blob → Prop)
    (Wallet : WalletCoins) (bc : Blockchain) (sk c : HashType)
    (v fee : Nat) (O : List Output) (pkM rhoM : HashType)
    (hwallet : Wallet sk c)
    (hcoin : CoinValid bc c v (H sk))
    (hchain : ChainValid bc)
    (hunforge : WalletUnforgeable Authored Wallet bc)
    (hhonest : HonestAuthored Authored Wallet bc)
    (hbalance : v = fee + (O.map (·.amount)).sum) :
    (∃ (h i : Nat) (beta : Blob),
        ValidlyAcceptedAt bc h i beta ∧ Authored sk beta ∧
        nullifier sk c ∈ nullifiersOf beta)
    ∨ (∃ (cb beta : Blob),
        WholeBlockValid (bc ++ [[cb, beta]]) [cb, beta] ∧
        (∀ (j : Nat) (o : Output), O[j]? = some o →
           CoinValid (bc ++ [[cb, beta]]) (cid beta.txid j o.rho) o.amount o.pk)) := by
  classical
  by_cases hex : ∃ (h i : Nat) (beta : Blob),
      AcceptedAt bc h i beta ∧ Authored sk beta ∧
      nullifier sk c ∈ nullifiersOf beta
  · left
    obtain ⟨h, i, beta, haccept, hauth, hnull⟩ := hex
    exact ⟨h, i, beta, ⟨hchain, haccept⟩, hauth, hnull⟩
  · right
    let cb := coinbaseBlob bc.length fee pkM rhoM
    let beta := spendBlob sk c v bc O (spendConflictList bc cb sk c v O)
    have hcbpairs : cb.pairs = [] := by simp [cb]
    have hnull : nullifiersOf beta = [nullifier sk c] := by simp [beta]
    have hinvalidate : ∀ cr ∈ Conflicts bc [cb] beta, Invalidates cr := by
      intro cr hcr
      obtain ⟨gamma, K, hrecord, hshare, rfl⟩ := mem_conflicts_iff.mp hcr
      have hn : nullifier sk c ∈ nullifiersOf gamma :=
        mem_nullifiersOf_of_shares hnull hshare
      change (gamma, K) ∈ recordHistory (bc.flatten ++ [cb]) at hrecord
      rw [recordHistory_snoc] at hrecord
      rcases List.mem_append.mp hrecord with hchainRecord | hcoinbaseRecord
      · by_cases hauth : Authored sk gamma
        · rcases hhonest.accepted_or_invalidates sk c gamma K
              hwallet hn hchainRecord hauth with ⟨h, i, haccept⟩ | hinv
          · exact False.elim (hex ⟨h, i, gamma, haccept, hauth, hn⟩)
          · exact hinv
        · have honchain : OnChain bc gamma :=
            onChain_of_mem_flatten (blob_mem_of_mem_recordHistory hchainRecord)
          exact lemma1_ii Authored bc sk c (hunforge sk c hwallet)
            gamma K honchain hn hauth
      · have hgamma : gamma = cb := by
          have hp : gamma = cb ∧ K = conflicts bc.flatten cb := by
            simpa only [List.mem_singleton, Prod.mk.injEq] using hcoinbaseRecord
          exact hp.1
        subst gamma
        rw [not_sharesNullifier_of_pairs_nil hcbpairs] at hshare
        simp at hshare
    refine ⟨cb, beta, ?_⟩
    exact spendability_ii_from_conflicts Authored bc sk c v fee O cb beta pkM rhoM
      hcbpairs (by rfl) (by rfl) hcoin hinvalidate hbalance

(*@\hypertarget{lean:wallet-multi-input-progress}{}@*)
/-- **Wallet multi-input progress.** Either one of the selected
    inputs has already been consumed by an accepted wallet-authorised spend,
    or the wallet can spend the entire selected list atomically.  Authored but
    contextually invalid on-chain attempts do not block the second outcome. -/
theorem wallet_multi_input_progress (Authored : HashType → Blob → Prop)
    (Wallet : WalletCoins) (bc : Blockchain) (inputs : List InputWitness)
    (fee : Nat) (O : List Output) (pkM rhoM : HashType)
    (hwallet : ∀ w ∈ inputs, Wallet w.sk w.cid)
    (hcoins : ∀ w ∈ inputs, CoinValid w.chain w.cid w.amount (H w.sk))
    (hinchain : ∀ w ∈ inputs, w.chain <+: bc)
    (hnullNodup : (inputs.map (fun w ↦ nullifier w.sk w.cid)).Nodup)
    (hchain : ChainValid bc)
    (hunforge : WalletUnforgeable Authored Wallet bc)
    (hhonest : HonestAuthored Authored Wallet bc)
    (hbalance : (inputs.map InputWitness.amount).sum =
      fee + (O.map Output.amount).sum) :
    (∃ (w : InputWitness), w ∈ inputs ∧
      ∃ (h i : Nat) (beta : Blob),
        ValidlyAcceptedAt bc h i beta ∧ Authored w.sk beta ∧
        nullifier w.sk w.cid ∈ nullifiersOf beta)
    ∨ (∃ (cb beta : Blob),
      WholeBlockValid (bc ++ [[cb, beta]]) [cb, beta] ∧
      (∀ (j : Nat) (o : Output), O[j]? = some o →
        CoinValid (bc ++ [[cb, beta]])
          (cid beta.txid j o.rho) o.amount o.pk)) := by
  classical
  by_cases hex : ∃ (w : InputWitness), w ∈ inputs ∧
      ∃ (h i : Nat) (beta : Blob),
        AcceptedAt bc h i beta ∧ Authored w.sk beta ∧
        nullifier w.sk w.cid ∈ nullifiersOf beta
  · left
    obtain ⟨w, hw, h, i, beta, haccept, hauth, hnull⟩ := hex
    exact ⟨w, hw, h, i, beta, ⟨hchain, haccept⟩, hauth, hnull⟩
  · right
    let cb := coinbaseBlob bc.length fee pkM rhoM
    let beta := spendBlobMany inputs O (spendConflictListMany bc cb inputs O)
    have hcbpairs : cb.pairs = [] := by simp [cb]
    have hnull : nullifiersOf beta =
        inputs.map (fun w ↦ nullifier w.sk w.cid) := by simp [beta]
    have hinvalidate : ∀ cr ∈ Conflicts bc [cb] beta, Invalidates cr := by
      intro cr hcr
      obtain ⟨gamma, K, hrecord, hshare, rfl⟩ := mem_conflicts_iff.mp hcr
      have hshare' : sharesNullifier gamma (spendBlobMany inputs O []) = true := by
        have heq : sharesNullifier gamma beta =
            sharesNullifier gamma (spendBlobMany inputs O []) := by
          simp only [sharesNullifier]
          rw [hnull, nullifiersOf_spendBlobMany]
        rw [← heq]
        exact hshare
      obtain ⟨w, hw, hn⟩ := input_of_shares_spendBlobMany hshare'
      change (gamma, K) ∈ recordHistory (bc.flatten ++ [cb]) at hrecord
      rw [recordHistory_snoc] at hrecord
      rcases List.mem_append.mp hrecord with hchainRecord | hcoinbaseRecord
      · by_cases hauth : Authored w.sk gamma
        · rcases hhonest.accepted_or_invalidates w.sk w.cid gamma K
              (hwallet w hw) hn hchainRecord hauth with ⟨h, i, haccept⟩ | hinv
          · exact False.elim (hex ⟨w, hw, h, i, gamma, haccept, hauth, hn⟩)
          · exact hinv
        · have honchain : OnChain bc gamma :=
            onChain_of_mem_flatten (blob_mem_of_mem_recordHistory hchainRecord)
          exact lemma1_ii Authored bc w.sk w.cid
            (hunforge w.sk w.cid (hwallet w hw)) gamma K honchain hn hauth
      · have hgamma : gamma = cb := by
          have hp : gamma = cb ∧ K = conflicts bc.flatten cb := by
            simpa only [List.mem_singleton, Prod.mk.injEq] using hcoinbaseRecord
          exact hp.1
        subst gamma
        rw [not_sharesNullifier_of_pairs_nil hcbpairs] at hshare
        simp at hshare
    refine ⟨cb, beta, ?_⟩
    exact spendability_many_from_conflicts bc inputs fee O cb beta pkM rhoM
      hcbpairs (by rfl) (by rfl) hcoins hinchain hnullNodup hinvalidate hbalance
end CompactShieldedCSV
\end{lstlisting}

\end{document}